\documentclass[11pt]{article}
\pdfoutput=1
\usepackage[utf8x]{inputenc}	
\usepackage[english]{babel}

\usepackage{placeins}

\usepackage{amssymb}
\usepackage{amsthm}
\usepackage{MnSymbol}
\usepackage{mathtools}		
\usepackage{dsfont}		
\usepackage{stmaryrd}		
\usepackage{cite}		
\usepackage[svgnames]{xcolor}
\usepackage{enumerate}
\usepackage{setspace}
\usepackage{booktabs}
\usepackage{longtable}

\usepackage{graphics}
\usepackage{todonotes}
\usepackage{paralist}
\usepackage[font={small,it},labelfont=bf]{caption} 
\usepackage{subcaption}
\usepackage[pdftex,breaklinks,colorlinks=true,urlcolor=RoyalBlue,linkcolor=blue,
citecolor=blue]{hyperref}		
\usepackage[a4paper,lmargin=2cm,rmargin=2cm,tmargin=2cm,bmargin=2cm]{geometry}

\usepackage{pdfpages}
\usepackage{graphicx}

\usepackage[ddmmyyyy,hhmmss]{datetime}

\usepackage{multirow}

\catcode`@=11
\renewcommand{\section}{\@startsection{section}{1}{0pt}{\medskipamount}
{\medskipamount}{\Large\bf}}
\numberwithin{equation}{section}
\catcode`@=12

\def\rank{\mathrm{rk}}

\def\dim{\mathrm{dim}}

\newcommand{\diff}{\mathrm{d}}
\newcommand{\HS}{\mathrm{HS}}
\newcommand{\HWG}{\mathrm{HWG}}

\newcommand{\PE}{\mathrm{PE}}
\newcommand{\PL}{\mathrm{PL}}

\newcommand{\C}{\mathbb{C}}
\newcommand{\R}{\mathbb{R}}

\newcommand{\HH}{\mathbb{H}}
  
\newcommand{\Z}{\mathbb{Z}}
\newcommand{\D}{\mathbb{D}}

\newcommand{\Coulomb}{\mathcal{C}}

\newcommand{\Higgs}{\mathcal{H}}

\newcommand{\orbit}[1]{\mathcal{O}_{#1}}
\newcommand{\clorbit}[1]{\overline{\mathcal{O}}_{#1}}
\newcommand{\height}[1]{\text{ht}(#1)}
\newcommand{\slice}[1]{\mathcal{S}_{#1}}

\newcommand{\Ncal}{\mathcal{N}}

\newcommand{\gfrak}{\mathfrak{g}}

\newcommand{\uo}{{ \mathrm{U}(1)}}

\newcommand{\urm}{{{\rm U}}}
\newcommand{\urmL}{{{\mathfrak u}}}
\newcommand{\surm}{{{\rm SU}}}
\newcommand{\surmL}{{{\mathfrak{su}}}}
\newcommand{\sorm}{{{\rm SO}}}
\newcommand{\orm}{{{\rm O}}}
\newcommand{\sormL}{{{\mathfrak{so}}}}

\newcommand{\sprm}{{{\rm Sp}}}
\newcommand{\sprmL}{{{\mathfrak{sp}}}}
\newcommand{\usprm}{{{\rm USp}}}

\newcommand{\Spin}{{\rm Spin}}

\newcommand{\Hh}{\mathrm{H}}

\newcommand{\GNOfrak}{\widehat{\mathfrak{g}}}

\newcommand{\ffour}{\mathfrak{f}_4}
\newcommand{\esix}{\mathfrak{e}_6}

\newcommand{\eeight}{\mathfrak{e}_8}

\newcommand{\gbal}{\mathfrak{g}_{\mathrm{balance}}}

\newtheorem{myStat}{Statement}

\newcommand{\NS}{\text{NS5}}

\newcommand{\Dthree}{\text{D3}}

\newcommand{\Dfive}{\text{D5}}
\newcommand{\Ds}{\text{D6}}
\newcommand{\Os}{\text{O6}}
\newcommand{\Osm}{\text{O6${}^{-}$}}
\newcommand{\Osmt}{\text{$\widetilde{\text{O6}}^{-}$}}
\newcommand{\Osp}{\text{O6${}^{+}$}}
\newcommand{\Ospt}{\text{$\widetilde{\text{O6}}^{+}$}}
\newcommand{\Dseven}{\text{D7}}
\newcommand{\De}{\text{D8}}

\newcommand{\Mf}{\text{M5}}

\newcommand{\magQuiv}{\mathsf{Q}}

\usepackage{xargs}

\newlength{\myline}
\newcommandx*{\doublearrow}[4][1=0, 2=1]{
  \draw[line width=\myline,double distance=3\myline,#3] #4;
}

\newcommandx*{\triplearrow}[4][1=0, 2=1]{
  \draw[line width=\myline,double distance=5\myline,#3] #4;
  \draw[line width=\myline,shorten <=#1\myline,shorten >=#2\myline,#3] #4;
}

\newcommandx*{\quadarrow}[4][1=0, 2=2.5]{
  \draw[line width=\myline,double distance=5\myline,#3] #4;
  \draw[line width=\myline,double distance=\myline,shorten <=#1\myline,shorten 
>=#2\myline,#3] #4;
}

\definecolor{myGreen1}{RGB}{53, 157, 42}
\definecolor{myGreen}{RGB}{54, 150, 45}

\newcommand{\ra}[1]{\renewcommand{\arraystretch}{#1}}
\allowdisplaybreaks

\begin{document}
\begin{titlepage}
\setcounter{page}{0}

\begin{flushright}
Imperial/TP/22/AH/02
\end{flushright}

\vskip 2cm

\begin{center}

{\Large\bf 
Magnetic quivers and negatively charged branes
}

\vspace{15mm}

{\large Amihay Hanany${}^{1}$} and \ 
{\large 
Marcus Sperling${}^{2}$} 
\\[5mm]
\noindent ${}^1${\em Theoretical Physics Group, Imperial College London\\
Prince Consort Road, London, SW7 2AZ, UK}\\
{Email: {\tt a.hanany@imperial.ac.uk}}
\\[5mm]
\noindent ${}^{2}${\em Shing-Tung Yau Center, Southeast University,}\\
{\em Xuanwu District, Nanjing, Jiangsu, 210096, China}\\
Email: {\tt msperling@seu.edu.cn}
\\[5mm]

\vspace{15mm}

\begin{abstract}
 The Higgs branches of the world-volume theories for multiple M5 branes on an $A_k$ or $D_k$-type ALE space are known to host a variety of fascinating properties, such as the small $E_8$ instanton transition or the discrete gauging phenomena. This setup can be further enriched by the inclusion of boundary conditions, which take the form of $\surm(k)$ or $\sorm(2k)$ partitions, respectively. Unlike the $A$-type case, $D$-type boundary conditions are eventually accompanied by negative brane numbers in the Type IIA brane realisation. While this may seem discouraging at first, we demonstrate that these setups are well-suited to analyse the Higgs branches via magnetic quivers. Along the way, we encounter multiple models with previously neglected Higgs branches that exhibit exciting physics and novel geometric realisations.
Nilpotent orbits, Słodowy slices, and symmetric products.
\end{abstract}

\end{center}

\end{titlepage}

{\baselineskip=12pt
{\footnotesize
\tableofcontents
}
}
  \section{Introduction}
The Higgs branch moduli spaces of 6d $\Ncal=(1,0)$ supersymmetric theories are substantially more intricate then is apparent from the tensor branch description. To be specific, focus on the class of $n$ \Mf\ branes on $\R\times\C^2\slash \Gamma_G$, where $\Gamma_G =\Z_k$ for 
$G=\surm(k)$ or  $\D_{k-2}$ for $G=\sorm(2k)$. The $6$d $\Ncal=(1,0)$ theory, denoted by $T_G^n$, has generically $G\times G$ global symmetry. This symmetry can be broken to subgroups by the Higgs mechanism, which is realised by some field that acquires a nilpotent vacuum expectation value (VEV) \cite{Gaiotto:2014lca} labelled by two $G$-partitions $\rho_{L,R}$. One may denote such theories by $T_G^{n}(\rho_L,\rho_R)$, with the convention $T_G^n 
(\rho_\mathrm{trivial},\rho_\mathrm{trivial}) = T_G^n$.
In the dual Type IIA frame \cite{Hanany:1997gh,Brunner:1997gk}, this modification of the $T_G^n$ theories is achieved by introducing \De\ branes that have D6 branes ending on them in a pattern described by $G$-partitions $\rho_{L,R}$.  In the F-theory frame, nilpotent VEVs correspond to residues for poles of Hitchin equations living on \Dseven\ branes \cite{DelZotto:2014hpa}, which is referred to as a special case of T-brane data, see \cite{Cecotti:2010bp,Anderson:2013rka}. An advantage of the F-theory approach is that it allows to study cases without a known Type IIA description, e.g.\ for example \Mf\ branes on $E$-type ALEs-singularities. In \cite{Heckman:2016ssk}, RG-flows between $6$d $\Ncal=(1,0)$ theories related by such nilpotent Higgsings have been considered for Type IIA \Ds -\De -\NS\ brane configurations with or without additional \Os\ orientifold planes. It has been observed that the case of $D$-type singularities can lead to brane systems with negative numbers of \Ds\ branes in between two adjacent \NS\ branes. Since this did not correspond to any conventional scenario, it has not been pursued further. However, the authors of \cite{Mekareeya:2016yal} demonstrated that the Type IIA configurations with negative charge branes can be used to derive consistent results for anomaly coefficients, cf.\ \cite{Hassler:2019eso}; see also \cite{Apruzzi:2017nck} for supergravity duals of such configurations. As a by-product, the  difference of the finite coupling Higgs branch dimensions for theories with different boundary conditions has been computed to be \cite{Mekareeya:2016yal}
\begin{align}
	\dim_\HH\ \Higgs^{6d}\left(  T_G^n 
	(\rho_\mathrm{trivial},\rho_\mathrm{trivial}) \right) - 
	\dim_\HH\ \Higgs^{6d} \left( T_G^n (\rho_L,\rho_R) \right)
	= \dim_\HH\ \clorbit{\rho_L} + \dim_\HH\ \clorbit{\rho_R} \,,
\end{align}
with $\clorbit{\rho}$ a nilpotent orbit closure of $G$, labelled by a partition $\rho$.
The next progress followed by the computation of the Higgs branch dimension at the origin of the tensor branch \cite{Mekareeya:2017sqh}
\begin{align}
	\dim_\HH\ \Higgs_{\infty}^{6d} (T_G^n ( \rho_L,\rho_R) )   = n + \dim\ G 
	-\dim_\HH\ \clorbit{\rho_L} -
	\dim_\HH\ \clorbit{\rho_R} \,.
\end{align}
Nonetheless, besides the jump in dimensions, not much else was known about the infinite coupling Higgs branches.

Recently, the magnetic quiver technique has been successful in providing a host of additional insights on the  Higgs branches of 6d $\Ncal=(1,0)$ theories \cite{Hanany:2018uhm,Hanany:2018vph,Hanany:2018cgo,Cabrera:2019izd,Cabrera:2019dob,Sperling:2021fcf}. For instance, \cite{Hanany:2018uhm} provides the magnetic quivers for $T_G^n$ theories at infinite coupling and conjectures a generalisation to T-brane theories $T_G^n(\rho_L,\rho_R)$. Shortly after, the brane constructions \cite{Cabrera:2019izd,Cabrera:2019dob} allowed to systematically derive the magnetic quivers for $T_G^n$ from brane systems. Besides the magnetic quiver itself, a host of additional information is accessible. Naturally, the magnetic quiver for each distinct tensor branch phase comes equipped with a Hilbert series, called monopole formula \cite{Cremonesi:2013lqa}. In terms of the 6d Higgs branches, this is a generating function for the Higgs branch operator spectrum in a given tensor branch phase. In addition, the phase structure of the Higgs branch is encoded in the phase (Hasse) diagram \cite{Bourget:2019aer}, which can be deduced from the magnetic quiver itself, from the brane system, or from geometric reasoning, depending on the circumstances.

The purpose of this note is to convey two points:
Firstly, brane system with negative numbers of \Ds\ branes should be taken seriously, because they allow to derive a host of consistent results, provide new predictions and new challenges.  For instance, hypermultiplets in spinor representations of $\sorm(n)$ gauge groups appear as well as the exceptional gauge group $G_2$. While this data is deduced from F-theory constructions, it nonetheless represents a useful construction within branes. More to the point, such non-standard matter and exceptional gauge algebras have previously only been constructed via branes in 5 dimensions, using the Higgs mechanism \cite{Hayashi:2018bkd,Hayashi:2019yxj}.

Secondly, the magnetic quiver constructions \cite{Cabrera:2019izd,Cabrera:2019dob} together with the acceptance of negative branes allow to derive all the infinite coupling magnetic quivers for the $T_G^n(\rho_L,\rho_R)$ theories with $G$ of type $A$ or $D$ (provided $\rho_{L,R}$ are special partitions, as explained below).
But most importantly, by focusing on a few explicit boundary conditions one is able to uncover exciting Higgs branch geometries such as nilpotent orbits for the exceptional groups $G_2$, $F_4$, and $E_6$ (the database and computations for nilpotent orbits of exceptional type \cite{Hanany:2017ooe} proves to be extremely useful for this purpose). For these nilpotent orbits, there are a host of tools available now: the brane system, the magnetic quiver techniques, and the phase diagram. In fact, it is an open problem to find a Coulomb branch quiver realisation for exceptional nilpotent orbits beyond height 2. The orbits found here are precisely and excitingly in this missing area.

The main results of this note are as follows:
\begin{compactitem}
\item In Section \ref{sec:SU3+Sp0} the 21 dimensional closure of the nilpotent orbit of $E_6$ with Bala-Carter label $A_2$ is realised as an infinite coupling 6d Higgs branch. We provide the explicit brane realisation, an exact hyper-K\"ahler quotient, the magnetic quivers, and the relation to the geometric Satake correspondence.
\item In Section \ref{sec:Hasse_curves_31} the infinite coupling Higgs branch phase diagram is derived for all 6d $G\times \sprm(0)$ quiver theories supported on $(-3)(-1)$ curves.
\item In Section \ref{sec:2231_curve_F4_orbit} the 20 dimensional closure of the nilpotent orbit of $F_4$ is also found to be an infinite coupling Higgs branch of a 6d theory. We detail the brane system and magnetic quivers; furthermore, we reconstruct the Higgs branch Hasse diagram using physical methods: 6d quivers, brane systems, magnetic quivers, and quiver subtraction.
\end{compactitem}

The remainder of the note is organised as follows: the Type IIA brane configurations dual to M5 branes on $A$ or $D$-type ALE spaces with non-trivial boundary conditions are reviewed in Section \ref{sec:M5_AD-sing}. An appetiser in Section \ref{sec:appetiser} shows how boundary condition can modify the Higgs branches into rich and sometimes under-appreciated geometries. Thereafter, Section \ref{sec:2M5s_RHS_trivial} starts exploring Higgs branches associated to 2 M5 branes on $\C^2 \slash D_6$, with the left boundary non-trivial and the right trivial, and ends with $\C^2 \slash D_4$. Here, a nilpotent orbit of $E_6$ emerges, which inspires the derivation of the phase diagram for a whole class of theories. In Section \ref{sec:3M5s_RHS_trivial} further $\sorm(8)$ boundary conditions are explored by enlarging the number of M5 branes. Here, a nilpotent orbit of $F_4$ emerges. These cases are naturally fitted into families of $\sorm(2k)$ boundary conditions. Subsequently, non-trivial boundaries are allowed on both sides in Sections \ref{sec:non-overlapping_bc} and \ref{sec:overlapping_bc}. Lastly, conclusions are provided in Section \ref{sec:conclusions}.

A number of appendices complement the main body. Appendix \ref{app:background} provides background information on brane configurations and global symmetries.  Appendix \ref{sec:examples} details explicit examples for M5 branes on an $A_3$ and $D_6$ ALE space, respectively.

\paragraph{Notation.}
For magnetic quivers, the global symmetry of the Coulomb branch is denoted by $G_J$, while the global symmetry algebra deduced from the balanced nodes is $\gbal$.
\section{M5 branes on A and D type singularities --- The brane system}
\label{sec:M5_AD-sing}
M5 branes on $A$ or $D$-type ALE singularities $\C^2 \slash \Gamma_{AD}$ admit dual Type IIA brane configurations. M5 branes become NS5 branes, the $\C^2\slash \Z_k$ singularity is dual to $k$ D6 branes filling the transverse space, while $\C^2\slash D_k$ is dual to $k$ full D6 branes on top of an O$6^-$ orientifold plane. The space-time occupations are summarised in Table \ref{tab:directions}. The stack of D6 branes is extended to $\pm \infty$ along the $x^6$ direction. These semi-infinite 6-branes can be terminated at finite $x^6$ position on D8 branes without breaking supersymmetry.
\begin{table}[t]
\centering
\begin{tabular}{c|ccccccccccc}
\toprule
Type IIA & $x^0$ &  $x^1$ & $x^2$ & $x^3$ & $x^4$ & $x^5$ & $x^6$ & $x^7$ & 
$x^8$ & $x^9$  & \\ \midrule 
\NS & $\times$ & $\times$ & $\times$ & $\times$ & $\times$ & $\times$ & & & & 
& \\
 \De & $\times$ & $\times$ & $\times$ & $\times$ & $\times$ & $\times$ & 
&$\times$ & $\times$ &$\times$ &   \\
\Ds, \Os & $\times$  & $\times$ & $\times$ & $\times$ & $\times$ & $\times$ & 
$\times$ &  &  & &   \\
\bottomrule
\end{tabular}
\caption{
Occupation of space-time directions by 
\NS, \De, \Ds, and \Os\ in Type IIA. }
\label{tab:directions}
\end{table}

\subsection{A-type singularity with boundary conditions}
\label{sec:A-type_boundary}
Consider $n$ \NS\ branes and $k$ \Ds\ branes as in Table \ref{tab:directions}. One may enrich the set-up by assigning boundary conditions of the \Ds\ ending on \De\ brane for very large positive and negative $x^6$. These boundary conditions can be cast in the
form of partitions $\rho_{L,R}$ of $k$. This input data determines how many D6 branes end on each D8 brane, see below or Table \ref{tab:SU4_examples} for examples.
Assuming that all \NS\ branes are well separated, one can read off the 
low-energy description in terms of a 6d electric quiver gauge theory 
\cite{Hanany:1997gh,Brunner:1997gk} which is labelled by 
$T_{\surm(k)}^n(\rho_L,\rho_R)$. 

In practice, the configuration is defined as follows (see also \cite{Hanany:1997gh,Cremonesi:2015bld} for a summary): Consider, for instance, $n\geq 6$ M5 branes on an $\C^2 \slash \Z_9$ singularity with boundary conditions $\rho_L= (4,2^2,1)$ and $\rho_R=(1^9)$. For the left boundary, one begins by placing one D8 brane in the first NS5 brane interval (counting from left), two D8 branes in the second interval, and one D8 brane in the fourth interval. Likewise, for the right boundary, one places nine D8 branes on the first NS5 brane interval from the right. Next, the D6 branes are added. In the centre of the configuration, far from the boundaries, the $\C^2\slash \Z_9$ singularity in M-theory dualises to 9 D6 branes with NS5 branes intersecting them. The central part of the brane configuration has vanishing cosmological constant $m$.
Next, consider intervals closer to the left boundary. Passing any of the D8 branes increases the cosmological constant by one unit \cite{Hanany:1997sa,Hanany:1997gh}. This change in cosmological constant affects how many D6 branes can end on the left and right of an NS5 brane.  In general, the difference between the number $\#(\mathrm{D6}_L)$ of D6s ending on the left and the number $\#(\mathrm{D6}_R)$ of D6s ending on the right is set by the value $m$ of the cosmological constant at the position of the NS5 brane. In short,   $m = \pm( \#(\mathrm{D6}_L) - \#(\mathrm{D6}_R))$ and the sign is a matter of convention. Hence, the presence of D8 branes lead to a decrease of D6 branes towards the boundaries. For the example considered, the brane system and the 6d quiver becomes
\begin{align}
    \raisebox{-.5\height}{
    \includegraphics[page=1]{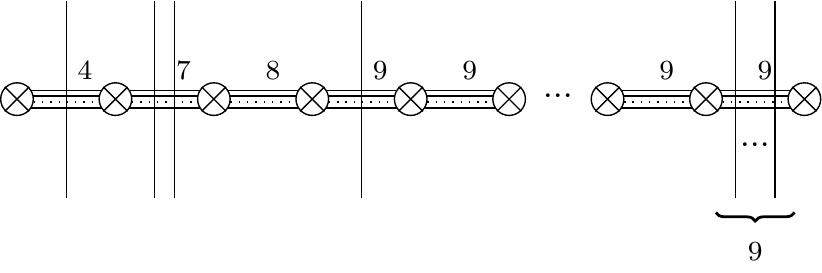}
	} \\
	\raisebox{-.5\height}{
	\includegraphics[page=2]{figures/figures_M5_ALE.pdf}
	} 
	\label{eq:6d_quiver_A-type}
\end{align}
Here and in the remainder of this note, \NS\ branes are denoted by $\bigotimes$, \Ds\ branes are horizontal solid lines, while \De\ branes are vertical solid lines. 
It is evident, that the changes induced from $\rho_L$ are obtainable from partial Higgs mechanism starting from the trivial partition $1^9$. The 6d theory \eqref{eq:6d_quiver_A-type} contains the usual quiver notation that encodes the hypermultiplets and vector multiplets. In addition, each vector multiplet is accompanied by a tensor multiplet. Below each gauge node, the F-theory curve of self-intersection $-n$ is indicated.

\paragraph{Magnetic quivers.}
Following the magnetic quiver construction for this class of theories \cite{Cabrera:2019izd}, it is straightforward to derive the magnetic quivers for each (singular or non-singular) point on the tensor branch. For instance, in \cite[eq.\ (4.2)]{Hanany:2018uhm} a conjecture for the Higgs branch of  $T_{\surm(k)}^{n}(\rho_L,\rho_R)$ at infinite gauge coupling has been put forward by using magnetic quivers. The reader is referred to Appendices \ref{app:A-type_bc} - \ref{app:Examples_SU4} for the exact form of the magnetic quivers and examples for $\surm(4)$.
%
%
%
\subsection{D-type singularity with boundary conditions}
Starting from a Type IIA set-up with $2n$ half \NS\ branes with $2k$ half \Ds\ branes on top of \Osm\ orientifolds, one may assign boundary conditions of the \Ds\ branes ending on the \De\ branes at $x^6 = \pm \infty$. Both sides can be labelled by $D$-type partitions $\rho_{L,R}$ of $2k$. The low-energy effective theory is labelled by $T_{\sorm(2k)}^{n}(\rho_L,\rho_R)$.  
As observed in \cite{Heckman:2016ssk,Mekareeya:2016yal}, $D$-type boundary condition inevitably include brane configurations wherein the number of D6 branes between two half NS5 branes can become negative (the lowest number is $-3$ \Ds\ on an O$6^+$ plane). It has been demonstrated in \cite{Mekareeya:2016yal} that despite this oddity, the brane systems can be used to correctly evaluate anomaly coefficients. 

To elaborate, the inclusion of $D$-type partition boundary data in these brane configuration proceeds as in the $A$-type case of Section \ref{sec:A-type_boundary}, except for the further subtlety of O6 orientifold planes. As recalled in Appendix \ref{app:orientifolds}, the orientifolds carry 6-brane charge. Thus, the difference between the  6-brane charge on the left and right hand side of an NS5 brane is given by the value of the cosmological constant. Each time a 8-brane is passed, the value changes, which enforces varying 6-brane numbers. As O$6^-$ and $\widetilde{\text{O}6}^-$ have negative 6-brane charge, there are naturally accompanied by more D6 branes than the positively charged O$6^+$ and $\widetilde{\text{O}6}^+$ planes. It is not surprising that the negative brane numbers are only encountered for O$6^+$/$\widetilde{\mathrm{O}6}^+$ planes. To be more specific, recall the algorithmic construction \cite{Mekareeya:2016yal} (see also \cite{Benini:2010uu,Cremonesi:2014uva}): denote the left boundary condition as $\rho_L\equiv \lambda$ and its transpose $\lambda^T=[\hat{\lambda}_1,\hat{\lambda}_2,\ldots,\hat{\lambda}_n ]$ with $\hat{\lambda}_1 \geq  \hat{\lambda}_2 \geq \ldots \geq \hat{\lambda}_n>0$. 
The position of the half D8 branes are determined by integers $\rho_i$ defined as 
\begin{align}
    \rho_i = \hat{\lambda}_i - \hat{\lambda}_{i+1} \;,\quad i=1,\ldots, n-1
    \qquad \text{and} \qquad 
    \rho_n = \hat{\lambda}_n \,.
\end{align}
The brane configuration and 6d quiver are given by
\begin{align}
\raisebox{-.5\height}{
\includegraphics[page=3]{figures/figures_M5_ALE.pdf}
	}
	\qquad  \longrightarrow\qquad
    \raisebox{-.5\height}{
    \includegraphics[page=4]{figures/figures_M5_ALE.pdf}
	} 
	\label{eq:branes+partition}
\end{align}
The numbers $r_j$ of half D6 branes are determined by $\lambda^T$ and are constrained by the anomaly cancellation conditions
\begin{align}
r_j = \begin{cases}
-8 + \sum_{i=1}^j \hat{\lambda}_i & j = \mathrm{odd} \,,\\
\sum_{i=1}^j \hat{\lambda}_i & j = \mathrm{even}\,,
\end{cases}
\qquad \text{with} \qquad 
\begin{cases}
r_{2j-1} &= \frac{1}{2} \left( r_{2j-2} + r_{2j} +\rho_{2j-1} \right) + 8 \,,\\ 
r_{2j} &= \frac{1}{2} \left( r_{2j-1} + r_{2j+1} +\rho_{2j} \right) - 8 \,.
\end{cases}
\end{align}
It follows that $r_{2j-1}$ becomes negative or zero if and only if $\sum_{i=1}^{2j-1} \hat{\lambda}_i \leq 8$. In particular, the Type IIA brane configuration has non-positive number of D6 branes suspended between adjacent half \NS\ if and only if the largest part of $\lambda^T$ is less than or equal to $8$. Moreover, the case of equality, i.e.\ $\hat{\lambda}_1$ equals 8, just has a vanishing number of D6; whereas genuine negative numbers of branes appear once the largest part of $\lambda^T$ is strictly less than 8.

Of course, the quiver diagram in \eqref{eq:branes+partition} is only meaningful for non-negative $r_i$; in contrast, as argued below, the brane configuration is a legitimate starting point for further studies. In this work, brane configurations with negatively charged branes are used to derive magnetic quivers for the Higgs branches at infinite coupling, see below and Appendix \ref{app:Examples_SO12}. Nonetheless, considering these brane systems in their own right leads to the open challenge of deducing certain matter representations from the brane system.

\paragraph{Classification of negative charge brane configurations.}
Based on the analysis of \cite{Mekareeya:2016yal}, one can simply compile a table with all brane configurations that contain negatively charged branes. To be specific, consider a $\C^2\slash D_k$ singularity with $D$-type partition $\rho_L=\lambda$ for the left boundary and the trivial partition $\rho_R=(1^{2k})$ for the right boundary. The possibilities are classified in terms of $\lambda^T$ as shown in Table \ref{tab:negative_branes}.

The prescription of Table \ref{tab:negative_branes} is applicable for brane configurations in which the \De\ branes coming from the left and right boundary conditions do not have to cross each other. In other words, one requires a sufficient number of \NS\ branes between them. For the case of overlapping boundary conditions, a detailed analysis has appeared in \cite{Hassler:2019eso} and some interesting examples are studied in detail below, see Sections \ref{sec:non-overlapping_bc} and \ref{sec:overlapping_bc}.

 \begin{center}
\begin{longtable}{lrc}
\toprule
 Brane configuration & 6d quiver & Partition$^T$ \\ \midrule
	\raisebox{-.5\height}{
\includegraphics[page=1]{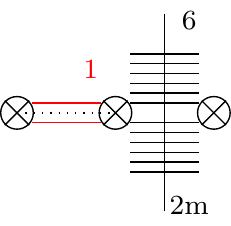}
	}
&
 	\raisebox{-.5\height}{
\includegraphics[page=2]{figures/figures_negative_branes.pdf}
}
& $\substack{(6,6,k,\ldots) \\ k=0,2,4,6 \\ 2m=6-k}$
\\ \midrule
	\raisebox{-.5\height}{
\includegraphics[page=3]{figures/figures_negative_branes.pdf}
	}
&
 	\raisebox{-.5\height}{
\includegraphics[page=4]{figures/figures_negative_branes.pdf}
}
& $\substack{(6,5,k,\ldots) \\ k=1,3,5 \\ 2m=5-k}$
\\ \midrule
	\raisebox{-.5\height}{
\includegraphics[page=5]{figures/figures_negative_branes.pdf}
	}
&
 	\raisebox{-.5\height}{
 \includegraphics[page=6]{figures/figures_negative_branes.pdf}
}
& $\substack{(6,4,k,\ldots) \\ k=0,2,4 \\ 2m=4-k}$
\\ \midrule
	\raisebox{-.5\height}{
	\includegraphics[page=7]{figures/figures_negative_branes.pdf}
	}
&
 	\raisebox{-.5\height}{
 	\includegraphics[page=8]{figures/figures_negative_branes.pdf}
}
& $\substack{(6,3,k,\ldots) \\ k=1,3 \\ 2m=3-k}$
\\ \midrule
	\raisebox{-.5\height}{
	\includegraphics[page=9]{figures/figures_negative_branes.pdf}
	}
&
 	\raisebox{-.5\height}{
 	\includegraphics[page=10]{figures/figures_negative_branes.pdf}
}
& $\scriptstyle{(6,2,2,\ldots)} $
\\ \midrule
	\raisebox{-.5\height}{
	\includegraphics[page=11]{figures/figures_negative_branes.pdf}
	}
&
 	\raisebox{-.5\height}{
 	\includegraphics[page=12]{figures/figures_negative_branes.pdf}
}
& $\scriptstyle{(6,1,1,\ldots)} $
\\ \midrule
	\raisebox{-.5\height}{
	\includegraphics[page=13]{figures/figures_negative_branes.pdf}
	}
&
 	\raisebox{-.5\height}{
 	\includegraphics[page=14]{figures/figures_negative_branes.pdf}
}
& $\substack{(4^3,k\ldots) \\ m=4-k } $
\\ \midrule
	\raisebox{-.5\height}{
	\includegraphics[page=15]{figures/figures_negative_branes.pdf}
	}
&
 	\raisebox{-.5\height}{
 	\includegraphics[page=16]{figures/figures_negative_branes.pdf}
}
& $\substack{(4^2,2,k\ldots) \\ m=2-k} $
\\ \midrule
	\raisebox{-.5\height}{
	\includegraphics[page=17]{figures/figures_negative_branes.pdf}
	}
&
 	\raisebox{-.5\height}{
 	\includegraphics[page=18]{figures/figures_negative_branes.pdf}
}
& $\substack{(4,3,3,k\ldots)\\ m=3-k}$
\\ \midrule
	\raisebox{-.5\height}{
	\includegraphics[page=19]{figures/figures_negative_branes.pdf}
	}
&
 	\raisebox{-.5\height}{
\includegraphics[page=20]{figures/figures_negative_branes.pdf}
}
& $\substack{(4,3,1,k,\ldots) \\ m=1-k}$
\\ \midrule
	\raisebox{-.5\height}{
\includegraphics[page=21]{figures/figures_negative_branes.pdf}
	}
&
 	\raisebox{-.5\height}{
 	\includegraphics[page=22]{figures/figures_negative_branes.pdf}
}
& $\substack{(4,2,2,k,\ldots)\\ m=2-k }$
\\ \midrule
	\raisebox{-.5\height}{
	\includegraphics[page=23]{figures/figures_negative_branes.pdf}
	}
&
 	\raisebox{-.5\height}{
 	\includegraphics[page=24]{figures/figures_negative_branes.pdf}
}
& $\scriptstyle{(4,1,1,\ldots)} $
\\ \midrule
	\raisebox{-.5\height}{
\includegraphics[page=25]{figures/figures_negative_branes.pdf}
	}
&
 	\raisebox{-.5\height}{
 	\includegraphics[page=26]{figures/figures_negative_branes.pdf}
}
& $\substack{(2^5,k\ldots) \\ m=2-k} $
\\ \midrule
	\raisebox{-.5\height}{
	\includegraphics[page=27]{figures/figures_negative_branes.pdf}
	}
&
 	\raisebox{-.5\height}{
 	\includegraphics[page=28]{figures/figures_negative_branes.pdf}
}
& $\scriptstyle{(2^3,1^2,\ldots)}$
\\ \midrule
	\raisebox{-.5\height}{
\includegraphics[page=29]{figures/figures_negative_branes.pdf}
}
&
 	\raisebox{-.5\height}{
 	\includegraphics[page=30]{figures/figures_negative_branes.pdf}
}
& $\scriptstyle{(2,1^{2n-2})}$
\\  \bottomrule
\caption{Brane configurations with negative charge branes. $\bigotimes$ denotes \NS\ branes, vertical solid lines denote half \De\ branes, horizontal black/red solid lines denote half \Ds\ branes (the positive/negative charge of the physical \Ds\ is written on top). The colour is black for positive and red for negative charge. O6 orientifolds are denoted as summarised in Appendix \ref{app:orientifolds}. Likewise, the 6d quiver is provided. Gauge groups are written explicitly below each round node $\circ$. The hypermultiplet matter content is encoded in the solid lines connecting nodes: black solid lines are bifundamental half-hypermultiplets, while red solid lines denote bi-spinor representations. The last column displays the transpose partition $\lambda^T$ that labels the distinct cases.}
\label{tab:negative_branes}
\end{longtable}
\end{center}

Before proceeding, let us pause and emphasise the status and the logic of Table \ref{tab:negative_branes}. The 6d quiver theories for any choice of boundary conditions are known from F-theory constructions. The brane systems with negative numbers are an analytic continuation of standard brane constructions. The approach taken in this note is to fit the brane system with the corresponding 6d quiver gauge theory (plus a suitable number of tensor multiplets). This is then the starting point for the new directions taken here: the brane systems enables us to derive a magnetic quiver and to study the Higgs branch moduli spaces in an unprecedented detail.

\paragraph{Magnetic quivers.}
In \cite[eq.\ (4.3)]{Hanany:2018uhm} a 
conjectural 
description of the Higgs branch at infinite coupling of 
$T_{\sorm(2k)}^{n}(\rho_L,\rho_R)$ has been presented as a  magnetic 
quiver. Building on the magnetic quiver construction \cite{Cabrera:2019dob} for this class of theories, it is now straightforward to derive these magnetic quivers from the brane configurations. The reader is referred to Appendices \ref{app:D-type_bc} - \ref{app:Examples_SO12} for the general form of the magnetic quivers and examples for $\sorm(12)$.

\section{Appetiser --- Nilpotent 6d Higgs branches}
\label{sec:appetiser}
\subsection{A nilpotent orbit via boundary conditions}
\begin{figure}[t]
	\centering
	\begin{subfigure}[b]{0.3\textwidth}
	\centering
	\includegraphics[page=30]{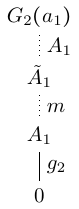}
\caption{$G_2(a_1)$ Hasse diagram}
\label{fig:Hasse_G2}
\end{subfigure}
\begin{subfigure}[b]{0.3\textwidth}
	\centering
	\includegraphics[page=31]{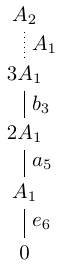}
\caption{$A_2$ of $E_6$ Hasse diagram}
\label{fig:Hasse_E6}
\end{subfigure}
\begin{subfigure}[b]{0.35\textwidth}
	\centering
	\includegraphics[page=32]{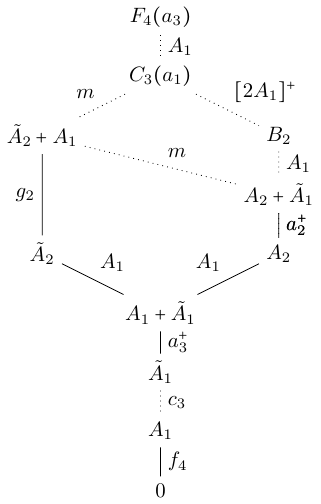}
\caption{$F_4(a_3)$ Hasse diagram}
\label{fig:Hasse_F4}
\end{subfigure}
\caption{The Hasse diagrams for three nilpotent orbits of $G_2$, $E_6$, and $F_4$, respectively, following the conventions of \cite{fu2017generic}. The orbits are denoted by their Bala-Carter labels. Nilpotent orbits that are contained in the same special piece are connected by a dotted line. Capital letters denote simple surface singularities, while lower-case letters stand for closures of minimal nilpotent orbits. The non-normal variety $m$ is detailed in \cite[Sec.\ 1.8.4.]{fu2017generic}. These three nilpotent orbits show up as Higgs branches of 6d $\Ncal=(1,0)$ supersymmetric theories. Note also that the special pieces (connected by dotted lines) have component group $S_3$, $S_2$, and $S_4$, respectively.}
\end{figure}

Based on the expositions in Section \ref{sec:M5_AD-sing}, the magnetic quivers for the theories with boundary conditions can be derived systematically. Now, it is time to demonstrate that interesting moduli spaces can arise.
Consider $3$ M5 branes on a $\C^2 \slash \Z_{2}$ singularity with boundary conditions $\rho_L=(2)$, $\rho_R=(1^2)$. The 6d quiver theory reads
\begin{align}
    \raisebox{-.5\height}{
    \includegraphics[page=1]{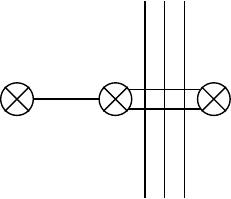}
    }
	\qquad 
	\longleftrightarrow \qquad 
	\raisebox{-.5\height}{
		\includegraphics[page=2]{figures/figures_appetiser.pdf}
		}
	\label{eq:6d_quiver_G2}
	\end{align}
	i.e. $\surm(2)$ gauge theory with effectively 4 fundamental hypermultiplets and 2 tensor multiplets. Its Higgs branch (at finite coupling) is captured by the following magnetic quiver
\begin{align}
(1^{3}): \quad 
\raisebox{-.5\height}{
\includegraphics[page=3]{figures/figures_appetiser.pdf}
	} 
	\qquad \begin{cases}
	G_J &= \sorm(8)  \\
	\dim\ \Coulomb &= 5
	\end{cases}
\end{align}
and the bouquet of $\urm(1)$ nodes is represented by the partition $(1^{3})$. This moduli space is known to be the minimal nilpotent orbit closure of $\sorm(8)$, which is fitting for the finite coupling Higgs branch of $\surm(2)$ with 4 fundamentals. As argued in \cite{Hanany:2018vph,Hanany:2018cgo,Cabrera:2019izd}, the collapse of $-2$ curve or, equivalently, taking one gauge coupling in \eqref{eq:6d_quiver_G2} to infinity is realised by discrete gauging. Collapsing a single $-2$ curve yields an $S_2$ discrete gauging (denoted by phase $(2,1)$), while the collapse of both $-2$ curves becomes an $S_3$ discrete gauging (denoted by phase $(3)$).
The magnetic quivers for these two Higgs branch phases are given by
\begin{align}
	(2,1): \quad 
	\raisebox{-.5\height}{
	\includegraphics[page=4]{figures/figures_appetiser.pdf}
	} 
	\qquad \begin{cases}
		G_J &= \sorm(7) \\
		\dim\ \Coulomb &= 5
	\end{cases}
\qquad 
(3): \quad 
\raisebox{-.5\height}{
\includegraphics[page=5]{figures/figures_appetiser.pdf}
	} 
	\qquad \begin{cases}
	G_J &= G_2 \\
	\dim\ \Coulomb &= 5
	\end{cases}
\end{align}
The Coulomb branch of $(3)$ is known to be the (quaternionic) $5$-dimensional sub-regular nilpotent orbit closure of $G_2$ \cite{Hanany:2017ooe}. Its Hasse diagram is displayed in Figure \ref{fig:Hasse_G2}. It is crucial to appreciate the appearance of the $G_2$ global symmetry at the origin of the tensor branch and not $\sorm(7)$. For $\surm(2)$ gauge theory with a single tensor, the infinite coupling point has $\sorm(7)$ global symmetry, because $\sorm(7)$ is the commutant of $S_2$ inside $\sorm(8)$. However, for $\surm(2)$ with two tensors, the infinite coupling Higgs branch has $G_2$, as $G_2$ is the commutant of $S_3$ inside $\sorm(8)$.
Even more is true, the statement extends beyond mere symmetry considerations. It is known \cite{Brylinski:1992} that the next-to-minimal orbit of $\sorm(7)$ is an $S_2$ quotient of the minimal orbit of $\sorm(8)$; likewise,
the sub-regular nilpotent orbit of $G_2$ is an $S_3$ quotient of the minimal orbit of $\sorm(8)$.

Moreover, it is instructive to keep track of the Hasse diagram changes for the three phases \cite{Bourget:2022ehw} 
 \begin{align}
	\label{eq:Hasse_appetiser}
	\raisebox{-.5\height}{
	\includegraphics[page=6]{figures/figures_appetiser.pdf}
	}
	\qquad \qquad 
	\raisebox{-.5\height}{
	\includegraphics[page=7]{figures/figures_appetiser.pdf}
	}
	\qquad \qquad 
\raisebox{-.5\height}{
\includegraphics[page=8]{figures/figures_appetiser.pdf}
}
\end{align}
where the change of global symmetry is clearly visible at the bottom transition.

\subsection{Hasse diagram for single gauge group factor}
\begin{figure}[t]
    \centering
    \begin{subfigure}[t]{0.24\textwidth}
    \centering
    \includegraphics[page=1]{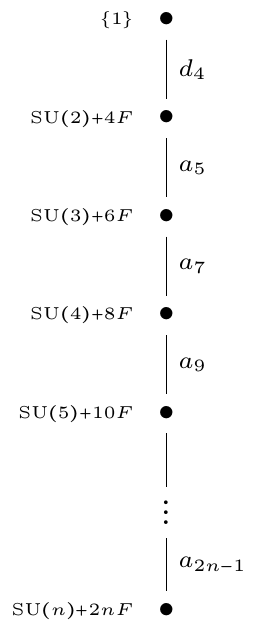}
    \caption{}
    \label{subfig:Hasse_2_SU}
    \end{subfigure}
    \begin{subfigure}[t]{0.24\textwidth}
    \centering
    \includegraphics[page=2]{figures/figures_Hasse_2_curve.pdf}
    \caption{}
    \label{subfig:Hasse_2_SO}
    \end{subfigure}
    \begin{subfigure}[t]{0.24\textwidth}
    \centering
    \includegraphics[page=3]{figures/figures_Hasse_2_curve.pdf}
    \caption{}
    \label{subfig:Hasse_2_SOp}
    \end{subfigure}
    \begin{subfigure}[t]{0.24\textwidth}
    \centering
    \includegraphics[page=4]{figures/figures_Hasse_2_curve.pdf}
    \caption{}
    \label{subfig:Hasse_2_E}
    \end{subfigure}
    \caption{The Higgs branch Hasse diagrams for the theories defined in a single $-2$ curve. \subref{subfig:Hasse_2_SU} contains the $\surm(n)$ type of theories. \subref{subfig:Hasse_2_SO} details the $\sorm(n)$ type theories, while  \subref{subfig:Hasse_2_SOp} shows the phase diagram for the family of theories related to $\sorm(12)$ with $6F+2\cdot \frac{1}{2} S(C)$. Lastly, \subref{subfig:Hasse_2_E} shows the Hasse diagram for the families that contain the exceptional theories. Here, $F$ denotes the  fundamental, $V$ the vector, $S$ the spinor, and $C$ the conjugate spinor representation. Each leaf is denoted by the 6d (electric) theory. The phase diagrams displayed are the finite coupling Higgs branch Hasse diagrams for the 6d theory at the bottom. The diagrams for the other theories are obtained by reduction.}
    \label{fig:Hasse_2_curve}
\end{figure}

\begin{figure}[t]
    \centering
    \begin{subfigure}[t]{0.45\textwidth}
    \centering
    \includegraphics[page=1]{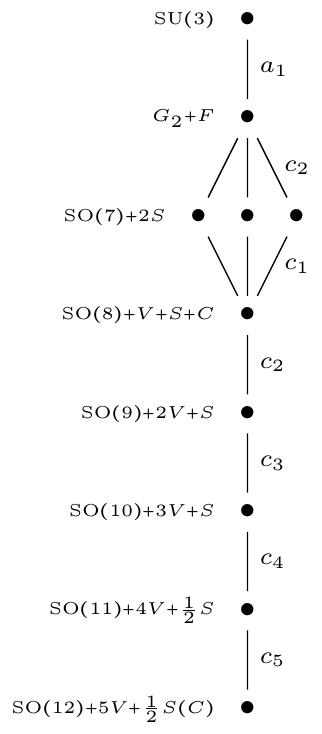}
    \caption{}
    \label{subfig:Hasse_3_SO}
    \end{subfigure}
    \begin{subfigure}[t]{0.45\textwidth}
    \centering
    \includegraphics[page=2]{figures/figures_Hasse_3_curve.pdf}
    \caption{}
    \label{subfig:Hasse_3_E}
    \end{subfigure}
\caption{The Higgs branch Hasse diagrams for the theories defined in a single $-3$ curve. \subref{subfig:Hasse_3_SO} details the $\sorm(n)$ type theories. Lastly, \subref{subfig:Hasse_3_E} shows the Hasse diagram for the families that contain the exceptional theories. Here, $F$ denotes the  fundamental, $V$ the vector, $S$ the spinor, and $C$ the conjugate spinor representation. Again, the phase diagrams displayed are the finite coupling Higgs branch Hasse diagrams for the 6d theory at the bottom. The diagrams for the other theories are obtained by reduction.}
    \label{fig:Hasse_3_curve}
\end{figure}

Consider the anomaly-free theories supported on a $-2$ curve. The partial Higgs mechanism between them gives rise to distinct sets, c.f.\ \cite{DelZotto:2018tcj}: (i) the $\surm(n)$ theories with $2n$ flavours, (ii) a set of $\sorm(n)$ theories with matter in the vector and spinor representations \cite{Danielsson:1997kt}, and (iii) a set that includes exceptional gauge groups. In Figure \ref{fig:Hasse_2_curve}, the Higgs branch Hasse diagram for each is displayed. Analogously, the Hasse diagrams for the anomaly-free theories on a $-3$ curve are summarised in Figure \ref{fig:Hasse_3_curve}.

For later purposes, briefly recall the conventions for a Higgs branch Hasse diagram \cite{Bourget:2019aer}. Each leaf is denoted by $\bullet$, and whenever two neighbouring leaves are partially ordered, they are connected by a line. The minimal slice between two partially ordered leaves $a,b$ with $a>b$, such that no third leaf $c$ with $a>c>b$ exists, is denoted either by $g$ for the minimal nilpotent  orbit closure of $G$, or by $A$, $D$, $E$ for Kleinian surface singularities $\C^2\slash \Gamma_{ADE}$. Other minimal transitions may appear and are referenced whenever they appear. 
Each leaf is denoted by the corresponding 6d electric theory, whose Higgs branch describes the slice to the top of the Hasse diagram.

\section{Search for interesting theories}
\label{sec:2M5s_RHS_trivial}
Considering a D-type singularity, the orthosymplectic magnetic quivers often suffer from ``bad'' magnetic gauge nodes which renders them incomputable with the monopole formula. In this Section, the aim is to search for computable magnetic quivers. Interestingly, one of the Higgs branches encountered in this search is a nilpotent orbit of $E_6$. We provide the explicit magnetic quiver, brane systems, and Hasse diagrams.

Consider the theory of 2 M5 branes on a $\C^2 \slash D_{6}$ singularity with boundary conditions $\rho_L=(2^6)$, $\rho_R=(1^{12})$. The Type IIA brane configuration leads to the following 6d quiver:
\begin{align}
\label{eq:SO12_on_3_a}
\raisebox{-.5\height}{
\includegraphics[page=1]{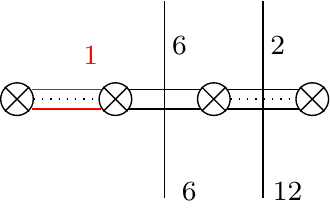}
	}
	\quad \longrightarrow \quad 
    \raisebox{-.5\height}{
    \includegraphics[page=2]{figures/figures_curve_31.pdf}
	} 
\end{align}
and the presence of an NS5 brane interval with negative brane number signals the quasi Higgs mechanism which trades the corresponding tensor multiplet for a number of hypermultiplets dictated by the gravitational anomaly cancellation \cite{Green:1984bx,RandjbarDaemi:1985wc,Dabholkar:1996zi}.
Thus, from the original 3 tensors (4 positions of half NS branes minus an overall position by translation invariance), only 2 remain --- consistent with two gauge couplings.
The $\sorm(12)$ theory is anomaly free with 1 half-hypermultiplets in the spinor representation of dimension 32 and 5 hypermultiplets in the vector representation. The $\sprm(2)$ gauge theory has 12 flavours, hence again is anomaly free. The 6 half D8 branes on the middle interval give rise to an $\sprm(3)$ global symmetry, while the 12 half D8 branes give rise to an $\sorm(12)$ global symmetry.

One can straight-forwardly Higgs the $\sprm(2)$ gauge node away. For instance, $\sprm(2) \to \sprm(1) \to \{1\}$ is a transparent Kraft-Procesi transition \cite{Cabrera:2017njm} in the brane configuration
\begin{align}
    \raisebox{-.5\height}{
    \includegraphics[page=1]{figures/figures_curve_31.pdf}
	}
	\quad \longrightarrow \quad 
	\raisebox{-.5\height}{
	\includegraphics[page=3]{figures/figures_curve_31.pdf}
	}
	\quad \longrightarrow \quad 
	\raisebox{-.5\height}{
	\includegraphics[page=4]{figures/figures_curve_31.pdf}
	}
\end{align}
and note that the rightmost four half D8 branes are fully decoupled. Hence, these are not kept for the subsequent discussion. Strictly speaking, the resulting theory is defined on the $-3$ curve coupled to the adjacent $-1$ curve. Based on the brane system, one can derive the following infinite coupling magnetic quiver 
\begin{align}
\label{eq:SO12_on_3_b}
    \raisebox{-.5\height}{
    \includegraphics[page=5]{figures/figures_curve_31.pdf}
	} 
	\quad \longrightarrow \quad 
	\raisebox{-.5\height}{
	\includegraphics[page=6]{figures/figures_curve_31.pdf}
	} 
\end{align}
which is not computable via the monopole formula as it has $\sprm(k)$ nodes with negative imbalance, depicted in grey (balanced nodes are depicted in red). Nevertheless, one can proceed and explore further  partial Higgs mechanisms. In terms of the brane system, the partial Higgsing $\sorm(12)\to\sorm(11)\to\sorm(10)$ cannot be separated into two processes
\begin{align}
\label{eq:SO12-SO11-SO10}
    	\raisebox{-.5\height}{
    	\includegraphics[page=7]{figures/figures_curve_31.pdf}
	}
	\quad \longrightarrow \quad
	\raisebox{-.5\height}{
	\includegraphics[page=8]{figures/figures_curve_31.pdf}
	}
		\quad \longrightarrow \quad
	\raisebox{-.5\height}{
	\includegraphics[page=9]{figures/figures_curve_31.pdf}
	}
\end{align}
because the last transition is not accompanied by creation or annihilation of physical branes. This effect has been denoted as \emph{collapse} in \cite{Cabrera:2017njm}, see also \cite[Fig.\ 8]{Feng:2000eq}.
Nonetheless, one can derive an infinite coupling magnetic quiver for the $\sorm(10)$ theory
\begin{align}
    \raisebox{-.5\height}{
    \includegraphics[page=10]{figures/figures_curve_31.pdf}
	} 
	\quad \longrightarrow \quad 
	\raisebox{-.5\height}{
	\includegraphics[page=11]{figures/figures_curve_31.pdf}
	} 
\end{align}
An encouraging sign is that there are two more balanced nodes and less grey nodes, indicating that if we proceed with Higgsing, then more nodes will turn balanced, and more importantly the quiver will become computable using the monopole formula.
Similarly, the brane system only sees the combined transition $\sorm(10)\to\sorm(9)\to\sorm(8)$ 
\begin{align}
\label{eq:SO10-SO9-SO8}
	\raisebox{-.5\height}{
	\includegraphics[page=9]{figures/figures_curve_31.pdf}
	}
	\quad \longrightarrow \quad
	\raisebox{-.5\height}{
	\includegraphics[page=12]{figures/figures_curve_31.pdf}
	}
		\quad \longrightarrow \quad
	\raisebox{-.5\height}{
	\includegraphics[page=13]{figures/figures_curve_31.pdf}
	}
\end{align}
because the last transition is, again, not accompanied by creation or annihilation of physical branes.
The infinite coupling magnetic quiver for the $\sorm(8)$ theory is obtained as 
\begin{align}
    \raisebox{-.5\height}{
    \includegraphics[page=14]{figures/figures_curve_31.pdf}
	} 
	\quad \longrightarrow \quad 
	\raisebox{-.5\height}{
	\includegraphics[page=15]{figures/figures_curve_31.pdf}
	} 
\end{align}
where 4 more nodes become balanced. Nevertheless, the quiver is still not computable.
Next, the transition $\sorm(8)\to \sorm(7)$ is visible in the brane system
\begin{align}
    \raisebox{-.5\height}{
    \includegraphics[page=13]{figures/figures_curve_31.pdf}
	}
	\quad \longrightarrow \quad 
	\raisebox{-.5\height}{
	\includegraphics[page=16]{figures/figures_curve_31.pdf}
	}
\end{align}
and the magnetic quiver reads
\begin{align}
    \raisebox{-.5\height}{
    \includegraphics[page=17]{figures/figures_curve_31.pdf}
	} 
	\quad \longrightarrow \quad 
	\raisebox{-.5\height}{
	\includegraphics[page=18]{figures/figures_curve_31.pdf}
	} 
\end{align}
and we happily hit a quiver with non negative imbalance, hence computable.
Lastly, the brane systems allows a combined transition $\sorm(7)\to G_2 \to \surm(3)$
\begin{align}
    \raisebox{-.5\height}{
    \includegraphics[page=16]{figures/figures_curve_31.pdf}
	}
	\; \longrightarrow \;
	    \raisebox{-.5\height}{
	    \includegraphics[page=19]{figures/figures_curve_31.pdf}
	}
		\; \longrightarrow \; 
	    \raisebox{-.5\height}{
	    \includegraphics[page=20]{figures/figures_curve_31.pdf}
	}
	\label{eq:Higgs_SO7-G2-SU3}
\end{align}
wherein the last two brane systems are indistinguishable in terms of Higgs branch degrees of freedom. Note also that the leftmost two half D8 branes have decoupled in the brane configuration of $G_2$ and $\surm(3)$.
Based on \eqref{eq:Higgs_SO7-G2-SU3}, one arrives at a proposal for the infinite coupling magnetic quiver of pure $\surm(3)$ on a $-3$ curve coupled to a $-1$ curve
\begin{align}
    \raisebox{-.5\height}{
    \includegraphics[page=21]{figures/figures_curve_31.pdf}
	} 
	\quad \longrightarrow \quad 
	\raisebox{-.5\height}{
	\includegraphics[page=22]{figures/figures_curve_31.pdf}
	} 
	\label{eq:magQuiv_SU3}
\end{align}
As there is a single gauge group and the infinite coupling of the $\sprm(0)$ involves a small $E_8$ instanton transition, we expect the moduli space to be the hyper-K\"ahler quotient of the closure of the minimal nilpotent orbit of $E_8$ by $\surm(3)$. An additional $S_2$ gauging acts on the Higgs branch when the $\surm(3)$ coupling is tuned to infinity. Furthermore, as the computations below indicate, the resulting moduli space is the closure of the 21 dimensional nilpotent orbit of $E_6$. This is a remarkable finding, as the studies of this brane system reveals a Coulomb branch construction for this nilpotent orbit. It is very uncommon to have Coulomb branch constructions for nilpotent orbits of exceptional type, hence this study reveals a new exciting result!

We proceed with a detailed discussion of these points.
\subsection{\texorpdfstring{$\surm(3)$ coupled to $\sprm(0)$}{SU3 coupled to Sp0}}
\label{sec:SU3+Sp0}
Consider $2$ M5 branes on $\C^2 \slash D_4$ with boundary conditions $\rho_L=(3^2,1^2)$, $\rho_R=(1^8)$.
The brane system for $\surm(3)$ coupled to $\sprm(0)$ at finite coupling is given by
\begin{align}
	    \raisebox{-.5\height}{
	    \includegraphics[page=23]{figures/figures_curve_31.pdf}
	}
	\qquad \longleftrightarrow \qquad
    \raisebox{-.5\height}{
    \includegraphics[page=21]{figures/figures_curve_31.pdf}
	} 
\end{align}
The $\surm(3)$ gauge theory is anomaly-free with zero hypermultiplets, while $\sprm(0)$ is anomaly-free with 16 half-hypermultiplets. Each gauge group comes with one tensor multiplet giving a total of two.

The brane system naively displays 3 intervals between 4 half NS5 branes. However, brane intervals with negatively charge branes do not give rise to a tensor multiplet; thus, there are only two effective tensor multiplets, i.e.\ two gauge couplings. The right-most brane interval is an O$6^+$ plane without D6 branes, which can be thought of as an $\sprm(0)$ gauge theory on a $-1$ curve. Next, the three left-most half NS5 branes conspire to yield a single pure $\surm(3)$ gauge theory, i.e.\ the $-3$ curve. Note that all the 12 half D8 branes are crucial for the understanding of the system; in particular, for the derivation of the magnetic quiver and the anomaly cancellation for the $\sprm(0)$ gauge group. Further note that the number of gauge nodes in the magnetic quiver is 9. This is given by 12, the number of half D8 branes, minus 3 (11 segments and two $\sprm(0)$ nodes at each end of the quiver).

Because this coupled system does not have any matter content, one can deduce the infinite coupling Higgs branch by the following reasoning. The collapse of a $-1$ curve is known to yield the small $E_8$ instanton transition \cite{Ganor:1996mu} --- the Higgs branch is a symplectic singularity (or hyper-K\"ahler moduli space) $\clorbit{E_8}^{\min}$ with global $E_8$ symmetry. Since, the system is still coupled to an $\surm(3)$ gauge theory on the $-3$ curve, the infinite coupling moduli space of the entire configuration is an $\surm(3)$ hyper-K\"ahler quotient $\clorbit{E_8}^{\min} /// \surm(3)$ of the minimal nilpotent orbit closure of $E_8$.

However, this is not the end of the story yet. The F-theory perspective suggests that the collapse of the $-1$ curve leads to the $E_8$ transition and, simultaneously, reduces the $-3$ to a $-2$ curve. The subsequent collapse of the $-2$ curve leads to the gauging of an $S_2$ permutation group \cite{Hanany:2018vph}.
In the brane configuration, collapsing the $-1$ and $-3$ curve means that the half NS5 branes need to merge pairwise on the orientifold plane. The attached D6 branes can reconnect, and the physical NS5 branes can split and move off the orientifold. 
The brane configuration in the phase where both pairs of half NS5 branes are away from the orientifold plane is given by
\begin{align}
(1^2): \quad 
  &\raisebox{-.5\height}{
  \includegraphics[page=24]{figures/figures_curve_31.pdf}
	} \label{eq:branes_SU3_separate} 
	\end{align}
where the number of physical D6 branes is denoted in each interval. It is apparent that there are two distinguished phases, the two pairs of NS5 brane are either separated or coincident. These are denoted by $(1^2)$ and $(2)$, respectively. As discussed in \cite{Hanany:2018vph,Hanany:2018cgo,Cabrera:2019izd,Cabrera:2019dob}, the difference between both is an $S_2$ action. Making the pairs coincident means the resulting $-2$ curve is collapsed, which in the brane system becomes
	\begin{align}
	(2):\quad 
  &\raisebox{-.5\height}{
  \includegraphics[page=25]{figures/figures_curve_31.pdf}
	}\label{eq:branes_SU3_coincident}
\end{align}
 The two infinite coupling moduli spaces obey a simple relation
\begin{align}
    \Higgs_{\infty}^{(1^2)} = \clorbit{E_8}^{\min} /// \surm(3) \;,\qquad  
    \Higgs_{\infty}^{(2)} = \Higgs_{\infty}^{(1^2)} /// \Z_2 \;.
\end{align}
For partition $(1^2)$, the Higgs branch is simply the $\surm(3)$ hyper-K\"ahler quotient; while for partition $(2)$, the origin of the tensor branch, the Higgs branch is the $\Z_2$ quotient thereof. Brane systems \eqref{eq:branes_SU3_separate} and \eqref{eq:branes_SU3_coincident} constitute the brane realisation of the geometric Satake correspondence which we turn to describe in Section \ref{sec:Satake}.
 
\paragraph{Hilbert series for the hyper-K\"ahler quotient.}
Starting from the $\clorbit{E_8}^{\min}$, the HWG is given by
\begin{align}
    \HWG_{\clorbit{E_8}^{\min}} = \PE \left[ \mu_7 t^2 \right]  
    \quad \longleftrightarrow \quad  \HS_{\clorbit{E_8}^{\min}} (\{x_i\}_{i=1}^8) 
\end{align}
and the Hilbert series depends on the $E_8$ fugacities $x_i$. Using branching $E_8 \to \surm(3)\times E_6$, with $E_6$ fugacities $\{y_i\}_{i=1}^6$  and $\surm(3)$ fugacities $\{z_{1,2}\}$, one performs a hyper-K\"ahler quotient with respect to $\surm(3)$. Denote the Haar measure by $\diff \mu_{\surm(3)}(z_{1,2})$ and recall the $\surm(3)$ F-terms $\Hh_F (z_{1,2}) = \PE[-\chi_{[1,1]}(z_1,z_2) \cdot t^2 ]$ with $\chi_{[1,1]}$ the character of the adjoint. The computation yields 
\begin{subequations}
\begin{align}
\label{eq:A2_hk}
\HS_{\mathrm{hK}} &= \int \diff \mu_{\surm(3)}(z_{1,2})\   \HS_{\clorbit{E_8}^{\min}} (\{y_i\}_{i=1}^6,\{z_{1,2}\})  \cdot \Hh_F(z_{1,2}) \\
&\stackrel{y_i\to1}{=}  
\frac{(1+t^2) 
}{(1-t^2)^{42}} 
\cdot 
\big(
1
+35\ t^2
+708\ t^4
+9121\ t^6
+78994\ t^8
+472618\ t^{10}
+1998110\ t^{12}
+6056837\ t^{14}\notag \\
&\qquad 
+13296080\ t^{16} 
+21263807\ t^{18}
+24858218\ t^{20}
+21263807\ t^{22}
+13296080\ t^{24}
+6056837\ t^{26}\notag \\
&\qquad 
+1998110\ t^{28} 
+472618\ t^{30}
+78994\ t^{32}
+9121\ t^{34}
+708\ t^{36}
+35\ t^{38}
+t^{40}
\big) \notag 
\end{align}
and the first few orders are given by
\begin{align}
\label{eq:HS_E6_Z2_cover}
    \HS_{\mathrm{hK}}&= 1+78 t^2+3158 t^4+86787 t^6+1797641 t^8+29702895 t^{10}+O\left(t^{11}\right) \,,\\
    \PL(\HS_{\mathrm{hK}})&= 78 t^2+77 t^4-1379 t^6+1223 t^8+116493 t^{10}+O\left(t^{11}\right) \,.
    \label{eq:PL_E6_Z2_cover}
\end{align}
\end{subequations}
This hyper-K\"ahler space has quaternionic dimension $21$ and $E_6\slash \Z_3$ global symmetry, as evident from the Hilbert series. Furthermore, it is evident from \eqref{eq:PL_E6_Z2_cover} that this moduli space is not a nilpotent orbit as there is a second adjoint valued generator at order $t^4$.
In fact, $E_6$ has a nilpotent orbit of the same dimension, denoted as the $A_2$ orbit, see its corresponding Hasse diagram in Figure \ref{fig:Hasse_E6}. It is instructive to compare \eqref{eq:A2_hk} against the known unrefined Hilbert series \cite[Tab.\ 12]{Hanany:2017ooe} of this $E_6$ orbit
\begin{subequations}
\label{eq:HS_E6_A2}
\begin{align}
    \HS_{\clorbit{E_6}^{A_2}} &=  \frac{(1+t^2)}{(1-t^2)^{42}}\cdot 
    \big(
    1
    +35\ t^2
    +630\ t^4
    +7120\ t^6
    +54640\ t^8
    +294385\ t^{10}
    +1139307\ t^{12} \\
    &\quad     
    +3216888\ t^{14}
    +6702843\ t^{16}
    +10382781 t^{18}
    +12008160\ t^{20}
    +10382781  t^{22}
    +6702843\ t^{24}\notag \\
    &\quad 
    +3216888\  t^{26}
    +1139307\  t^{28}
    +294385\ t^{30}
    +54640\ t^{32}
    +7120\  t^{34}
    +630\  t^{36}
    +35\ t^{38}
    +t^{40}
    \big)\notag 
    \end{align}
where the first few orders in perturbative expansion read    \begin{align}
    \HS_{\clorbit{E_6}^{A_2}} &=  1+78 t^2+3080 t^4+81432 t^6+1613534 t^8+25483029 t^{10}+O\left(t^{11}\right) \;,\\
    \PL(\HS_{\clorbit{E_6}^{A_2}})&= 78 t^2-t^4-650 t^6+3575 t^8+3003 t^{10}+O\left(t^{11}\right) \,.
\end{align} 
\end{subequations}
Computing the limit of the following ratio 
\begin{align}
    \lim_{t\to1} 
    \frac{ \HS_{\mathrm{hK}} }{ \HS_{\clorbit{E_6}^{A_2}} } =2
\end{align}
suggests that the $\surm(3)$ hyper-K\"ahler quotient is the $\Z_2$ cover of the $A_2$ orbit of $E_6$.

\paragraph{SQCD with 27 flavours.}
Returning to the picture that $\surm(3)$ is gauged inside $E_8$ naturally leads to an $\surm(3)$ SQCD.
Consider the decomposition of the adjoint representation of $E_8$ to representations of its subgroup $\surm(3)\times E_6$. We have $\mu_6 + \nu_1 \mu_5+\nu_2\mu_1+\nu_1\nu_2$, where $\nu$ are $\surm(3)$ fugacities and $\mu$ are $E_6$ fugacities.
The adjoint of $E_6$ survives and is the first contribution to the HWG at order $t^2$. The adjoint of $\surm(3)$ gets projected out by the quotient, and we are left with 27 fundamental hypermultiplets. For getting invariants from this representation, it is convenient to think about SQCD with 27 flavors with the embedding of $E_6$ in $\surm(27)$ that projects the fundamental to the fundamental. One should take into account that the ``quarks" arise at order $t^2$ which compares with SQCD in which quarks arise at order $t$, hence there is a rescaling $t\to t^2$ for representations arising in this way. The HWG for SQCD is well known \cite{Benvenuti:2010pq,Ferlito:2016grh}, and has a PL
\begin{align}
\mu_1 \mu_{26} t^2 + \mu_2 \mu_{25} t^4 + \mu_3 t^3 + \mu_{24} t^3\;.
\end{align}
The first term is the adjoint representation of $\surm(27)$ which gives a contribution $\mu_6+\mu_1\mu_5$ of $E_6$ at order $t^4$.
The baryons and antibaryons both project into the $\mu_3$ of $E_6$. Similarly $\mu_{26, 25, 24}$ of $\surm(27)$ project into $\mu_{5,4,3}$ of $E_6$.
Taking into account the rescaling in $t$ and the projection of these 4 terms we get the expression for the HWG
\begin{align}
\label{eq:HWG_E6_A2}
\HWG_{(1^2)} &= \mathrm{PE}\left[ 
\mu_6 t^2
+\mu_6 t^4
+\mu_1 \mu_5 t^4
+2 \mu_3 t^6
+\mu_2 \mu_4 t^8 \right] .
\end{align}
Reverting this into an unrefined Hilbert series yields back \eqref{eq:A2_hk}, showing that the HWG computes the hyper-K\"ahler quotient. It is important to note that while the derivation of the HWG is based on heuristic arguments, equation \eqref{eq:HWG_E6_A2} is an exact expression for the moduli space $\Higgs_{\infty}^{(1^2)} = \clorbit{E_8}^{\min} /// \surm(3)$.
The HWG \eqref{eq:HWG_E6_A2} also shows that all generators are invariant under the $\Z_3$ centre of $E_6$ such that the global symmetry group of the Higgs branch $\Higgs_{\infty}^{(1^2)}$ is $E_6\slash \Z_3$.

For 6d gauge theories on a single $-2$ curve, there exists a finite and infinite coupling Higgs branch, both being related by an $S_2$ gauging. More precisely, the $\Higgs_f$ is the $\Z_2$ cover of $\Higgs_\infty$. Therefore, \eqref{eq:HWG_E6_A2} can be associated to the phase \eqref{eq:branes_SU3_separate} of the brane system.

To deduce $\HWG_{(2)}$ from $\HWG_{(1^2)}$, one employs the following $\Z_2$ action
\begin{align}
 \frac{1}{ \left(1-\mu_6 t^4\right)} 
 \to \frac{1}{ \left(1+\mu_6 t^4\right)}
  \qquad 
  \frac{1}{\left(1-\mu_3 t^6\right)^2 } 
  \to 
  \frac{1}{\left(1-\mu_3 t^6\right) \left(1+\mu_3 t^6\right)}  \,.
\end{align}
To motivate this, the discrete gauging that translates finite to infinite coupling of $\surm(n)$ theories with $2n$ flavours acts in two ways \cite[eq.\ (2.4) and (2.29)]{Hanany:2018vph}: first, there are two baryonic invariants, which are permuted by $S_2$. Thus, one $\mu_3\to + \mu_3$ and the other $\mu_3 \to - \mu_3$. Second, the $\Z_2$ acts on the $\urm(1)$ part of the global symmetry. The analogue here is the second adjoint $\mu_6$ at order $t^4$. After gauging the $\Z_2$ the HWG is 
\begin{align}
     \HWG_{(2)} &=
     \mathrm{PE}\left[  \mu_6 t^2+  \mu_1 \mu_5 t^4+\mu_3 t^6 + \mu_2 \mu_4 t^8+ \mu_6^2 t^8 +\mu_3 \mu_6 t^{10} + \mu_3^2 t^{12} -\mu_3^2 \mu_6^2 t^{20}
     \right] \label{eq:HWG_SQCD_Z2} 
\end{align}
and deriving the unrefined Hilbert series returns \eqref{eq:HS_E6_Z2_cover}. Again, this confirms the alternative $\surm(3)\times \Z_2$ hyper-K\"ahler quotient.
\begin{table}[t]
	\ra{1.25}
	\centering
	\begin{tabular}{crl}
		\toprule
		phase   &  \multicolumn{2}{c}{quantity} \\ \midrule
		$(2)$  &  $\Hh=$& $ 1 + 78\ t^2 + 3080\ t^4 + 81432\ t^6 +O(t^7)$ \\
		& $ \HS_{\mathbb{Z}} =$  &  $  1+46 t^2 + 1608\ t^4 + 41208\ t^6 +O(t^7)$\\
		& $ \HS_{\mathbb{Z}+\frac{1}{2}} =$ &  $ 32 t^2 + 1472\ t^4 + 40224\ t^6 +O(t^7)$  \\
		& $\PL =$&  $ 78\ t^2 - t^4 - 650 t^6 +O(t^7)$ \\ \midrule
		$(1^2)$ & $\Hh=$ & $1 + 78\ t^2 + 3158\ t^4 + 86787\ t^6  +O(t^7)$ \\
		& $\HS_{\mathbb{Z}} = $ &  $ 1+46 t^2 + 1654\ t^4 + 43971\ t^6  +O(t^7)$\\
		& $\HS_{\mathbb{Z}+\frac{1}{2}} =$&  $ 32 t^2 + 1504\ t^4 + 42816\ t^6 +O(t^7)$\\
		& $\PL= $&$78\ t^2 + 77 t^4 - 1379 t^6 +O(t^7) $  \\ \bottomrule
	\end{tabular}
	\caption{Perturbative Hilbert series for the different phases \eqref{eq:magQuiv_SU3_C2} and \eqref{eq:magQuiv_SU3_C1C1}.}
	\label{tab:HS_SU3_with_Sp0}
\end{table}

\paragraph{Magnetic quivers.}
Starting from configuration \eqref{eq:branes_SU3_separate}, the magnetic quiver reads
\begin{align}
    	(1^2): \quad \raisebox{-.5\height}{
\includegraphics[page=33]{figures/figures_curve_31.pdf}
	} 
	\qquad 
	\begin{cases}
	\dim\ \Coulomb &= 21 \,,  \\
	\quad 
	\gbal&=\sormL(10) \,,
	\end{cases}
	\label{eq:magQuiv_SU3_C1C1}
\end{align}
and the monopole formula is evaluated as in Table \ref{tab:HS_SU3_with_Sp0}. One recognises that $78=\dim\ E_6$ and $45=\dim\ \sormL(10)$. In other words, the integer lattice contribution is consistent with an $\sormL(10)\times \urmL(1)$ global symmetry, which is a maximal subalgebra of $\esix$. Whereas the full Hilbert series is consistent with an $E_6\slash \Z_3$ global symmetry group. The perturbative expansion agrees with \eqref{eq:HS_E6_Z2_cover}, i.e.\ the quiver \eqref{eq:magQuiv_SU3_C1C1} realises the $\Z_2$ cover of the $A_2$ orbit closure of $E_6$ as an orthosymplectic quiver. The same space has a known unitary magnetic quiver via the geometric Satake correspondence, which is discussed below.

Similarly, for the phase \eqref{eq:branes_SU3_separate} of the brane system, the magnetic quiver is 
\begin{align}
(2): \quad 
    	\raisebox{-.5\height}{
\includegraphics[page=22]{figures/figures_curve_31.pdf}
	} 
	\qquad
	\begin{cases}
	\dim\ \Coulomb &= 21 \,, \\
	\quad 
	\gbal&=\sormL(10) \,,
	\end{cases}
	\label{eq:magQuiv_SU3_C2}
\end{align}
and the monopole formula is summarised in Table \ref{tab:HS_SU3_with_Sp0}. Again this has an $E_6\slash \Z_3$ global symmetry and the perturbative expansion agrees with \eqref{eq:HS_E6_A2}. Excitingly, \eqref{eq:magQuiv_SU3_C2} is the first quiver realisation of the $E_6$ orbit closure with Bala-Carter label $A_2$.

\subsection{Relation to geometric Satake}
\label{sec:Satake}
As discussed above $\Higgs_{\infty}^{(1^2)} = \clorbit{E_8}^{\min} /// \surm(3)$ is the $\Z_2$ cover of the $E_6$ nilpotent orbit $[000002]$ (Bala-Carter label $A_2$). Following the work on geometric Satake and small representations \cite{reeder2002small,achar2013geometric}, this nilpotent orbit is a top member of a special (Reeder) piece in the Hasse diagram for the nilpotent cone of $E_6$, whereas the non special member in this piece is the nilpotent orbit $[001000]$. Setting these labels of the orbit to be flavour data of an $E_6$ quiver, and requiring all gauge nodes to be balanced, we get \cite{MagQuivSatake}
\begin{align}
    \raisebox{-.5\height}{
    \includegraphics[page=26]{figures/figures_curve_31.pdf}
	}  
	\label{eq:unitary_quiver_E6}
\end{align}
By construction, the Coulomb branch of this quiver is a slice in the affine Grassmanian of $E_6$ \cite{Kamnitzer:2012} and its HWG is given by \eqref{eq:HWG_E6_A2}. Furthermore, by the geometric Satake correspondence, this moduli space is the $\Z_2$ cover of the closure of the special nilpotent orbit in the special piece, namely the orbit $[000002]$. The Hasse diagram for the $A_2$ orbit is recalled in Figure \ref{fig:Hasse_E6}. Note that the quiver \eqref{eq:unitary_quiver_E6} allows to recover the Hasse diagram via quiver subtraction \cite{Cabrera:2018ann}, see \eqref{eq:Hasse_Satake}. First, one identifies a $d_4$ transition. After rebalancing, an $a_5$ transition becomes apparent, which leaves an affine $E_6$ Dynkin quiver. Hence, the final transition is an $e_6$. Lastly, the first transition splits into $A_1$ and $b_3$ due to the discrete $\Z_2$. Therefore, the collapse of a $-2$ curve affects the Hasse diagram as follows \cite{Bourget:2019aer}
\begin{align}
\label{eq:Hasse_Satake}
\raisebox{-.5\height}{
\includegraphics[page=28]{figures/figures_curve_31.pdf}
}
\qquad\xrightarrow[]{\Z_2} \qquad 
    \raisebox{-.5\height}{
    \includegraphics[page=29]{figures/figures_curve_31.pdf}
}
\end{align}

\begin{figure}[t]
    \centering
    \includegraphics[page=27]{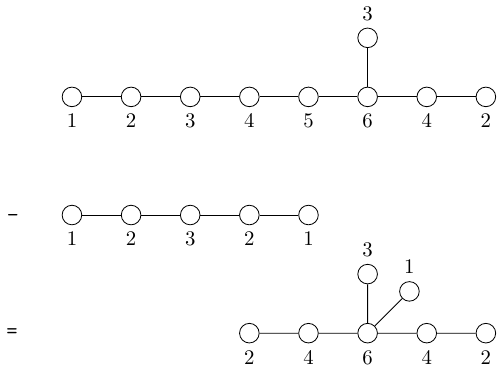}
    \caption{Quiver subtraction from the affine $E_8$ Dynkin quiver, with Coulomb branch $\clorbit{E_8}^{\min}$, which results in the quiver \eqref{eq:unitary_quiver_E6}.}
    \label{fig:quiver_sub}
\end{figure}

\paragraph{hyper-K\"ahler quotient and quiver subtraction.}
As a side remark, which is not related to the geometric Satake correspondence, it is an observation that the quiver \eqref{eq:unitary_quiver_E6} can also be obtained by subtraction: The starting point is the affine $E_8$ Dynkin quiver as this has $\clorbit{E_8}^{\min}$ as Coulomb branch. To realise the $\surm(3)$ hyper-K\"ahler quotient, one takes the quiver $(1)-(2)-(3)-(2)-(1)$ and aligns it with the $E_8$ Dynkin diagram such that the remaining set of balanced nodes after subtraction is the desired $E_6$ Dynkin diagram, see Figure \ref{fig:quiver_sub}. After rebalancing the quiver, one obtains \eqref{eq:unitary_quiver_E6}. 

\subsection{\texorpdfstring{$\sorm(7)$ with 2 spinors coupled to $\sprm(0)$}{SO7 coupled to Sp0}}
\label{sec:SO7+Sp0}
Consider $2$ M5 branes on $\C^2 \slash D_4$ with boundary conditions $\rho_L=(3,1^5)$, $\rho_R=(1^8)$.
The brane system for the $\sorm(7)$ gauge theory with 2 hypermultiplets in the spinor representation and coupled to an $\sprm(0)$ reads 
\begin{align}
\raisebox{-.5\height}{
\includegraphics[page=16]{figures/figures_curve_31.pdf}
	}
    \qquad 	\longleftrightarrow	\qquad 
		\raisebox{-.5\height}{
		\includegraphics[page=17]{figures/figures_curve_31.pdf}
	} 
\end{align}
The field theory is well-defined as $\sprm(0)$ perceives 16 half-hypermultiplets and $\sorm(7)$ is anomaly-free with 32  half-hypermultiplets, as the spinor representation is 8 dimensional. Each gauge group is accompanied by a tensor multiplet, cancelling potential gauge anomalies.

The brane system has three intervals between 4 half \NS\ branes. However, the interval with negatively charged \Ds\ branes does not contribute a tensor multiplet, as this tensor has already been traded for a number of hypermultiplets. Hence, there are only two tensor multiplets. The $\sorm(7)$ and $\sprm(0)$ gauge groups are a simple consequence of \Ds\ brane on top of O6 planes, see Appendix \ref{app:orientifolds}. Moreover, one notes that all \De\ branes are necessary for anomaly cancellation. It is not known how to deduce the matter transforming in the spinor representation from the brane configuration; instead, the results of Table \ref{tab:negative_branes} provide guidance.

By analogous reasoning as in Section \ref{sec:SU3+Sp0}, the brane system admits two infinite coupling Higgs branch phases. Once the pairs of half \NS\ brane have left the orientifold, they can either be separated or coincident, denoted by $(1^2)$ and $(2)$ respectively. The $(1^2)$ phase is associated with the collapse of the $(-1)$ curve. Hence, the appearance of a small $E_8$ instanton. Since the theory has an $\sorm(7)$ gauge symmetry, the non-abelian global symmetry of the infinite coupling Higgs branches receive a contribution from the commutant of $\sorm(7)$ inside $E_8$, which is $\sorm(9)$. In addition, the finite coupling Higgs branch global symmetry $\sprm(2)\cong \sorm(5)$ is still present at infinite coupling too. However, due to the presence of the non-trivial matter content, the infinite coupling Higgs branch is not simply a $\sorm(7)$ hyper-K\"ahler quotient of $\clorbit{E_8}^{\min}$.

The $(2)$ phase, the origin of the tensor branch, is reached after collapsing the remaining curve. Due to the collapse of the $-1$ curve, the $-3$ curve becomes a $-2$ curve, whose collapse is known to lead to an $S_2$ discrete gauging. Thus, the relation $\Higgs_{\infty}^{(2)} = \Higgs_{\infty}^{(1^2)} /// \Z_2$ holds, but now $\Higgs_{\infty}^{(1^2)}$ is not a simple space.

\paragraph{Magnetic quivers.}
The magnetic quiver for the phase with separated pairs of \NS\ branes reads
\begin{align}
(1^2):\qquad 
    \raisebox{-.5\height}{
    \includegraphics[page=34]{figures/figures_curve_31.pdf}
	} 	\qquad 
	\begin{cases}
	\dim\ \Coulomb &= 24 \,,  \\
	\gbal&=\sormL(9)\oplus \sormL(5) \,,
	\end{cases}
	\label{eq:magQuiv_SO7_C1C1}
\end{align}
and one straightforwardly evaluates the monopole formula as summarised in Table \ref{tab:HS_SO7_2spinors}. The computed dimension of the global symmetry is consistent with $\sormL(9)\oplus\sormL(5)$, which has dimension $46=36+10$. The PL reveals that the moduli space is rather complex. The generators transform in the adjoint representations at order $t^2$, and in the $\sorm(9)\times \sorm(5)$ bispinor representation at orders $t^3$, $t^4$, and $t^5$. While the adjoints are invariant under the centre symmetries, the bispinor is invariant under the diagonal $\Z_2$ centre symmetry. This suggests that the symmetry group is $(\Spin(9)\times\Spin(5))\slash \Z_2^{\mathrm{diag}}$.

The magnetic quiver for the Higgs branch phase at the origin of the tensor branch becomes
\begin{align}
(2):\qquad 
    \raisebox{-.5\height}{
    	\includegraphics[page=18]{figures/figures_curve_31.pdf}
	} 	\qquad 
	\begin{cases}
	\dim\ \Coulomb &= 24 \,, \\ 
	\gbal&=\sormL(9)\oplus \sormL(5) \,,
	\end{cases}
	\label{eq:magQuiv_SO7_C2}
\end{align}
and the Hilbert series is provided in Table \ref{tab:HS_SO7_2spinors}. Again, the $t^2$ coefficient is reflecting the $\sormL(9)\oplus \sormL(5)$ global symmetry. The $\sormL(9)$ factor is an infinite coupling enhancement, while the $\sprmL(2)$ factor is the finite coupling flavour symmetry. The symmetry group is $(\Spin(9)\times\Spin(5))\slash \Z_2^{\mathrm{diag}}$.

\begin{table}[t]
	\ra{1.25}
	\centering
	\begin{tabular}{crl}
		\toprule
		phase   &  \multicolumn{2}{c}{quantity} \\ \midrule
		$(2)$  &  $\Hh=$& $ 1 +46\ t^2 + 64\ t^3 + 1135\ t^4+ 2944\ t^5 + 21631\ t^6 +71744\ t^7 +O(t^8)$ \\
		& $ \HS_{\mathbb{Z}} =$  &  $ 1 + 46\ t^2 + 1135\ t^4  + 21631\ t^6 +O(t^8)$\\
		& $ \HS_{\mathbb{Z}+\frac{1}{2}} =$ &  $ 64\ t^3 + 2944\ t^5  + 71744\ t^7 +O(t^8) $  \\
		& $\PL =$&  $46\ t^2 + 64\ t^3 + 54\ t^4  - 229\ t^6  - 896\ t^7 +O(t^8)$ \\ \midrule
		$(1^2)$ & $\Hh=$ & $1 + 46\ t^2 + 64\ t^3 + 1145\ t^4 + 3008\ t^5 + 22271\ t^6 + 75200\ t^7 +O(t^8)$ \\
		& $\HS_{\mathbb{Z}} = $ &  $1 + 46\ t^2 + 1145\ t^4 +22271\ t^6 +O(t^8)$\\
		& $\HS_{\mathbb{Z}+\frac{1}{2}} =$&  $  64\ t^3 + 3008\ t^5  + 75200\ t^7+O(t^8)$\\
		& $\PL= $&$46\ t^2 + 64\ t^3 + 64\ t^4 + 64\ t^5 - 49\ t^6  -1024\ t^7+O(t^8) $  \\ \bottomrule
	\end{tabular}
	\caption{Perturbative Hilbert series for the different phases \eqref{eq:magQuiv_SO7_C1C1} and \eqref{eq:magQuiv_SO7_C2}.}
	\label{tab:HS_SO7_2spinors}
\end{table}

\subsection{Hasse diagram}
\label{sec:Hasse_curves_31}
\begin{figure}[t]
    \centering
\includegraphics[page=14]{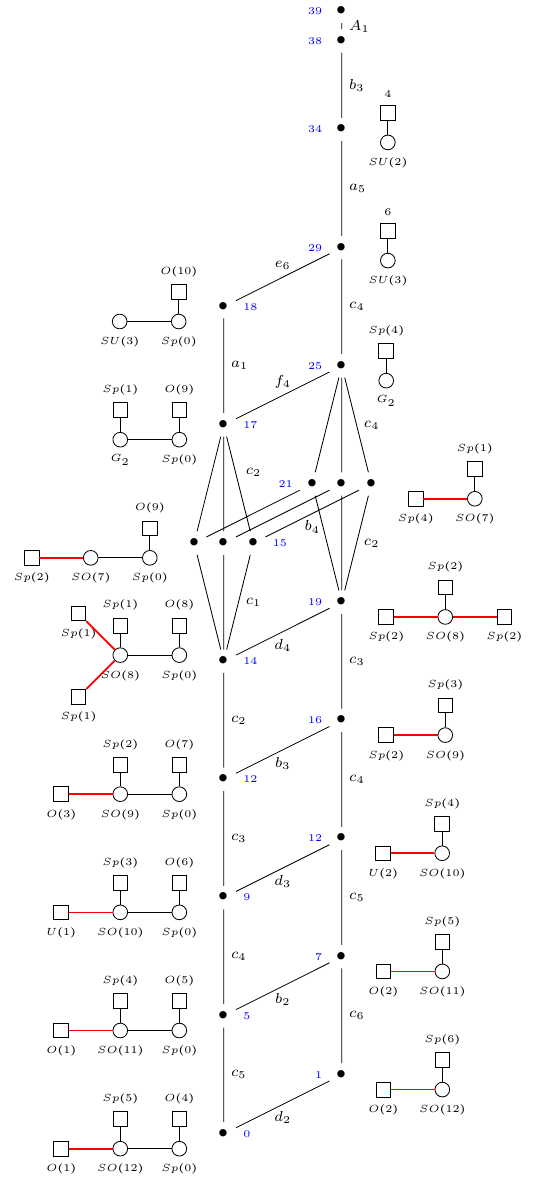}
\caption{Infinite coupling Higgs branch Hasse diagram for the $\sorm(n)\times \sprm(0)$-type theories. The leaves are denoted by the slice to the top, i.e.\ the leaves are denoted by the 6d (electric) theory. More precisely, this is the Hasse diagram for the infinite coupling Higgs branch of the $\sorm(12)\times\sprm(0)$ theory. The phase diagrams for the other theories are simply obtained by suitable reduction.}
    \label{fig:SO_theories_on_3}
\end{figure}

\begin{figure}[t]
    \centering
\includegraphics[page=15]{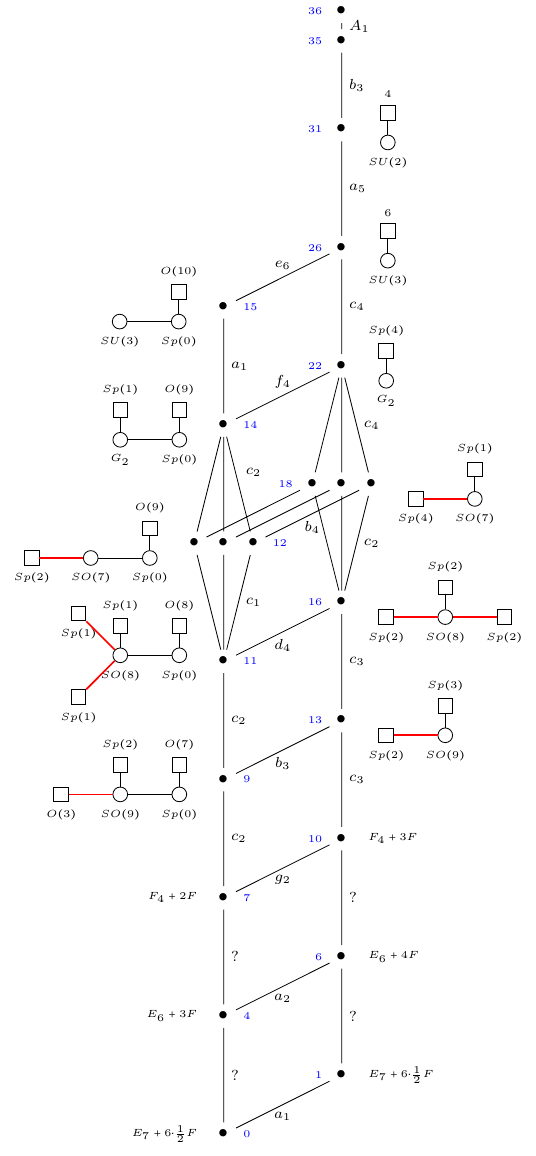}
\caption{Infinite coupling Higgs branch Hasse diagram for the set of $G\times \sprm(0)$ theories including exceptional factors $G=F_4,E_6,E_7$. Again, the leaves are denoted by the 6d (electric) theory. Concretely, this is the Hasse diagram for the infinite coupling Higgs branch of the $E_7$ theory with six $\frac{1}{2}F$. The phase diagrams for the other theories are simply obtained by suitable reduction.}
    \label{fig:E_theories_on_3}
\end{figure}

Even though we point out above that many models do not have a computable magnetic quiver, we are still able to provide a guess for the Higgs branch phase diagram. 
Based on the observations of this Section, one can conjecture the Hasse diagram for a theory of the form
\begin{align}
\label{eq:quiver_G_Sp0}
    \raisebox{-.5\height}{
    	\includegraphics[page=35]{figures/figures_curve_31.pdf}
	} \,,
\end{align}
where $F$ encodes the total global symmetry (possibly containing different factors from hypermultiplets in various representations).
To obtain the phase diagram, recall that there are two possible transitions: (i) If $G$ is not the minimal $\surm(3)$ theory, then one can higgs the $G$ gauge theory according to the Hasse diagram of the $-3$ curve, see Figure \ref{fig:Hasse_3_curve}.  (ii) One could decide to collapse the $-1$ curve, which leaves behind a $-2$ curves with the same gauge group $G$, but some modified matter content. The Hasse diagram for this theory is detailed in Figure \ref{fig:Hasse_2_curve}. Thus, all that is left to do is to specify the transition of the collapsing $-1$ curve coupled to a $G$ gauge theory on the $-3$ curve. The symmetry of this transition is simply the commutant $H= C_{E_8}(G)$ of $G$ inside $E_8$, as observed in Sections \ref{sec:SU3+Sp0} and \ref{sec:SO7+Sp0}. Further intuition is gained by inspecting the (infinite coupling) Higgs branch dimensions on a $-2$ curve and the $(-3)(-1)$ curve, see Table \ref{tab:dim_and_commutant}. As a result, the transition can be identified with $h=\clorbit{H}^{\min}$ --- the minimal nilpotent orbit closure of the commutant $H$.
The full Hasse diagram of \eqref{eq:quiver_G_Sp0} is a combination of the Hasse diagram of the $G$ gauge theory on the $-2$ and $-3$ curves, where the transitions in between are of type $h=\clorbit{H}^{\min}$. Figures \ref{fig:SO_theories_on_3} and \ref{fig:E_theories_on_3} display the infinite coupling phase diagrams. 

\begin{table}[t]
    \centering
    \ra{1.25}
    \begin{tabular}{ccc|ccc|ccc}
    \toprule
    \multicolumn{3}{c|}{$-2$ curve} & \multicolumn{3}{c|}{$(-3)(-1)$ curves} & \multicolumn{3}{c}{transition} \\ 
        $G$ & matter & $\dim\ \Higgs_\infty$ & $G$ & matter & $\dim\ \Higgs_\infty$ & $\Delta \dim\ \Higgs_\infty$  &$C_{E_8}(G)$ & $\clorbit{H}^{\min}$  \\\midrule
        $\surm(3)$ &  $6F$ & $10$ &  $\surm(3)$ &  --- & $21$ & $11$ & $E_6$ & $e_6$\\
        $G_2$ & $4F$ & $14$ & $G_2$  &$F$ &$22$ &  $8$ & $F_4$ & $f_4$ \\
        $\sorm(7)$ & $1F+4S$ & $18$ & $\sorm(7)$ & $2S$ & $24$ & $6$& $\sorm(9)$ & $b_4$  \\
        $\sorm(8)$ & $2F+2S+2C$ & $20$ & $\sorm(8)$ & $F+S+C$ & $25$ & $5$ & $\sorm(8)$ & $d_4$\\
        $\sorm(9)$ & $3F+2S$ & $23$  & $\sorm(9)$ & $2F+S$ & $27$ & $4$  & $\sorm(7)$ & $b_3$ \\
        $\sorm(10)$ & $4F+2S $ & $27$ & $\sorm(10)$ & $3F+1S $ & $30$ & $3$ &  $\sorm(6)$  & $d_3$ \\
        $\sorm(11)$ & $5F+S$ & $32$ &  $\sorm(11)$ & $4F+\frac{1}{2}S$ & $34$ &  $2$ & $\sorm(5)$ &$b_2$ \\
        $\sorm(12)$ & $6F+2\cdot \frac{1}{2}S$ & $38$ & $\sorm(12)$ & $5F+ \frac{1}{2}S$ & $39$ &$1$ & $\sorm(4)$ & $d_2$ \\ \midrule
        $F_4$ & $3F$ & $26$  & $F_4$ & $2F$ & $29$ & $3$ & $G_2$ & $g_2$\\
        $E_6$ & $4F$ & $30$ & $E_6$& $3F$ & $32$ & $2$ & $\surm(3)$ & $a_2$ \\
        $E_7$ & $6\cdot \frac{1}{2}F$ & $35$ & $E_7$& $5\cdot \frac{1}{2}F$  & $36$ & $1$ & $\surm(2)$ & $a_1$ \\ \bottomrule
    \end{tabular}
    \caption{Higgs branch dimension and commutant $C_{E_8}(G)$. The anomaly-free matter content for the $\sorm(n)$ gauge groups can be taken from \cite{Danielsson:1997kt}. Note that $\dim \Higgs_\infty$ for the $(-3)(-1)$ curves crucially depends on the $-1$ curve, due to the small $E_8$ instanton transition.}
    \label{tab:dim_and_commutant}
\end{table}

Following the transitions in \eqref{eq:SO12-SO11-SO10} and \eqref{eq:SO10-SO9-SO8}, transitions from $D$-type to $B$-type gauge groups are often not visible in the brane system, despite the clear field theory description. In view of Table \ref{tab:dim_and_commutant}, theories with $G=\sorm(2k+1)$ ($k=3,4,5$) have a $B$-type commutant $H=C_{E_8}(G)=\sorm(15-2k)$. The corresponding $b_{7-k}$ is generically deducible in the brane setting.
Likewise, the brane system \eqref{eq:Higgs_SO7-G2-SU3} for $G_2$ is not distinguishable from the $\surm(3)$ phase. Then the non-simply laced $f_4$ transition is also not visible from the branes. The observation is that transitions that are not visible in the brane system are consistent with non-simply laced algebras.

\FloatBarrier

\FloatBarrier

\section{\texorpdfstring{Multiple M5s on $\C^2\slash D_k$ with one non-trivial boundary condition}{Multiple M5s on D-type ALE with one non-trivial boundary condition}}
\label{sec:3M5s_RHS_trivial}
The case of 2 M5 branes in Section \ref{sec:2M5s_RHS_trivial} has already revealed exciting Higgs branch geometries related to the theories on a $-3$ curve. However, one is limited to certain boundary conditions that can be realised by the number of brane intervals.  In this section, further boundary conditions are explored by increasing the number of M5 branes. We start with the  minimal $D_4$ singularity and proceed to higher $D_k$.

\subsection{\texorpdfstring{$(k-1)$ M5s with $\rho_L=(2k-3,3)$, $\rho_R=(1^{2k})$}{k-1 M5s}}
\subsubsection{\texorpdfstring{3 M5s on $D_4$ singularity:  $\sprm(1)\times G_2$ coupled to $\sprm(0)$}{3 M5s on D4 singularity: Sp1 x G2 coupled to Sp0}}
\label{sec:Sp1+G2_coupled_to_Sp0}
Consider 3 M5 branes on a $\C^2 \slash D_4$ singularity with boundary conditions $\rho_L=(5,3)$, $\rho_R=(1^8)$, for which the brane system and 6d quiver become
\begin{align}
    \raisebox{-.5\height}{
\includegraphics[page=1]{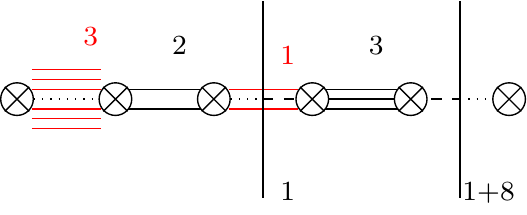}
	}
	\qquad \longleftrightarrow  \qquad 
	\raisebox{-.5\height}{
	 \includegraphics[page=2]{figures/figures_curve_231.pdf}
	} 
	\label{eq:Sp1+G2_curves_231}
\end{align}
To begin with, the field theory is anomaly-free, because $\sprm(1)=\surm(2)$ has 8 half-hypermultiplets, $G_2$ has one fundamental flavour, and $\sprm(0)$ perceives 16 half-hypermultiplets. Each gauge group is accompanied by one tensor multiplet, in total three.

The brane system contains 5 brane intervals between six half \NS\ brane, two of which have negatively charged branes. Consequently, the tensor multiplets of those intervals are traded for a certain number of hypermultiplets, dictated by the gravitational anomaly cancellation condition. The remaining three dynamical tensor multiplets are associated to the intervals with positive or vanishing \Ds\ brane number. The gauge algebras and matter content are identified by using Table \ref{tab:negative_branes}.

The infinite coupling phases are reached as before. The half \NS\ branes lift pairwise off the orientifold. This first step is understood as collapse of the $-1$ curve, signalling a small $E_8$ instanton. Thereafter, the three pairs of half \NS\ admit distinct phases, labelled by partitions of $3$: all pairs are pairwise separated $(1^3)$, two pairs are coincident and the third is separated $(2,1)$, and all three pairs are coincident $(3)$.

The $(1^3)$ phase is directly reached after the collapse of the $-1$ curve, which simultaneously reduces the remaining curves from $(-2)(-3)$ to $(-2)(-2)$. Again, vanishing $-2$ curves results in discrete gauging: one collapse yields an $S_2$ gauging of $(1^3)$ to $(2,1)$, and collapsing both curves yields an $S_3$ gauging of $(1^3)$ to $(3)$.

While the finite coupling Higgs branch of \eqref{eq:Sp1+G2_curves_231} is a point, the infinite coupling Higgs branches are non-trivial and are expected to exhibit an exceptional $F_4$ global symmetry. This is because the small instanton transition of the $-1$ curve is coupled to a $G_2$ gauge theory and the commutant of $G_2$ inside $E_8$ is $F_4$. The discrete gauging transitions of the $-2$ curves do not affect this global symmetry.

\paragraph{Magnetic quiver.}
The magnetic quivers for the phases $(1^3)$ and  $(3)$ are given by
\begin{align}
(1^3):\qquad 	&\raisebox{-.5\height}{
	        	\includegraphics[page=3]{figures/figures_curve_231.pdf}
	} 
	\qquad 
	\begin{cases}
	\dim\ \Coulomb &= 20 \;,\\
	\gbal &= \sormL(9) \;,
	\end{cases}
		\label{eq:magQuiv_Sp1+SO7_111}
 \\
(3):\qquad 	&\raisebox{-.5\height}{
	        	\includegraphics[page=4]{figures/figures_curve_231.pdf}
	} 
	\qquad 
	\begin{cases}
	\dim\ \Coulomb &= 20 \;,\\
	\gbal &= \sormL(9) \;.
	\end{cases}
	\label{eq:magQuiv_Sp1+SO7_3}
\end{align}
To understand these Coulomb branches, one notices that the phase $(1^3)$ has been evaluated in \cite{Bourget:2021zyc}. The Hilbert series is given by
\begin{subequations}
\begin{align}
   \HS\eqref{eq:magQuiv_Sp1+SO7_111}&= 1 + 52\ t^2 + 1455\ t^4 + 28834\ t^6 +
449122\ t^8 + 5793780\ t^{10} + 63853945\ t^{12}  +
613989328\ t^{14}  \notag \\ &\quad + 5232181818\ t^{16} +
40010832518\ t^{18}  + 277431116267\ t^{ 20} +
O\left(t^{22} \right) \,,  \\
\PL\eqref{eq:magQuiv_Sp1+SO7_111}&= 52\ t^2 + 77\ t^4 + 26\ t^6 − 2394\ t^8 −
5442\ t^{10} +O\left( t^{12}\right) \,,
\end{align}
\end{subequations}
which displays a global $F_4$ symmetry. The Coulomb branch is the $S_4$ cover of the closure of the 20-dimensional nilpotent orbit of $F_4$ with Bala-Carter label $F_4(a_3)$, see also Figure \ref{fig:Hasse_F4}. Explicitly,
\begin{align}
    \Higgs_{\infty}^{(1^3)} = \Coulomb \eqref{eq:magQuiv_Sp1+SO7_111} \qquad \text{and} \qquad 
    \clorbit{F_4(a_3)} = \Coulomb \eqref{eq:magQuiv_Sp1+SO7_111} /// S_4 \,.
\end{align}
From this, the phase $(3)$ is easily understood. The difference between \eqref{eq:magQuiv_Sp1+SO7_111} and \eqref{eq:magQuiv_Sp1+SO7_3} is an $S_3$ quotient \cite{Hanany:2018cgo}. Because $S_3$ is a normal subgroup of $S_4$ with quotient group $\Z_4$, the moduli space satisfies
\begin{align}
     \Higgs_{\infty}^{(3)} = \Coulomb \eqref{eq:magQuiv_Sp1+SO7_3} = \Coulomb \eqref{eq:magQuiv_Sp1+SO7_111} ///S_3  \qquad \text{and} \qquad 
    \clorbit{F_4(a_3)} = \Coulomb \eqref{eq:magQuiv_Sp1+SO7_3} /// \Z_4 \,.
\end{align}
Since the $F_4$ commutes with the permutation groups, the Coulomb branch of \eqref{eq:magQuiv_Sp1+SO7_3} is also expected to have a global symmetry $F_4$. Note that the subset of balanced nodes in \eqref{eq:magQuiv_Sp1+SO7_111} and \eqref{eq:magQuiv_Sp1+SO7_3} makes an $\sormL(9)$ manifest and $\sormL(9)\subset \ffour$ is a maximal subalgebra. The phase $(2,1)$ can be analysed by the same reasoning, starting from an $S_2$ quotient of \eqref{eq:magQuiv_Sp1+SO7_111}.

Further confirmation is obtained by computing the monopole formula for \eqref{eq:magQuiv_Sp1+SO7_3}, which results in
\begin{subequations}
\begin{align}
    \mathrm{H}\eqref{eq:magQuiv_Sp1+SO7_3}&= 1+52 t^2+1403 t^4+26078 t^6+O\left(t^7\right) \;,\\
    \mathrm{HS}_{\mathbb{Z}} &= 1+36 t^2+811 t^4+13902 t^6+O\left(t^7\right) \;,\\
    \mathrm{HS}_{\mathbb{Z}+\tfrac{1}{2}} &=16 t^2+592 t^4+12176 t^6+O\left(t^7\right) \;,\\
    \mathrm{PL}\eqref{eq:magQuiv_Sp1+SO7_3}&=52 t^2+25 t^4-26 t^6+O\left(t^7\right) \;.
    \label{eq:HS_Sp1+SO7_3_PL}
\end{align}
\end{subequations}
The integer lattice Hilbert series $\mathrm{HS}_{\mathbb{Z}}$ seems to have an $\sormL(9)$ global symmetry with dimension $36$. While it is clear from \eqref{eq:HS_Sp1+SO7_3_PL} that the infinite coupling Higgs branch moduli space of \eqref{eq:Sp1+G2_curves_231} is not a nilpotent orbit of $F_4$, we encounter a close cousin in Section \ref{sec:2231_curve_F4_orbit} that is in fact the closure of the  nilpotent orbit $F_4(a_3)$.  

\subsubsection{\texorpdfstring{$(k-1)$ M5s on $D_k$ singularity}{k-1 M5s on Dk singularity}}
The above case admits a simple generalisation: 
consider $k-1\geq 3$ M5 branes on a $\C^2 \slash D_{k}$ singularity with boundary conditions $\rho_L=(2k-3,3)$, $\rho_R=(1^{2k})$. The 6d quiver is now given by
\begin{align}
    \raisebox{-.5\height}{
\includegraphics[page=5]{figures/figures_curve_231.pdf}
	} 
\end{align}
which has non-abelian $\sorm(2k+1)$ flavour symmetry. The infinite coupling Higgs phases are conceptually similar to the discussion above. However, in the generic case there is no expectation on symmetry enhancement at the conformal fixed point. This is because the $\sprm(0)$ node is coupled to a $G_2$ and $\sorm(9)$, which reduces the $E_8$ global symmetry to at most a discrete subgroup. Likewise, the $-1$ curve at the right-hand side is coupled to a $-4$ curve which supports a minimal $\sormL(8)$ gauge algebra. Thus, the global symmetry factor is not exceptional.

The Higgs branch at the conformal fixed point is described by the following magnetic quiver:
\begin{align}
	\raisebox{-.5\height}{
	 	\includegraphics[page=6]{figures/figures_curve_231.pdf}
	} 
	\qquad 
	\begin{cases}
	\dim\ \Coulomb &= k(k+1)\;,\\
	\gbal &= \sormL(2k+1)\;. 
	\end{cases}
\end{align}
In contrast to the minimal case $k=4$, the generic case is not expected to have an exotic global symmetry. The symmetry group is $\Spin(2k+1)$.

From the magnetic quiver in conjunction with the monopole formula, this can be understood as follows: the integer lattice contribution gives rise to the global symmetry $\sormL(2k+1)$, visible at order $t^2$. The first contribution for the half-integer lattice is the spinor representation for $\sormL(2k+1)$ at order $t^\Delta$. One observes, see for instance \cite[Tab.\ 3]{Bourget:2021zyc}, that $\Delta=\frac{1}{2} (k-4)^2 + \frac{5}{2} (k-4) +2$ such that only the spinor of $\sormL(9)$ has the suitable R-charge to contribute to the global symmetry. This leads to the enhancement $\sorm(9)\to F_4$, while in all higher $k>4$ cases, the R-charge of the spinor representation is too high.
%
\subsection{\texorpdfstring{$(k-1)$ M5s with $\rho_L=(2k-3,1^3)$, $\rho_R=(1^{2k})$}{k-1 M5s}}
\subsubsection{\texorpdfstring{3 M5s on $D_4$ singularity: $\sprm(1)\times \sorm(7)$ coupled to $\sprm(0)$}{3 M5s on D4 singularity:Sp1 x SO7 coupled to Sp0}}
\label{sec:3M5_Sp1xSO7}
In similar spirit, consider 3 M5 branes on a $\C^2 \slash D_4$ singularity with boundary conditions $\rho_L=(5,1^3)$, $\rho_R=(1^8)$. The brane system and the 6d quiver are 
\begin{align}
    \raisebox{-.5\height}{
    \includegraphics[page=7]{figures/figures_curve_231.pdf}
	}
	\qquad \longleftrightarrow \qquad 
	\raisebox{-.5\height}{
	  	\includegraphics[page=8]{figures/figures_curve_231.pdf}
	} 
\end{align}
and most of the discussion of Section \ref{sec:Sp1+G2_coupled_to_Sp0} is straightforwardly applied here. For instance, the field theory is well-defined and there are only three tensor multiplets. The differences is the non-abelian flavour symmetry: an $\sprm(1)$ for the bi-spinor hypermultiplet of $\sorm(7)$.
At infinite coupling, one expects an enhancement of the global symmetry by the commutant $C_{E_8}(\sorm(7))$ of $\sorm(7)$ inside $E_8$, which is $\sorm(9)$.

\paragraph{Magnetic quiver.}
The infinite coupling magnetic quiver is readily derived to be
\begin{align}
	\raisebox{-.5\height}{
	   \includegraphics[page=9]{figures/figures_curve_231.pdf}
	} 
	\qquad 
	\begin{cases}
	\dim\ \Coulomb &= 21 \;,\\
	\gbal &= \sormL(9) \oplus \sormL(3)\;.
	\end{cases}
	\label{eq:magQuiv_Sp1+G2}
\end{align}
The expected global symmetry has a classical factor of $\sprmL(1)\cong \sormL(3)$. The remainder is expected as infinite coupling contribution. Note that the $\sormL(9)$ could simply be the commutant of the $\sormL(7)$ gauge algebra on the $-3$ curve inside $E_8$. Further verification is obtained via the monopole formula
\begin{subequations}
\begin{align}
    \mathrm{H}&= 1+39 t^2+32 t^3+789 t^4+1248 t^5+11536 t^6+O\left(t^7\right) \;,\\
    \mathrm{HS}_{\mathbb{Z}} &=1+39 t^2+789 t^4+11536 t^6+O\left(t^7\right) \;,\\
    \mathrm{HS}_{\mathbb{Z}+\tfrac{1}{2}} &=32 t^3+1248 t^5+O\left(t^7\right) \;,\\
    \mathrm{PL}&= 39 t^2+32 t^3+9 t^4-3 t^6+O\left(t^7\right) \;. 
\label{eq:PL_Sp1+G2}
\end{align}
\end{subequations}
The global symmetry dimension is consistent with the symmetry obtained from the balanced nodes, i.e.\ $\sormL(9)\oplus \sormL(3)$ with dimension $36+3=39$. The global symmetry is $\left(\surm(2)\times \mathrm{Spin}(9)\right) \slash \Z_2$ with $\Z_2 $ the diagonal combination of the $\Z_2$ centre symmetries of $\sorm(9)$ and $\surm(2)$. To see this, one identifies the generators from the PL \eqref{eq:PL_Sp1+G2}: order $t^2$ transforms in the adjoint representations of $\surm(2)\times\sorm(9)$, order $t^4$ transforms in the $\surm(2)\times\sorm(9)$ bispinor representation, while order $t^4$ transforms in the $\sorm(9)$ vector representation. The adjoint and vector representations are invariant under the centre symmetries, whereas the bispinor is only invariant under the diagonal combination.

\subsubsection{\texorpdfstring{$(k-1)$ M5s on $D_k$ singularity}{k-1 M5s on Dk singularity}}
Again, one recognises a pattern. Consider $k-1\geq 3$ M5 branes on a $\C^2 \slash D_{k}$ singularity with boundary conditions $\rho_L=(2k-3,1^3)$, $\rho_R=(1^{2k})$, i.e.\
\begin{align}
    \raisebox{-.5\height}{
	\includegraphics[page=10]{figures/figures_curve_231.pdf}
	} 
\end{align}
and the associated infinite coupling magnetic quiver becomes
\begin{align}
	\raisebox{-.5\height}{
	  \includegraphics[page=11]{figures/figures_curve_231.pdf}
	} 
	\qquad 
	\begin{cases}
	\dim\ \Coulomb &= k(k+1)+1 \;,\\
	\gbal &= \sormL(2k+1) \oplus \sormL(3)\;.
	\end{cases}
\end{align}
The symmetry group is $(\surm(2)\times\Spin(2k+1))\slash \Z_2$ with $\Z_2$ the diagonal combination of the two centre symmetries.

\subsection{\texorpdfstring{$k$ M5s with $\rho_L=(2k-1,1)$, $\rho_R=(1^{2k})$}{k M5s}}
\label{sec:4M5s_RHS_trivial}
\subsubsection{\texorpdfstring{$4$ M5s on $D_4$ singularity: $\sprm(1)\times G_2$ coupled to $\sprm(0)$}{4 M5s on D4 singularity:Sp1 x G2 coupled to Sp0}}
\label{sec:2231_curve_F4_orbit}
Consider 4 M5 branes on a $\C^2 \slash D_4$ singularity with boundary conditions $\rho_L=(7,1)$, $\rho_R=(1^8)$. The brane configuration and the 6d quiver are given by
\begin{align}
\label{eq:branes_223_curve}
    \raisebox{-.5\height}{
  \includegraphics[page=1]{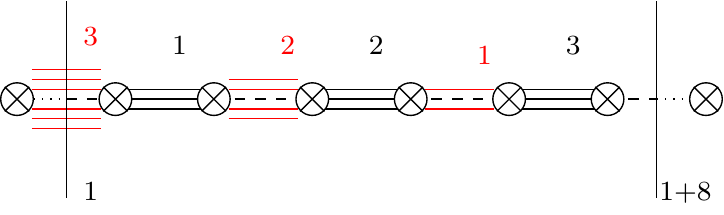}
	}
	\qquad \longleftrightarrow \qquad 
	    \raisebox{-.5\height}{
\includegraphics[page=2]{figures/figures_curve_2231.pdf}
	} 
\end{align}
and one first verifies that the field theory is anomaly-free: $\sprm(1)\cong \surm(2)$ has 8 half-hypermultiplets, $G_2$ has one fundamental, and $\sprm(0)$ has 16 half-hypermultiplets. Also, each gauge group factor as accompanied by one tensor multiplet. Moreover, there is one additional tensor.

The brane configuration contains 7 brane intervals between 8 half \NS\ branes. However, three negative brane intervals reduce the number of tensor multiplets to 4 by converting 3 tensor into a certain number of hypermultiplets. Again, the gauge groups and matter content are not obvious from the brane system, but are taken from Table \ref{tab:negative_branes}.

One notes that the gauge theory data is identical to \eqref{eq:Sp1+G2_curves_231}, but here there exists an additional tensor multiplet, reminiscent of the situation in Section \ref{sec:appetiser}.

The infinite coupling Higgs branch phases are addressed as above. The \NS\ pairs join pairwise along the O$6$ plane and can lift off. This corresponds to the collapsing $-1$ curve, which is accompanied by the small $E_8$ instanton. Since the $-1$ is coupled to a $G_2$ gauge theory, one expects a global symmetry of $F_4$, which is the commutant of $G_2$ inside $E_8$. Again, there are several infinite coupling phases labelled by partitions of $4$, since there are 4 pairs of \NS\ branes which can coincide in the pattern of a given partition. The transition between $(1^4)$ and any other transition is realised by discrete $S_d$ gauging. As by now familiar, this can be understood in terms of curves, because the collapse of the $-1$ curve leaves behind three $-2$ curves.

\paragraph{Magnetic quiver.}
The Higgs branch at the conformal fixed point is captured by the magnetic quiver of partition $(4)$, which reads
\begin{align}
	\raisebox{-.5\height}{
	\includegraphics[page=3]{figures/figures_curve_2231.pdf}
	} 
	\qquad 
	\begin{cases}
	\dim\ \Coulomb &= 20 \,, \\
	\gbal &= \sormL(9) \,.
	\end{cases}
	\label{eq:magQuiv_Sp1+G2_223}
\end{align}
The set of balanced nodes suggest at least an $\sormL(9)$ global symmetry. However, recall that the commutant $G_2 \subset E_8$ is $F_4$, with $\sormL(9)\subset \ffour$ being a maximal subalgebra. Such a symmetry group was observed in \cite[Tab.\ 3]{Bourget:2021zyc}, where the quiver is an $S_4$ cover of \eqref{eq:magQuiv_Sp1+G2_223}, i.e. the phase $(1^4)$.
The Higgs branch $\Higgs_\infty$, at the origin of the tensor branch, that is captured by \eqref{eq:magQuiv_Sp1+G2_223} is identified with the closure of the nilpotent orbit of $F_4$ of dimension $20$. The unrefined Hilbert series \cite{Hanany:2017ooe} is known to be 
\begin{subequations}
\label{eq:HS_F4_dim20}
\begin{align}
    \Hh&=  \frac{(1+t^2) }{(1-t^2)^{40}}  \cdot 
    \big(  1 
    + 10 t^2 
    +56 t^4
    +230 t^6
    +745 t^8 
    + 1946 t^{10}  
    + 4112 t^{12} 
    + 7028 t^{14}
    +9692 t^{16} \\
    &
    +10782 t^{18} 
    +10782  t^{20}
    +9692 t^{22}
    + 7028 t^{24}
    + 4112  t^{26}
    + 1946  t^{28}
    +745  t^{30}
    +230  t^{32}
    +56  t^{34}
    + 10 t^{36}
    +t^{38}
    \big) \notag
\end{align}
and perturbative expansion yields
\begin{align}
    \Hh=1+52 t^2+1377 t^4+24752 t^6+338951 t^8+O\left(t^9\right) \;, \quad 
    \PL= 52 t^2-t^4-726 t^8+O\left(t^9\right)\,.
\end{align}
\end{subequations}
The expectations on the moduli space can be verified on the level of Hilbert series by evaluating the monopole formula for \eqref{eq:magQuiv_Sp1+G2_223}. One finds
\begin{subequations}
\begin{align}
    \Hh&=  1+52 t^2+1377 t^4+24752 t^6+O\left(t^7\right) \,,\\
    \HS_{\mathbb{Z}} &=1+36 t^2+801 t^4+13296 t^6+O\left(t^7\right) \,,\\
    \HS_{\mathbb{Z}+\tfrac{1}{2}} &= 16 t^2+576 t^4+11456 t^6+O\left(t^7\right) \,,\\
    \PL&=52 t^2-t^4+O\left(t^7\right) \,,
\end{align}
\end{subequations}
and the expected $F_4$ symmetry is consistent with the dimension $52$ term at order $t^2$. Again, the integer lattice displays the $\sormL(9)$ maximal subalgebra. The result agrees with \eqref{eq:HS_F4_dim20} at the given order of expansion.
The identification of the moduli space as the closure of the nilpotent orbit $F_4(a_3)$ of $F_4$ allows us to compute the Hasse diagram for this theory, see Figure \ref{fig:Hasse_F4}. 

\paragraph{Hasse diagram.}
The identification of the infinite coupling Higgs branch with an $F_4$ orbit leads to the intriguing question whether this can be understood from 6d theory and its magnetic quiver.

\begin{figure}[t]
    \centering
    \includegraphics[page=10]{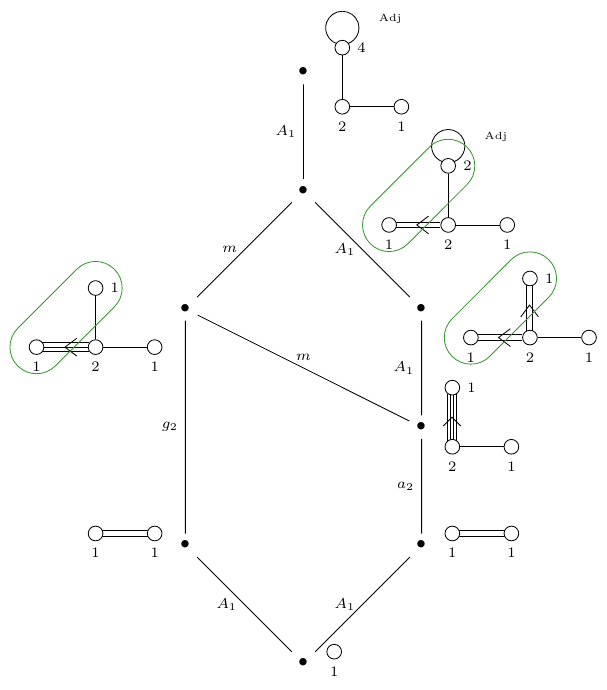}
    \caption{Hasse diagram for the infinite coupling magnetic quiver of the 6d $\surm(2)\times \surm(2)$ quiver theory \eqref{eq:Higgs_SU1xSU2xSU3_to_SU1xSU2xSU2} with 3 tensor multiplets obtained via quiver subtraction. In some quivers, certain notes are grouped together by a green line, called ``decoration'' in \cite{Bourget:2022ehw}. For those magnetic quivers, the evaluation of the monopole formula is not clear; however, quiver subtraction together with decoration does allow to provide a guess for the Hasse diagram. The guess is consistent with Figure \ref{fig:Hasse_F4}, up to the $[2A_1]$ transition, which here only appears as $A_1$.}
    \label{fig:SU1xSU2xSU2_Hasse}
\end{figure}

To begin with, recall that the theory in \eqref{eq:branes_223_curve} cannot be higgsed any further. The only possible transition is the collapse of the $-1$ curves, through which a single tensor multiplet is converted into a number of hypermultiplets. The $G_2$ gauge theory on the $-3$ becomes a $G_2$ theory on a $-2$ curve, i.e.\
\begin{align}
     \raisebox{-.5\height}{
\includegraphics[page=2]{figures/figures_curve_2231.pdf}
	}
	\qquad \longrightarrow
\qquad
	 \raisebox{-.5\height}{
\includegraphics[page=4]{figures/figures_curve_2231.pdf}
	}
	\cong
	 \raisebox{-.5\height}{
\includegraphics[page=5]{figures/figures_curve_2231.pdf}
	}
\end{align}
and as argued above in Section \ref{sec:Hasse_curves_31}, this is an $f_4$ transition. The resulting 6d quiver theory is defined on a chain of three $-2$ curves, and the $\sprm(3)$ flavour indicates a clear partial Higgs mechanism for $G_2\to \surm(3)$. In detail,
\begin{align}
     \raisebox{-.5\height}{
\includegraphics[page=5]{figures/figures_curve_2231.pdf}
	}
	\qquad \longrightarrow\qquad 
	 \raisebox{-.5\height}{
\includegraphics[page=6]{figures/figures_curve_2231.pdf}
	}
\end{align}
which is a $c_3$ transition. More to the point, the resulting theory is equivalently realised by 4 M5 branes on an $\C^2\slash \Z_4$ singularity with boundary conditions $\rho_L=(4)$, $\rho_R=(1^4)$, see Table \ref{tab:SU4_examples}. The infinite coupling magnetic quiver is given by
\begin{align}
     \raisebox{-.5\height}{
\includegraphics[page=7]{figures/figures_curve_2231.pdf}
	}
		\qquad 
	\begin{cases}
	\dim\ \Coulomb &= 9 \,,\\
	\gbal &= \surmL(4) \,.
	\end{cases}
\end{align}
The next step is the partial Higgs mechanism for the fundamental flavours on the $\surm(3)$ gauge group
\begin{align}
     \raisebox{-.5\height}{
\includegraphics[page=6]{figures/figures_curve_2231.pdf}
	}
	\qquad \longrightarrow\qquad 
	 \raisebox{-.5\height}{
\includegraphics[page=8]{figures/figures_curve_2231.pdf}
	}
	\label{eq:Higgs_SU1xSU2xSU3_to_SU1xSU2xSU2}
\end{align}
which is a $a_3$ transition. The  infinite coupling magnetic quiver for the $\surm(2)\times\surm(2)$ theory with 3 tensor multiplets is given by: 
\begin{align}
     \raisebox{-.5\height}{
\includegraphics[page=9]{figures/figures_curve_2231.pdf}
	}
		\qquad 
	\begin{cases}
	\dim\ \Coulomb &= 6 \,, \\
	\gbal &= \surmL(2)\oplus\surmL(2) \,.
	\end{cases}
\end{align}
Since this quiver is small enough, one can attempt a direct evaluation of the Hasse diagram via quiver subtraction. The result is displayed in Figure \ref{fig:SU1xSU2xSU2_Hasse}. It follows that composing this phase diagram with a sequence of $a_3$, $c_3$, and $f_4$ transitions at the bottom yields the $F_4$ Hasse diagram of Figure \ref{fig:Hasse_F4}. The only deviation occurs for the expected $[2A_1]$ transition, which here only appears as single $A_1$. By using a combination of techniques --- 6d quiver theories, brane systems, magnetic quivers, and quiver subtraction --- we reconstructed the Hasse diagram of a rather non-trivial nilpotent orbit of $F_4$.

\subsubsection{\texorpdfstring{$k$ M5s on $D_k$ singularity}{k M5s on Dk singularity}}
The setup can be generalised to beyond the minimal $D$-type singularity.
Consider $k\geq 4$ M5 branes on a $\C^2 \slash D_k$ singularity with boundary conditions $\rho_L=(2k-1,1)$, $\rho_R=(1^{2k})$. The brane configuration is a simple generalisation of \eqref{eq:branes_223_curve} and the 6d quiver theory reads
\begin{align}
\raisebox{-.5\height}{
\includegraphics[page=11]{figures/figures_curve_2231.pdf}
	} 
\end{align}
which can be analysed in the same fashion as above. Most importantly, the finite coupling non-abelian global symmetry $\sormL(2k+1)$ is not expected to enhance at infinite coupling.

The magnetic quiver for the Higgs branch at the conformal fixed point is readily evaluated to be
\begin{align}
	\raisebox{-.5\height}{
	\includegraphics[page=12]{figures/figures_curve_2231.pdf}
	} 
	\qquad 
	\begin{cases}
	\dim\ \Coulomb &= k(k+1) \,,\\
	\gbal &= \sormL(2k+1) \,,
	\end{cases}
	\label{eq:magQuiv_Sp1+G2_223_general}
\end{align}
and the global symmetry is $\sormL(2k+1)$ as indicated by the balanced set of notes, see also\footnote{Note that the quivers appearing in \cite[Tab.\ 3]{Bourget:2021zyc} are in a work titled ``hyper-K\"ahler implosions'', but they are a failed attempt to describe the hyper-K\"ahler implosion of the nilpotent cone of $\sormL(2k+1)$. Thus, \eqref{eq:magQuiv_Sp1+G2_223_general} is not related to the implosion of a maximal orbit closure. Instead these quivers turn out to have a natural physical interpretation as magnetic quivers for brane systems of the type \eqref{eq:branes_223_curve}, which describe Higgs branches of 6d theories.} \cite[Tab.\ 3]{Bourget:2021zyc}. The symmetry group is $\Spin(2k+1)$.
As in Section \ref{sec:3M5_Sp1xSO7}, the integer lattice gives rise to generator in the adjoint of $\sorm(2k+1)$ at order $t^2$, while the half-integer lattice contributes a
generator in the spinor representation of $\sorm(2k+1)$ at order $t^\Delta$. Crucially, $\Delta=2$ only for $k=4$, leading to the enhancement $\sorm(9)\to F_4$.

\section{Combining boundary conditions}
\label{sec:non-overlapping_bc}
The brane systems considered above always have a trivial boundary condition on one side, and a non-trivial boundary condition on the other side.  In this Section we continue the exploration to non-trivial boundaries conditions on both sides.
\subsection{\texorpdfstring{Two $\surm(3)$ coupled to $\sprm(0)$}{Two SU3 coupled to Sp0}}
In Section \ref{sec:SU3+Sp0}, the boundary conditions for 2 M5 branes on $\C^2 \slash D_4$ are $\rho_L=(3^2,1^2)$, $\rho_R=(1^8)$. One can go ahead and assign identical non-trivial boundary conditions to both sides of 3 M5 branes on $\C^2 \slash D_4$, i.e.\ $\rho_L=\rho_R=(3^2,1^2)$. The brane configuration and 6d quiver become
\begin{align}
\raisebox{-.5\height}{
\includegraphics[page=1]{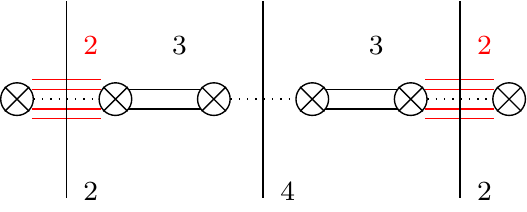}
	}
	\qquad \longleftrightarrow \qquad 
    	\raisebox{-.5\height}{
\includegraphics[page=2]{figures/figures_curve_313.pdf}
	}
\end{align}
The infinite coupling Higgs branches can be discussed by the same reasoning as above. Lifting the half \NS\ brane pairwise off the O6 leads to the collapse of the $-1$ curve. The associated nilpotent orbit closure $\clorbit{E_8}^{\min}$ is now coupled to two $\surm(3)$ gauge groups. This suggests that the resulting global symmetry is the commutant of $\surm(3)\times \surm(3)$ inside $E_8$, which is again a $\surm(3) \times \surm(3)$. Thereafter, the infinite coupling phases are labelled by partitions of $3$, indicating how many pairs of \NS\ branes are coincident.

\paragraph{hyper-K\"ahler quotient.}
The infinite coupling Higgs branch phase follows the same reasoning as in Section \ref{sec:SU3+Sp0}. The collapse of the $-1$ curves leads to the small $E_8$ transitions, but the theory is still coupled to two $\surm(3)$ gauge theories. Because these have no matter, the Higgs branch is simply a $\surm(3)\times \surm(3)$ hyper-K\"ahler quotient of the closure of the minimal nilpotent orbit of $E_8$, i.e.
\begin{align}
    \Higgs_{\infty}^{(1^3)} = \clorbit{E_8}^{\min} /// \left( \surm(3)\times \surm(3) \right) \,. \label{eq:A2xA2_hKQ}
\end{align}
The Hilbert series is evaluated in two steps: the first $\surm(3)$ hyper-K\"ahler quotient has been computed in \eqref{eq:A2_hk}. The resulting moduli space has $E_6$ global symmetry, and one embeds the second $\surm(3)$ into any one of the three $\surm(3)$ subgroups for the maximal subalgebra $\surmL(3)\times \surmL(3)\times\surmL(3)$ of $E_6$. A direct computation results in
\begin{subequations}
\label{eq:HS_A2xA2_hKQ}
\begin{align}
    \HS_{\mathrm{hK}}^{A_2\times A_2} &=    \int \diff \mu_{\surm(3)}(x_{1,2}) 
    \int \diff \mu_{\surm(3)}(z_{1,2})\   \HS_{\clorbit{E_8}^{\min}} (\{y_i\}_{i=1}^4,\{x_{1,2}\},\{z_{1,2}\})  
    \cdot \Hh_F(z_{1,2})
    \cdot \Hh_F(x_{1,2}) \notag \\
    &\stackrel{\eqref{eq:A2_hk}}{=} \int \diff \mu_{\surm(3)}(x_{1,2})\
     \HS_{\mathrm{hK}}(\{y_i\}_{i=1}^4,\{x_{1,2}\})  
    \cdot \Hh_F(x_{1,2}) \\
  &\stackrel{y_i\to1}{=}  \frac{1}{\left(1-t^2\right)^{10} \left(1-t^4\right)^{13} \left(1-t^6\right)^3} 
  \cdot 
  \bigg( 
  1 
 +6 t^2
  +104 t^4
  +700 t^6
  +5084 t^8
 +25706 t^{10}
  +115525 t^{12} \notag\\
  &\qquad +417585 t^{14}
  +1307923 t^{16}
 +3463261 t^{18}
  +7987946 t^{20}
 +15943916 t^{22}
 +27958179 t^{24}\notag\\
  &\qquad
 +42969861 t^{26} 
 +58390228 t^{28}
 +70007697 t^{30}
 +74452240 t^{32}
 +70007697 t^{34}
 +58390228 t^{36}\notag\\
  &\qquad
+42969861 t^{38}
 +27958179 t^{40}
+15943916 t^{42}
 +7987946 t^{44}
+3463261 t^{46}
 +1307923 t^{48}\notag\\
  &\qquad
+417585 t^{50}
+115525 t^{52}
+25706 t^{54}
 +5084 t^{56}
+700 t^{58}
+104 t^{60}
 +6 t^{62}
 +t^{64}
  \bigg) \notag
\end{align}
and a perturbative evaluations yields
\begin{align}
    \HS_{\mathrm{hK}}^{A_2\times A_2}&= 1+16 t^2+232 t^4+2501 t^6+22825 t^8+176140 t^{10}+1183373 t^{12}+O\left(t^{13}\right) \;,\\
    \PL(\HS_{\mathrm{hK}}^{A_2\times A_2})&= 16 t^2+96 t^4+149 t^6-1147 t^8-8412 t^{10}+10774 t^{12}+O\left(t^{13}\right)\,.
\end{align}
\end{subequations}
The coefficient at order $t^2$ reflects the $\surmL(3)\oplus \surmL(3)$ global symmetry algebra. Based on the $\PL$, the centre of each $\surm(3)$ acts trivial on the generators such that the global symmetry group is $(\surm(3)\times \surm(3))/ (\Z_3 \times \Z_3) =\mathrm{PSU}(3) \times \mathrm{PSU}(3)$.

The infinite coupling Higgs branch, phase (3), is the $S_3$ quotient of \eqref{eq:A2xA2_hKQ}.

\paragraph{Magnetic quiver.}
To exemplify, the Higgs branches in the phases $(1^3)$ and $(3)$ are captured by the following magnetic quivers 
\begin{align}
    (1^3):\qquad &\raisebox{-.5\height}{
\includegraphics[page=3]{figures/figures_curve_313.pdf}
	} 
	\quad 
\begin{cases}
    \dim\ \Coulomb &= 13 \,, \\ 
	\gbal&=\sormL(4) \,,
\end{cases} 
\label{eq:magQuiv_SU3xSp0xSU3_111}\\
       (3):\qquad &\raisebox{-.5\height}{
\includegraphics[page=4]{figures/figures_curve_313.pdf}
	} 
	\quad 
\begin{cases}
    \dim\ \Coulomb &= 13 \,, \\ 
	\gbal&=\sormL(4) \,.
\end{cases}
\label{eq:magQuiv_SU3xSp0xSU3_3}
\end{align}
The subset of balanced nodes suggests that the global symmetry is at least $\sormL(4)$. A more refined analysis is provided evaluating the monopole formula as in Table \ref{tab:SU3xSp0xSU3}. One finds that the $t^2$ coefficient indeed confirms a global symmetry of dimension $16$, consistent with $\surmL(3)\oplus \surmL(3)$. Moreover, the monopole formula for \eqref{eq:magQuiv_SU3xSp0xSU3_111} agrees precisely with the $\surm(3)\times\surm(3)$ hyper-K\"ahler quotient \eqref{eq:HS_A2xA2_hKQ} of $\clorbit{E_8}^{\min}$, with global symmetry $\mathrm{PSU}(3) \times \mathrm{PSU}(3)$. Thus \eqref{eq:magQuiv_SU3xSp0xSU3_111} is a new quiver realisation for such a non-trivial quotient.
\begin{table}[t]
\ra{1.25}
    \centering
    \begin{tabular}{crl}
    \toprule
phase   &  \multicolumn{2}{c}{quantity} \\ \midrule
$(3)$  &  $\Hh=$& $1+16 t^2+200 t^4+1864 t^6+14629 t^8+98284 t^{10}+582712 t^{12}+O\left(t^{13}\right)$ \\
 & $ \HS_{\mathbb{Z}} =$  &$1+8 t^2+104 t^4+928 t^6+7349 t^8+49108 t^{10}+291552 t^{12}+O\left(t^{13}\right)$\\
 & $ \HS_{\mathbb{Z}+\frac{1}{2}} =$ & $8 t^2+96 t^4+936 t^6+7280 t^8+49176 t^{10}+291160 t^{12}+O\left(t^{13}\right)$ \\
  & $\PL =$& $16 t^2+64 t^4+24 t^6-415 t^8-884 t^{10}+4428 t^{12}+O\left(t^{13}\right)$ \\ \midrule
 $(2,1)$ & $\Hh = $ & $ 1+16 t^2+216 t^4+2182 t^6+18667 t^8+136080 t^{10}+869924 t^{12}+O\left(t^{13}\right)$\\
 & $\HS_{\mathbb{Z}} = $ & $ 1+8 t^2+112 t^4+1086 t^6+9371 t^8+67992 t^{10}+435188 t^{12}+O\left(t^{13}\right)$ \\
 & $\HS_{\mathbb{Z}+\frac{1}{2}} = $ & $8 t^2+104 t^4+1096 t^6+9296 t^8+68088 t^{10}+434736 t^{12}+O\left(t^{13}\right)$ \\
 & $\PL =$ & $16 t^2+80 t^4+86 t^6-705 t^8-3840 t^{10}+6103 t^{12}+O\left(t^{13}\right)$ \\ \midrule
 $(1^3)$ & $\Hh=$ & $1+16 t^2+232 t^4+2501 t^6+22825 t^8+176140 t^{10}+1183373 t^{12}+O\left(t^{13}\right)$ \\
  & $\HS_{\mathbb{Z}} = $ & $1+8 t^2+120 t^4+1245 t^6+11449 t^8+88012 t^{10}+591909 t^{12}+O\left(t^{13}\right)$ \\
  & $\HS_{\mathbb{Z}+\frac{1}{2}} =$&  $8 t^2+112 t^4+1256 t^6+11376 t^8+88128 t^{10}+591464 t^{12}+O\left(t^{13}\right)$ \\
  & $\PL= $& $ 16 t^2+96 t^4+149 t^6-1147 t^8-8412 t^{10}+10774 t^{12}+O\left(t^{13}\right)$ \\ \bottomrule
    \end{tabular}
    \caption{Perturbative Hilbert series for the different phases of\eqref{eq:magQuiv_SU3xSp0xSU3_111} and \eqref{eq:magQuiv_SU3xSp0xSU3_3}.}
    \label{tab:SU3xSp0xSU3}
\end{table}

\subsection{\texorpdfstring{ $\surm(3)$ and $\sorm(7)$ coupled to $\sprm(0)$}{SU3 and SO7 coupled to Sp0}}
For 3 M5 branes on $\C^2 \slash D_4$ with boundary conditions $\rho_L=(3,1^5)$ and $\rho_R=(3^2,1^2)$, the brane configuration and 6d quiver become
\begin{align}
    \raisebox{-.5\height}{
\includegraphics[page=5]{figures/figures_curve_313.pdf}
	}
		\qquad \longleftrightarrow \qquad 
	    	\raisebox{-.5\height}{
\includegraphics[page=6]{figures/figures_curve_313.pdf}
	}
\end{align}
The finite coupling Higgs branch has a $\sorm(5)\cong \sprm(2)$ global symmetry from the hypermultiplets transforming in the spinor of $\sorm(7)$. The infinite coupling Higgs branch is expected to exhibit further enhancement. Since there is an $E_8$ instanton coupled to $\sorm(7)$ and $\surm(3)$, one looks at the  commutant of $\surmL(3)\oplus \sormL(7)\subset \eeight$. Briefly, the commutant $\sormL(7)\subset \eeight$ is $\sormL(9)$. Then $\sormL(9) \supset \surmL(4)\oplus \surmL(2)$, such that $\surmL(4)\supset \surmL(3)\oplus \urmL(1)$; hence, one might argue that the commutant $\surmL(3)\subset \sormL(9)$ is $\surmL(2)\oplus\urmL(1)$. In total, the expected symmetry is $\sprmL(2)\oplus \surmL(2) \oplus \urmL(1)$. Next, compare this to the magnetic quiver analysis. The magnetic quiver at the conformal fixed point is given by
\begin{align}
    \raisebox{-.5\height}{
\includegraphics[page=7]{figures/figures_curve_313.pdf}
	} 	\qquad 
	\begin{cases}
	\dim\ \Coulomb &= 16 \,,\\  
	\gbal&=\sormL(5) \oplus \sormL(3)   \,, 
	\end{cases}
	\label{eq:magQuiv_SO7xSp0xSU3}
\end{align}
and the subset of balanced nodes suggest at least a $\sprmL(2)\oplus \surmL(2)\cong \sormL(5) \oplus \sormL(3)$ global symmetry.
A more robust verification is given by perturbatively evaluating monopole formula are summarised in Table \ref{tab:SO7xSp0xSU3}. The order of the $t^2$ coefficient is consistent with a $\sprmL(2)\oplus \surmL(2) \oplus \urmL(1)$ symmetry, which has dimension $10+3+1=14$. 

\begin{table}[t]
\ra{1.25}
    \centering
    \begin{tabular}{crl}
    \toprule
phase   &  \multicolumn{2}{c}{quantity} \\ \midrule
$(3)$  &  $\Hh=$& $1+14 t^2+16 t^3+135 t^4+272 t^5+1147 t^6+O\left(t^7\right)$ \\
 & $ \HS_{\mathbb{Z}} =$  & $1+14 t^2+135 t^4+1147 t^6+O\left(t^7\right)$\\
 & $ \HS_{\mathbb{Z}+\frac{1}{2}} =$ &  $16 t^3+272 t^5+O\left(t^7\right)$ \\
  & $\PL =$&  $14 t^2+16 t^3+30 t^4+48 t^5+31 t^6+O\left(t^7\right)$ \\ \midrule
 $(2,1)$ & $\Hh = $ & $1+14 t^2+16 t^3+139 t^4+288 t^5+1231 t^6+O\left(t^7\right)$\\
 & $\HS_{\mathbb{Z}} = $ & $1+14 t^2+139 t^4+1231 t^6+O\left(t^7\right)$  \\
 & $\HS_{\mathbb{Z}+\frac{1}{2}} = $ & $16 t^3+288 t^5+O\left(t^7\right)$ \\
 & $\PL =$ & $14 t^2+16 t^3+34 t^4+64 t^5+59 t^6+O\left(t^7\right)$ \\ \midrule
 $(1^3)$ & $\Hh=$ & $1+14 t^2+16 t^3+143 t^4+304 t^5+1315 t^6+O\left(t^7\right)$ \\
  & $\HS_{\mathbb{Z}} = $ &  $1+14 t^2+143 t^4+1315 t^6+O\left(t^7\right)$\\
  & $\HS_{\mathbb{Z}+\frac{1}{2}} =$& $16 t^3+304 t^5+O\left(t^7\right)$ \\
  & $\PL= $& $14 t^2+16 t^3+38 t^4+80 t^5+87 t^6+O\left(t^7\right)$ \\ \bottomrule
    \end{tabular}
    \caption{Perturbative Hilbert series for the different phases of \eqref{eq:magQuiv_SO7xSp0xSU3}.}
    \label{tab:SO7xSp0xSU3}
\end{table}

\subsection{\texorpdfstring{Two $\sorm(7)$ coupled to $\sprm(0)$}{Two SO7 coupled to Sp0}}
For 3 M5 branes on $\C^2 \slash D_4$ with boundary conditions $\rho_L=\rho_R=(3,1^5)$, the brane configuration and 6d quiver become
\begin{align}
    \raisebox{-.5\height}{
    \includegraphics[page=8]{figures/figures_curve_313.pdf}
	}
		\qquad \longleftrightarrow \qquad 
		\raisebox{-.5\height}{
\includegraphics[page=9]{figures/figures_curve_313.pdf}
	}
\end{align}
The 6d quiver has two flavour symmetry factors of $\sormL(5)\cong \sprmL(2)$. Moving towards infinite coupling, the enhancement of the Higgs branch symmetry is expected to arise from the commutant of $\sormL(7)\oplus \sormL(7)\subset \eeight$. Using that the commutant of $\sormL(7) \subset \eeight$ is $\sormL(9)$ and that the commutant of $\sormL(7)\subset \sormL(9)$ is $\sormL(2)$, one expects an additional $\urmL(1)$ factor. 

The magnetic quiver for the Higgs branch at the conformal fix point reads
\begin{align}
    \raisebox{-.5\height}{
\includegraphics[page=10]{figures/figures_curve_313.pdf}
	} 
		\qquad 
		\begin{cases}
	\dim\ \Coulomb &= 19 \,, \\
	\gbal&=\sormL(5) \oplus \sormL(5) \oplus \sormL(2) \,,
		\end{cases}
		\label{eq:magQuiv_SO7xSp0xSO7}
\end{align}
and the balanced subset nodes does indicate an $\sormL(5) \oplus \sormL(5) \oplus \sormL(2)$ Coulomb branch symmetry. Upon evaluating the monopole formula, as summarised in Table \ref{tab:SO7xSp0xSO7}, one finds a $t^2$ coefficient of $21$. This is consistent with expected global symmetry of dimension $10+10+1$.

\begin{table}[t]
\ra{1.25}
    \centering
    \begin{tabular}{crl}
    \toprule
phase   &  \multicolumn{2}{c}{quantity} \\ \midrule
$(3)$  &  $\Hh=$& $1+21 t^2+288 t^4+3071 t^6+O\left(t^7\right)$ \\
 & $ \HS_{\mathbb{Z}} =$  &  $1+21 t^2+256 t^4+2335 t^6+O\left(t^7\right)$\\
 & $ \HS_{\mathbb{Z}+\frac{1}{2}} =$ &  $ 32 t^4+736 t^6+O\left(t^7\right)$  \\
  & $\PL =$&  $21 t^2+57 t^4+103 t^6+O\left(t^7\right)$ \\ \midrule
 $(2,1)$ & $\Hh = $ & $1+21 t^2+291 t^4+3179 t^6+O\left(t^7\right)$ \\
 & $\HS_{\mathbb{Z}} = $ & $ 1+21 t^2+259 t^4+2411 t^6+O\left(t^7\right)$ \\
 & $\HS_{\mathbb{Z}+\frac{1}{2}} = $ & $32 t^4+768 t^6+O\left(t^7\right)$  \\
 & $\PL =$ & $21 t^2+60 t^4+148 t^6+O\left(t^7\right)$ \\ \midrule
 $(1^3)$ & $\Hh=$ & $1+21 t^2+294 t^4+3287 t^6+O\left(t^7\right)$ \\
  & $\HS_{\mathbb{Z}} = $ &  $1+21 t^2+262 t^4+2487 t^6+O\left(t^7\right) $\\
  & $\HS_{\mathbb{Z}+\frac{1}{2}} =$&  $32 t^4+800 t^6+O\left(t^7\right)$\\
  & $\PL= $&$21 t^2+63 t^4+193 t^6+O\left(t^7\right)$  \\ \bottomrule
    \end{tabular}
    \caption{Perturbative Hilbert series for the different phases of \eqref{eq:magQuiv_SO7xSp0xSO7}.}
    \label{tab:SO7xSp0xSO7}
\end{table}

\section{Boundary conditions for theories on -2 curve}
\label{sec:overlapping_bc}
Above, boundary conditions have either involved one trivial and one non-trivial partition, or two non-trivial partitions. In all of these cases, the 6d quiver contained an $\sprm(0)$ node. In this section, exemplary cases for boundary conditions without an $\sprm(0)$ node in the 6d quiver are considered. The identification of the 6d quiver theory follows from the results of \cite{Hassler:2019eso} rather than Table \ref{tab:negative_branes}.

\subsection{\texorpdfstring{ $\surm(4)$ with 8 fundamentals}{SU4}}
For 2 M5 branes on $\C^2 \slash D_4$ with boundary conditions $\rho_L=\rho_R=(3,1^5)$, the brane configuration and 6d quiver become
\begin{align}
    \raisebox{-.5\height}{
    \includegraphics[page=1]{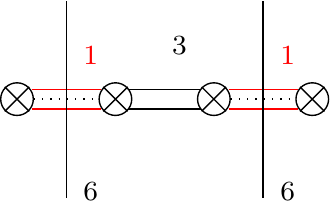}
	}
		\qquad \longleftrightarrow \qquad 
	\raisebox{-.5\height}{
	\includegraphics[page=2]{figures/figures_curve_2.pdf}
	}
	\cong
	    	\raisebox{-.5\height}{
	\includegraphics[page=3]{figures/figures_curve_2.pdf}
	}
\end{align}
and the $6$ half \De\ branes on each side are understood as $5+1$, i.e. $5$ \De\ originating from the $1^5$ part of one partition and the remaining brane comes from the part $3$ of the other partition.
Using the intuition gained, the brane system can be understood as follows: each red brane interval contributes 16 hypermultiplets in a bispinor representation of $\sorm(6)\times \sorm(6)$. The 6 \De\ flavour branes give rise to an $\sorm(6)$ global symmetry, while the brane interval with 3 \Ds\ branes leads to an $\sorm(6)$ gauge group. As the $\sorm(6)$ spinor is 4-dimensional, the bispinor leads to 16 hypermultiplets. Combining two of such bispinors, and  since the spinor is a complex representation, we get an $\surm(8)$ global symmetry rather than $\sorm(12)$.

The brane system contains two intervals with negative brane number, for which one does not have a tensor multiplet.  Thus, one tensor multiplet remains, which sets the gauge coupling of the $\surm(4)$ theory. The system has only two phases: either the two \NS\ pairs are off the O6 plane and separated or they are coincident. 

The finite coupling Higgs branch has global symmetries $\surm(8)\times \urm(1)$, where the abelian factor is due to the anomalous baryon number. At infinite coupling, only the $\surm(8)$ Higgs branch isometry remains.
The Higgs branch at finite coupling, phase $(1^2)$, and infinite coupling, phase $(2)$, are captured by the following magnetic quivers 
\begin{subequations}
\label{eq:magQuiv_OSp_SU4}
\begin{align}
    (1^2): \qquad &\raisebox{-.5\height}{
\includegraphics[page=4]{figures/figures_curve_2.pdf}
	} 
		\qquad 
		\begin{cases}
	\dim\ \Coulomb &= 17 \,, \\
	\gbal&=\sormL(6)^2 \oplus \sormL(2)^2 \,,
		\end{cases}
		\label{eq:magQuiv_SU4_finite} \\
(2): \qquad 		    &\raisebox{-.5\height}{
	\includegraphics[page=5]{figures/figures_curve_2.pdf}
	} 
		\qquad 
		\begin{cases}
	\dim\ \Coulomb &= 17  \,,\\
	\gbal&=\sormL(6) \oplus \sormL(6) \,,
		\end{cases}
		\label{eq:magQuiv_SU4_infty}
\end{align}
\end{subequations}
which are related by an $S_2$ discrete gauging \cite{Hanany:2018cgo}.
The monopole formula for \eqref{eq:magQuiv_SU4_finite} and \eqref{eq:magQuiv_SU4_infty} has been evaluated in \cite[App.\ C.1]{Bourget:2020xdz} and shown to be consistent with an $\surm(8)$ global symmetry. Moreover, the balanced set of nodes in phase $(1^2)$ display an $\sormL(6)^2\oplus\sormL(2)^2$ global symmetry, which can be understood as maximal subalgebra $\sormL(6)^2 \oplus \sormL(2)\cong \surmL(4)^2\oplus \urmL(1) \subset \surmL(8)$ of the non-abelian global symmetry and the anomalous $\urm(1)$ baryon symmetry. The balanced set of nodes in the infinite coupling phase $(2)$ only displays the non-abelian factor $\sormL(6)^2$; nonetheless, the $\surmL(8)$ symmetry is manifest in the monopole formula.

The electric theory can also be realised by 2 M5 branes on a $\C^2 \slash \Z_4$ singularity with trivial boundary conditions \cite{Hanany:2018vph,Cabrera:2019izd}. The magnetic quiver simply reads
\begin{subequations}
\label{eq:magQuiv_Uni_SU4}
\begin{align}
(1^2):\qquad 
&\raisebox{-.5\height}{
	\includegraphics[page=6]{figures/figures_curve_2.pdf}
	} 
			\qquad 
		\begin{cases}
	\dim\ \Coulomb &= 17 \,, \\
	\gbal&=\surmL(8) \oplus \urmL(1)  \,,
		\end{cases}
	\\
	(2):\qquad 
&\raisebox{-.5\height}{
	\includegraphics[page=7]{figures/figures_curve_2.pdf}
	}
				\qquad 
		\begin{cases}
	\dim\ \Coulomb &= 17 \,, \\
	\gbal&=\surmL(8) \,.
		\end{cases}
\end{align}
\end{subequations}
Recall that the finite coupling and infinite coupling HWG \cite{Hanany:2018vph} are given by
\begin{subequations}
\begin{align}
    \HWG_{(1^2)} &= \PE\left[ \sum_{i=1}^{3}\mu_i \mu_{8-i} t^{2i} + t^2 + 2\mu_4  t^4\right] \,,\\
    \HWG_{(2)} &= \PE\left[ \sum_{i=1}^{4}\mu_i \mu_{8-i} t^{2i}  + \mu_4  t^4 + t^4 + \mu_4 t^{6}  - \mu_4^2 t^{12}\right] \,,
\end{align}
\end{subequations}
with $\mu_i$ $\surm(8)$ highest weight fugacities. Converting the HWG into Hilbert series shows agreement with the monopole formula of \eqref{eq:magQuiv_OSp_SU4} and \eqref{eq:magQuiv_Uni_SU4}. The HWG shows that all generators are invariant under a $\Z_4$ subgroup of the $\Z_8$ centre symmetry, such that the symmetry group is $\surm(8)\slash\Z_4$. All powers of $t$ are even, implying that the R-symmetry is $\sorm(3)_R$ rather than $\surm(2)_R$.

The finite and infinite coupling Hasse diagram can be derived to read
\begin{align}
\raisebox{-.5\height}{
	\includegraphics[page=8]{figures/figures_curve_2.pdf}
}
\qquad\xrightarrow[]{\Z_2} \qquad 
    \raisebox{-.5\height}{
    	\includegraphics[page=9]{figures/figures_curve_2.pdf}
}
\end{align}
and again, the difference is simply given by a $S_2$ gauging.

\subsection{\texorpdfstring{ $\surm(2)$ with 4 fundamentals}{SU2}}
For 2 M5 branes on $\C^2 \slash D_4$ with boundary conditions $\rho_L=(3^2,1^2)$ $\rho_R=(3,1^5)$, the brane configuration and 6d quiver become
\begin{align}
    \raisebox{-.5\height}{
        	\includegraphics[page=10]{figures/figures_curve_2.pdf}
	}
		\qquad \longleftrightarrow \qquad 
			    	\raisebox{-.5\height}{
	\includegraphics[page=11]{figures/figures_curve_2.pdf}
	}
	\cong
	    	\raisebox{-.5\height}{
	\includegraphics[page=12]{figures/figures_curve_2.pdf}
	}
\end{align}
The intuitive understanding of the brane system is as follows: the left-most brane interval with $2$ negatively charged branes modifies the naive $\sorm(5)$ gauge group in the adjacent brane interval into an $\surm(2)$. The right-most brane interval has $1$ negative brane, this leads to bi-spinor matter of the $\surm(2)$ with the $\sorm(7)$ flavour symmetry from the $7$ half \De\ branes.
As the brane system has two intervals with negative branes such that there is only one tensor multiplet. The $\surm(2)$ theory is anomaly-free with 8 half-hypermultiplets. As in the case above, there are only two phases, indicated by whether the pairs of \NS\ branes are away from the O6 and separated or coincident.
The magnetic quivers for the finite coupling and infinite coupling Higgs branch are given by
\begin{align}
    (1^2):\qquad &\raisebox{-.5\height}{
    \includegraphics[page=13]{figures/figures_curve_2.pdf}
	} 
		\qquad 
		\begin{cases}
	\dim\ \Coulomb &= 13 \,, \\
	\gbal&=\sormL(6) \oplus \sormL(3) \oplus \sormL(2)^2 \,,
		\end{cases}
		\label{eq:magQuiv_SU2_finite} \\
		(2):\qquad &\raisebox{-.5\height}{
	\includegraphics[page=14]{figures/figures_curve_2.pdf}
	} 
		\qquad 
		\begin{cases}
	\dim\ \Coulomb &= 13 \,, \\
	\gbal&=\sormL(6) \oplus \sormL(3) \,,
		\end{cases}
		\label{eq:magQuiv_SU2_infty}
\end{align}
and the monopole formula is evaluated as summarised in Table \ref{tab:HS_SU2_OSp}. The PL shows in both phases the appearance of $16$ complex free moduli, which come with a global symmetry of $\sprm(8)$. One can simply remove this contribution and obtain the Hilbert series $\widetilde{\Hh}$ for the interacting part. 

The finite coupling Higgs branch of $\surm(2)$ for 4 fundamentals is the closure of the minimal nilpotent orbit of $\sorm(8)$. Its unrefined Hilbert series \cite[Tab.\ 15]{Hanany:2016gbz} reads 
\begin{align}
  \Hh_{\clorbit{\sorm(8)}^{\mathrm{min}}} &=    \frac{\left(1+t^2\right)
  \left(1+17 t^2 +48 t^4+17 t^6 +t^8\right)
  }{\left(1-t^2\right)^{10}}
\end{align}
and the perturbative expansion and PL agree with the results for $\widetilde{\Hh}\eqref{eq:magQuiv_SU2_finite}$ in Table \ref{tab:HS_SU2_OSp}.
Likewise, the infinite coupling Higgs branch is known to be the closure of the next-to-minimal nilpotent orbit of $\sorm(7)$. Hence, one may compare $\widetilde{\Hh}\eqref{eq:magQuiv_SU2_infty}$ against the known unrefined Hilbert series \cite[Tab.\ 10]{Hanany:2016gbz}
\begin{align}
    \Hh_{\clorbit{\sorm(7)}^{\mathrm{n-t-min}}} &= 
    \frac{\left(1+t^2\right) 
    \left(1 +10 t^2+20 t^4  +10 t^6 + t^8\right)
    }{\left(1-t^2\right)^{10}} 
\end{align}
and finds that perturbative expansion and PL agree with the results for $\widetilde{\Hh}\eqref{eq:magQuiv_SU2_infty}$, see Table \ref{tab:HS_SU2_OSp}.

Alternatively, the exact HWG for finite and infinite coupling are known \cite{Hanany:2018vph}
\begin{align}
    \HWG_{(1^2)} = \PE\left[ \mu_2 t^2 \right] \qquad \text{and} \qquad 
      \HWG_{(2)} =\PE\left[ \nu_2 t^2 +\nu_1^2  t^4\right]
\end{align}
where $\mu_i$ and $\nu_i$ denote $\sorm(8)$ and $\sorm(7)$ highest weight fugacities, respectively. The symmetry group is $\Spin(8)\slash(\Z_2\times \Z_2)$ in phase $(1^2)$ and $\Spin(7)\slash \Z_2$ for $(2)$.
 The Higgs branch Hasse diagram for both phases is readily available 
 \begin{align}
 \raisebox{-.5\height}{
 	\includegraphics[page=15]{figures/figures_curve_2.pdf}
 }
 \qquad\xrightarrow[]{\Z_2} \qquad 
     \raisebox{-.5\height}{
     	\includegraphics[page=16]{figures/figures_curve_2.pdf}
 }
 \end{align}

Observe that the orthosymplectic magnetic quivers \eqref{eq:magQuiv_SU2_finite} and \eqref{eq:magQuiv_SU2_infty} have Coulomb branch dimension $13$, but after removing the $8$ quaternionic free moduli, the dimension reduces to $5$, as appropriate for $\surm(2)$ SQCD. The appearance of these $8$ moduli has also been observed in \cite[Tab.\ 2]{Hassler:2019eso}.

\begin{table}[t]
\ra{1.25}
    \centering
    \begin{tabular}{crl}
    \toprule
phase   &  \multicolumn{2}{c}{quantity} \\ \midrule
$(2)$  &  $\Hh\eqref{eq:magQuiv_SU2_infty}=$& \parbox{12.5cm}{$1+16 t+157 t^2+1152 t^3+6927 t^4+35760 t^5+163335 t^6+673728 t^7+2547854 t^8+O\left(t^9\right) $} \\
 & $ \HS_{\mathbb{Z}} =$  &  $ 1+157 t^2+6927 t^4+163335 t^6+2547854 t^8+O\left(t^9\right) $\\
 & $ \HS_{\mathbb{Z}+\frac{1}{2}} =$ &  $ 16 t +1152 t^3 +35760 t^5 +673728 t^7 +  O\left(t^9\right)$  \\
  & $\PL =$&  $16t+21 t^2-36 t^4+140 t^6-784 t^8+O\left(t^9\right) $ \\ \vspace{5pt}
  & $\widetilde{\Hh}\eqref{eq:magQuiv_SU2_infty}=$ &  $(1-t)^{16} \cdot \Hh\eqref{eq:magQuiv_SU2_infty}=  1+21 t^2+195 t^4+1155 t^6+5096 t^8+O\left(t^9\right) $ \\
   & $\PL(\widetilde{\Hh})=$ & $ 21 t^2-36 t^4+140 t^6-784 t^8+O\left(t^9\right)$\\
  \midrule
 $(1^2)$ & $\Hh\eqref{eq:magQuiv_SU2_finite}=$ &\parbox{12.5cm}{ $1+16 t+164 t^2+1264 t^3+7984 t^4+43152 t^5+205517 t^6+880256 t^7+3443224 t^8+O\left(t^9\right) $} \\
  & $\HS_{\mathbb{Z}} = $ &  $1+164 t^2+7984 t^4+205517 t^6+3443224 t^8+O\left(t^9\right) $\\
  & $\HS_{\mathbb{Z}+\frac{1}{2}} =$&  $16 t+1264 t^3+43152 t^5+880256 t^7+O\left(t^9\right)  $\\
  & $\PL= $&$ 16 t+28 t^2-106 t^4+833 t^6-8400 t^8+O\left(t^9\right)$  \\  \vspace{5pt}
  & $\widetilde{\Hh}\eqref{eq:magQuiv_SU2_finite}=$ &  $ (1-t)^{16} \cdot \Hh\eqref{eq:magQuiv_SU2_finite} = 1+28 t^2+300 t^4+1925 t^6+8918 t^8+O\left(t^9\right) $ \\
   & $\PL(\widetilde{\Hh})=$ & $28 t^2-106 t^4+833 t^6-8400 t^8+O\left(t^9\right)$\\ \bottomrule
    \end{tabular}
    \caption{Perturbative Hilbert series for the different phases \eqref{eq:magQuiv_SU2_finite} and \eqref{eq:magQuiv_SU2_infty}.}
    \label{tab:HS_SU2_OSp}
\end{table}
\subsection{\texorpdfstring{ $\surm(4)\times \surm(2)$ quiver}{SU4xSU2 quiver}}
For 3 M5 branes on $\C^2 \slash D_4$ with boundary conditions $\rho_L=(3,1^5)$ $\rho_R=(5,1^3)$, the brane configuration and 6d quiver become
\begin{align}
    \raisebox{-.5\height}{
    \includegraphics[page=17]{figures/figures_curve_2.pdf}
	}
		\qquad \longleftrightarrow \qquad 
	    	\raisebox{-.5\height}{
	  	\includegraphics[page=18]{figures/figures_curve_2.pdf}
	}
\end{align}
The Higgs branch at finite coupling (phase $(1^3)$) and at the fixed point (phase $(3)$) are captured by the following magnetic quivers
\begin{align}
(1^3):\qquad    &\raisebox{-.5\height}{
	  	\includegraphics[page=19]{figures/figures_curve_2.pdf}
	} 
		\qquad 
		\begin{cases}
	\dim\ \Coulomb &= 14 \,,  \\
	\gbal&=\sormL(6) \oplus \sormL(3)  \oplus \sormL(2)^3 \,,
		\end{cases}
		\label{eq:magQuiv_SU4SU2_finite} \\
(3):\qquad    &\raisebox{-.5\height}{
	  \includegraphics[page=20]{figures/figures_curve_2.pdf}
	} 
		\qquad 
		\begin{cases}
	\dim\ \Coulomb &= 14 \,, \\
	\gbal&=\sormL(6) \oplus \sormL(3) \,.
		\end{cases}
		\label{eq:magQuiv_SU4SU2_infty}
\end{align}
Table \ref{tab:HS_SU4SU2_OSp} summarises the perturbative monopole formula. One observes that phase $(3)$ is compatible with an $\surm(6)$ global symmetry, while the finite coupling phase has $\surm(6) \times \urm(1)^3$ moduli space isometry.

\begin{table}[t]
\ra{1.25}
    \centering
    \begin{tabular}{crl}
    \toprule
phase   &  \multicolumn{2}{c}{quantity} \\ \midrule
$(3)$  &  $\Hh\eqref{eq:magQuiv_SU4SU2_infty}=$& $1+35 t^2+660 t^4+8743 t^6+90244 t^8+O\left(t^9\right) $ \\
 & $ \HS_{\mathbb{Z}} =$  &  $1+19 t^2+340 t^4+4391 t^6+45220 t^8+O\left(t^9\right) $\\
 & $ \HS_{\mathbb{Z}+\frac{1}{2}} =$ &  $ 16 t^2+320 t^4+4352 t^6+45024 t^8+O\left(t^9\right)$  \\
  & $\PL =$&  $ 35 t^2+30 t^4-77 t^6-241 t^8+O\left(t^9\right)$ \\ \midrule
 $(1^3)$ & $\Hh\eqref{eq:magQuiv_SU4SU2_finite}=$ & $1+37 t^2+792 t^4+12180 t^6+145838 t^8+O\left(t^9\right) $ \\
  & $\HS_{\mathbb{Z}} = $ &  $ 1+21 t^2+408 t^4+6116 t^6+73022 t^8+O\left(t^9\right)$\\
  & $\HS_{\mathbb{Z}+\frac{1}{2}} =$&  $16 t^2+384 t^4+6064 t^6+72816 t^8+O\left(t^9\right) $\\
  & $\PL= $&$37 t^2+89 t^4-252 t^6-2800 t^8+O\left(t^9\right) $  \\ \bottomrule
    \end{tabular}
    \caption{Perturbative Hilbert series for the different phases \eqref{eq:magQuiv_SU4SU2_finite} and \eqref{eq:magQuiv_SU4SU2_infty}.}
    \label{tab:HS_SU4SU2_OSp}
\end{table}

The electric theory can also be realised by 3 M5 branes on a $\C^2 \slash \Z_4$ singularity with non-trivial boundary conditions, cf.\ Table \ref{tab:SU4_examples}. The magnetic quiver simply reads
\begin{align}
\label{eq:magQuiv_SU4SU2_uni_finite}
(1^3):\qquad 
&\raisebox{-.5\height}{
\includegraphics[page=21]{figures/figures_curve_2.pdf}
	}
			\qquad 
		\begin{cases}
	\dim\ \Coulomb &= 14 \,, \\
	\gbal&=\surmL(6) \oplus \urmL(1)^2 \,,
		\end{cases}\\
	(3):\qquad 
&\raisebox{-.5\height}{
\includegraphics[page=22]{figures/figures_curve_2.pdf}
	}
			\qquad 
		\begin{cases}
	\dim\ \Coulomb &= 14  \,,\\
	\gbal&=\surmL(6) \,,
		\end{cases}
		\label{eq:magQuiv_SU4SU2_uni_infty}
\end{align}
and a straightforward computation shows that the Coulomb branch Hilbert series of the unitary magnetic quiver \eqref{eq:magQuiv_SU4SU2_uni_finite} and \eqref{eq:magQuiv_SU4SU2_uni_infty} agrees with \eqref{eq:magQuiv_SU4SU2_finite} and \eqref{eq:magQuiv_SU4SU2_infty}, respectively.
The $\PL$ for \eqref{eq:magQuiv_SU4SU2_uni_infty} shows a generator at $t^2$ in the adjoint representation of $\surm(6)$ and a generator in the second anti-symmetric representation (plus conjugate) at $t^4$. Both generators are invariant under a $\Z_2$ subgroup of the $\Z_6$ centre; thus, suggesting a $\surm(6)\slash \Z_2$ global symmetry. 

The information on the moduli space can be encoded in the Hasse diagram (see also \cite{Bourget:2019aer})
 \begin{align}
 	\label{eq:Hasse_SU4xSU2}
	\raisebox{-.5\height}{
	\includegraphics[page=23]{figures/figures_curve_2.pdf}
	}
	\qquad\xrightarrow[]{S_3} \qquad 
	\raisebox{-.5\height}{
	\includegraphics[page=24]{figures/figures_curve_2.pdf}
	}
\end{align}
and note that the infinite coupling transitions only modify the top of the diagram. This confirms that the non-abelian Higgs branch isometry does not change between the phases.

\subsection{\texorpdfstring{ $\surm(3)\times \surm(2)$ quiver}{SU3xSU2 quiver}}
For 3 M5 branes on $\C^2 \slash D_4$ with boundary conditions $\rho_L=(3,1^5)$ $\rho_R=(5,3)$, the brane configuration and 6d quiver become
\begin{align}
    \raisebox{-.5\height}{
    	\includegraphics[page=25]{figures/figures_curve_2.pdf}
	}
		\qquad \longleftrightarrow \qquad 
	    	\raisebox{-.5\height}{
	   \includegraphics[page=26]{figures/figures_curve_2.pdf}
	}
\end{align}
The magnetic quivers for the finite and infinite coupling Higgs branch are given by
\begin{align}    
(1^3): \qquad 
    &\raisebox{-.5\height}{
    \includegraphics[page=27]{figures/figures_curve_2.pdf}
	} 
		\qquad 
		\begin{cases}
	\dim\ \Coulomb &= 13 \,,  \\
	\gbal&=\sormL(6) \oplus \sormL(2)^4 \,,
		\end{cases}
		\label{eq:magQuiv_SU3_SU2_finite} 
		\\
		(3): \qquad &\raisebox{-.5\height}{
		\includegraphics[page=28]{figures/figures_curve_2.pdf}
	} 
		\qquad 
		\begin{cases}
	\dim\ \Coulomb &= 13 \,, \\
	\gbal&=\sormL(6) \oplus \sormL(2) \,.
		\end{cases}
		\label{eq:magQuiv_SU3_SU2_infinite}
\end{align}
The monopole formula results are summarised in Table \ref{tab:HS_SU3_SU2_OSp}. 
The PL shows $8$ complex free moduli, which can be removed to study the interacting part.
For phase $(1^3)$, the $t^2$ coefficient is compatible with the dimension of the $\surmL(4)\times \urmL(1)^3$ symmetry. For phase $(3)$, the $t^2$ coefficient is consistent with the $\surmL(4)\times \urmL(1)$ symmetry.

\begin{table}[t]
\ra{1.25}
    \centering
    \begin{tabular}{crl}
    \toprule
phase   &  \multicolumn{2}{c}{quantity} \\ \midrule
$(3)$  &  $\Hh\eqref{eq:magQuiv_SU3_SU2_infinite}=$&\parbox{12.5cm}{ $ 1+8 t+52 t^2+264 t^3+1182 t^4+4720 t^5+17321 t^6+58984 t^7+188673 t^8+O\left(t^9\right) $} \\
 & $ \HS_{\mathbb{Z}} =$  &  $1+52 t^2+1182 t^4+17321 t^6+188673 t^8+O\left(t^9\right)   $\\
 & $ \HS_{\mathbb{Z}+\frac{1}{2}} =$ &  $ 8 t+264 t^3+4720 t^5+58984 t^7+O\left(t^9\right) $  \\
  & $\PL =$&  $ 8 t+16 t^2+16 t^3+12 t^4-8 t^5-51 t^6-72 t^7+O\left(t^9\right) $ \\ \vspace{5pt}
  & $\widetilde{\Hh}\eqref{eq:magQuiv_SU3_SU2_infinite}=$ & \parbox{12.5cm}{ $(1-t)^{8} \cdot \Hh\eqref{eq:magQuiv_SU3_SU2_infinite}=  1+16 t^2+16 t^3+148 t^4+248 t^5+1093 t^6+2168 t^7+6818 t^8+O\left(t^9\right) $} \\
   & $\PL(\widetilde{\Hh})=$ & $ 16 t^2 +16 t^3 +12 t^4 -8 t^5 -51 t^6 -72 t^7 +O\left(t^9\right) $\\
  \midrule
 $(1^3)$ & $\Hh\eqref{eq:magQuiv_SU3_SU2_finite}=$ & \parbox{12.5cm}{$ 1+8 t+54 t^2+296 t^3+1440 t^4+6296 t^5+25257 t^6+93840 t^7+325958 t^8+O\left(t^9\right) $} \\
  & $\HS_{\mathbb{Z}} = $ &  $ 1+54 t^2+1440 t^4+25257 t^6+325958 t^8+O\left(t^9\right) $\\
  & $\HS_{\mathbb{Z}+\frac{1}{2}} =$&  $ 8 t+296 t^3+6296 t^5+93840 t^7+O\left(t^9\right)  $\\
  & $\PL= $&$ 8 t+18 t^2+32 t^3+35 t^4-32 t^5-305 t^6-672 t^7-59 t^8+O\left(t^9\right) $  \\  \vspace{5pt}
  & $\widetilde{\Hh}\eqref{eq:magQuiv_SU3_SU2_finite}=$ & \parbox{12.5cm}{ $ (1-t)^{8} \cdot \Hh\eqref{eq:magQuiv_SU3_SU2_finite} =1+18 t^2+32 t^3+206 t^4+544 t^5+1993 t^6+5344 t^7+15531 t^8+O\left(t^9\right)   $} \\
   & $\PL(\widetilde{\Hh})=$ & $ 18 t^2+32 t^3+35 t^4-32 t^5-305 t^6-672 t^7-59 t^8+O\left(t^9\right) $\\ \bottomrule
    \end{tabular}
    \caption{Perturbative Hilbert series for the different phases \eqref{eq:magQuiv_SU3_SU2_finite} and \eqref{eq:magQuiv_SU3_SU2_infinite}.}
    \label{tab:HS_SU3_SU2_OSp}
\end{table}

The electric theory admits a realisation via 3 M5 branes on a $\C^2 \slash \Z_3$ singularity with non-trivial boundary conditions.  The corresponding magnetic quivers are given by
\begin{align}
(1^3):\qquad 
&\raisebox{-.5\height}{
	\includegraphics[page=29]{figures/figures_curve_2.pdf}
	}
			\qquad 
		\begin{cases}
	\dim\ \Coulomb &= 9  \\
	\gbal&=\surmL(4) \oplus \urmL(1)^3 
		\end{cases} 
		\label{eq:magQuiv_SU3_SU2_uni_finite}\\
	(3):\qquad 
&\raisebox{-.5\height}{
		\includegraphics[page=30]{figures/figures_curve_2.pdf}
	}
		\qquad 
		\begin{cases}
	\dim\ \Coulomb &= 9  \\
	\gbal&=\surmL(4) \oplus \urmL(1) 
		\end{cases}
		\label{eq:magQuiv_SU3_SU2_uni_infinite}
\end{align}
and it is straightforward to verify that the perturbative monopole formula for \eqref{eq:magQuiv_SU3_SU2_uni_finite} and \eqref{eq:magQuiv_SU3_SU2_uni_infinite} agree with the results for $\widetilde{\Hh}\eqref{eq:magQuiv_SU3_SU2_finite}$ and $\widetilde{\Hh}\eqref{eq:magQuiv_SU3_SU2_infinite}$, respectively. Note also, that the Coulomb branch dimensions of the orthosymplectic quivers are $13$, while the unitary magnetic quivers have Coulomb branch of dimension $9$. The difference are precisely the $4$ quaternionic free moduli encountered in Table \ref{tab:HS_SU3_SU2_OSp}. The number of free moduli has also been observed in \cite[Tab.\ 2]{Hassler:2019eso}.

Note that the Coulomb branch of  \eqref{eq:magQuiv_SU3_SU2_uni_finite} is the hyper-K\"ahler implosion of the nilpotent cone of $\surmL(4)$ \cite{Dancer:2020wll}, which here finds a  natural realisation in brane systems.

The Higgs branch Hasse diagram is given by (see also \cite{Bourget:2019aer})
 \begin{align}
 	\label{eq:Hasse_SU3xSU2}
	\raisebox{-.5\height}{
	\includegraphics[page=31]{figures/figures_curve_2.pdf}
	}
	\qquad\xrightarrow[]{S_3} \qquad 
	\raisebox{-.5\height}{
	\includegraphics[page=32]{figures/figures_curve_2.pdf}
	}
\end{align}
and one observes that it is a subdiagram of \eqref{eq:Hasse_SU4xSU2}. This follows simply because $\surm(4)\times \surm(2)$ can be Higgsed to $\surm(3)\times \surm(2)$.

\subsection{\texorpdfstring{ $\surm(2)\times \surm(2)$ quiver}{SU2xSU2 quiver}}
For 3 M5 branes on $\C^2 \slash D_4$ with boundary conditions $\rho_L=(3^2,1^2)$, $\rho_R=(5,1^3)$, the brane configuration and 6d quiver become
\begin{align}
    \raisebox{-.5\height}{	\includegraphics[page=33]{figures/figures_curve_2.pdf}
	}
		\qquad \longleftrightarrow \qquad 
	    	\raisebox{-.5\height}{
	   \includegraphics[page=34]{figures/figures_curve_2.pdf}
	}
\end{align}
The finite coupling Higgs branch is expected to have $\sorm(4)\times \surm(2)\times \sorm(4) \cong \surm(2)^5$ global symmetry, while the infinite coupling Higgs branch admits an $\surm(2)^3$ isometry.
\begin{align}
    (1^3):\qquad &\raisebox{-.5\height}{
\includegraphics[page=35]{figures/figures_curve_2.pdf}
	} 
		\qquad 
		\begin{cases}
	\dim\ \Coulomb &= 10  \\
	\gbal&= \sormL(3)^2 \oplus \sormL(2)^3 
		\end{cases}
		\label{eq:magQuiv_SU2_SU2_finite} \\
    (3):\qquad &\raisebox{-.5\height}{    	   \includegraphics[page=36]{figures/figures_curve_2.pdf}
	} 
		\qquad 
		\begin{cases}
	\dim\ \Coulomb &= 10  \\
	\gbal&=\sormL(3) \oplus \sormL(3) 
		\end{cases}
		\label{eq:magQuiv_SU2_SU2_infty}
\end{align}
The perturbatively evaluated monopole formula is summarised in Table \ref{tab:HS_SU2_SU2_OSp}.
Again, the PL indicates $8$ complex free moduli, which are acted on by an $\sprm(4)$ symmetry. After removing those, the Coulomb branch Hilbert series of the interacting part is denoted by $\widetilde{\Hh}$. The finite coupling phase $(1^3)$ displays a dimension $15$ global symmetry, consistent with the $\surm(2)^5$ expectation. Likewise, the infinite coupling phase $(3)$ has a global symmetry of dimension $9$, which reflects the expected $\surm(2)^3$ symmetry.
\begin{table}[t]
\ra{1.25}
    \centering
    \begin{tabular}{crl}
    \toprule
phase   &  \multicolumn{2}{c}{quantity} \\ \midrule
$(3)$  &  $\Hh\eqref{eq:magQuiv_SU2_SU2_infty}=$& \parbox{12.5cm}{ $1+8 t+45 t^2+208 t^3+831 t^4+2968 t^5+9692 t^6+29344 t^7+83267 t^8+223288 t^9 +569535 t^{10}+1389120 t^{11}+3254165 t^{12}+O\left(t^{13}\right) $} \\
 & $ \HS_{\mathbb{Z}} =$  &  $ 1+45 t^2+831 t^4+9692 t^6+83267 t^8+569535 t^{10}+3254165 t^{12}+O\left(t^{13}\right)$\\
 & $ \HS_{\mathbb{Z}+\frac{1}{2}} =$ &  $8 t+208 t^3+2968 t^5+29344 t^7+223288 t^9+1389120 t^{11}+O\left(t^{13}\right) $  \\
  & $\PL =$&  \parbox{12.5cm}{ $8 t+9 t^2+16 t^3+4 t^4-16 t^5-39 t^6-8 t^7+100 t^8+176 t^9-54 t^{10}-768 t^{11}-1059 t^{12}+O\left(t^{13}\right) $} \\\vspace{5pt}
  & $\widetilde{\Hh}\eqref{eq:magQuiv_SU2_SU2_infty}=$ & \parbox{11cm}{ $ (1-t)^{8} \cdot \Hh\eqref{eq:magQuiv_SU2_SU2_infty} = 1+9 t^2+16 t^3+49 t^4+128 t^5+298 t^6+632 t^7+1402 t^8+2728 t^9+5324 t^{10}+9944 t^{11}+17946 t^{12}+O\left(t^{13}\right)$}   \\
   & $\PL(\widetilde{\Hh})=$ & \parbox{12.5cm}{ $ 9 t^2+16 t^3+4 t^4-16 t^5-39 t^6-8 t^7+100 t^8+176 t^9-54 t^{10}-768 t^{11}-1059 t^{12}+O\left(t^{13}\right)$}\\ \midrule
 $(1^3)$ & $\Hh\eqref{eq:magQuiv_SU2_SU2_finite}=$ & \parbox{12.5cm}{$1+8 t+51 t^2+272 t^3+1242 t^4+5024 t^5+18361 t^6+61480 t^7+190857 t^8+554464 t^9+1518870 t^{10}+3948304 t^{11}+9791797 t^{12}+O\left(t^{13}\right)$}  \\
  & $\HS_{\mathbb{Z}} = $ &  $1+51 t^2+1242 t^4+18361 t^6+190857 t^8+1518870 t^{10}+9791797 t^{12}+O\left(t^{13}\right) $\\
  & $\HS_{\mathbb{Z}+\frac{1}{2}} =$&  $ 8 t+272 t^3+5024 t^5+61480 t^7+554464 t^9+3948304 t^{11}+O\left(t^{13}\right)$\\
  & $\PL= $& \parbox{12.5cm}{$8 t+15 t^2+32 t^3-4 t^4-128 t^5-285 t^6+320 t^7+2719 t^8+3520 t^9-14048 t^{10}-61440 t^{11}-20985 t^{12}+O\left(t^{13}\right) $}  \\ \vspace{5pt}
  & $\widetilde{\Hh}\eqref{eq:magQuiv_SU2_SU2_finite}=$ &  \parbox{11cm}{$ (1-t)^{8} \cdot \Hh\eqref{eq:magQuiv_SU2_SU2_finite} =  1+15 t^2+32 t^3+116 t^4+352 t^5+863 t^6+2112 t^7+4854 t^8+10176 t^9+20851 t^{10}+40736 t^{11}+76009 t^{12}+O\left(t^{13}\right)$}   \\
   & $\PL(\widetilde{\Hh})=$ & \parbox{12.5cm}{$  15 t^2+32 t^3-4 t^4-128 t^5-285 t^6+320 t^7+2719 t^8+3520 t^9-14048 t^{10}-61440 t^{11}-20985 t^{12}+O\left(t^{13}\right)$} \\ \bottomrule
    \end{tabular}
    \caption{Perturbative Hilbert series for the different phases of \eqref{eq:magQuiv_SU2_SU2_finite} and \eqref{eq:magQuiv_SU2_SU2_infty}}
    \label{tab:HS_SU2_SU2_OSp}
\end{table}

The same theory admits a realisation via 3 M5 branes on a $\C^2 \slash \Z_2$ singularity.  The corresponding magnetic quivers are given by \cite{Hanany:2018vph,Cabrera:2019izd}
\begin{align}
(1^3):\qquad 
&\raisebox{-.5\height}{
\includegraphics[page=37]{figures/figures_curve_2.pdf}
	}
	\qquad 
		\begin{cases}
	\dim\ \Coulomb &= 6  \\
	\gbal&=\surmL(2)^5  
		\end{cases}
		\label{eq:magQuiv_SU2_SU2_uni_finite}\\
		(3):\qquad 
&\raisebox{-.5\height}{
\includegraphics[page=38]{figures/figures_curve_2.pdf}
	}
	\qquad 
		\begin{cases}
	\dim\ \Coulomb &= 6  \\
	\gbal&=\surmL(2)^3
		\end{cases}
		\label{eq:magQuiv_SU2_SU2_uni_infty}
\end{align}
A straightforward computation confirms that the monopole formula of \eqref{eq:magQuiv_SU2_SU2_uni_finite} and \eqref{eq:magQuiv_SU2_SU2_uni_infty} agrees with $\widetilde{\Hh}\eqref{eq:magQuiv_SU2_SU2_finite}$ and $\widetilde{\Hh}\eqref{eq:magQuiv_SU2_SU2_infty}$, respectively.
In fact, the exact HWG for \eqref{eq:magQuiv_SU2_SU2_uni_finite} is known to be \cite{Hanany:2010qu}
\begin{align}
    \HWG_{(1^3)} =
    \PE \left[ \left( \sum_{i=1}^3 \nu_i^2 + \mu_1^2 +\mu_2^2 \right) t^2 
    + \mu_1 \mu_2 \prod_{i=1}^3 \nu_i (t^3+t^5) 
    + t^4
    - (\mu_1 \mu_2 \prod_{i=1}^3 \nu_i)^2 t^{10}
    \right]
\end{align}
where $\nu_i$ with $i=1,2,3$ denotes the $\surm(2)_i$ highest weight fugacities for the three balanced $\urm(1)$ nodes on top in \eqref{eq:magQuiv_SU2_SU2_uni_finite}. The remaining $\mu_{1,2}$ are the $\surm(2)$ highest weight fugacities for the left and right balanced $\urm(1)$. The reason for the $2+3$ split lies in the nature of the $\urm(1)$ nodes: three originate from \NS\ branes, while the other two from \Ds.

Again, the Coulomb branch dimension of the unitary magnetic quivers is $6$, while the dimension of the orthosymplectic quivers is $10$. The difference is accounted by the $4$ quaternionic free moduli observed in Table \ref{tab:HS_SU2_SU2_OSp}. This is consistent with the findings of \cite[Tab.\ 2]{Hassler:2019eso}.

The Hasse diagram of the 6d $\surm(2)\times \surm(2)$ theory is given by (see also \cite{Bourget:2019aer})
 \begin{align}
 	\label{eq:Hasse_SU2xSU2}
 \raisebox{-.5\height}{
 \includegraphics[page=39]{figures/figures_curve_2.pdf}
 }
 \qquad\xrightarrow[]{S_3} \qquad 
     \raisebox{-.5\height}{
     \includegraphics[page=40]{figures/figures_curve_2.pdf}
}
 \end{align}
which is a subdiagram of \eqref{eq:Hasse_SU3xSU2}.

\section{Conclusion}
\label{sec:conclusions}
Despite numerous studies on 6d $\Ncal=(1,0)$ theories, their Higgs branches are far from being fully understood. In this note, further advances have been made to work out selected cases in detail.

Starting from $n$ M5 branes on the minimal $D$-type singularity $\C^2 \slash D_4$, non-trivial Higgs branches can be accessed by trading tensor multiplets, corresponding to $\sprm(0)$ gauge factors, for a number of hypermultiplets. The simplest and cleanest configuration is that of an $\surm(3)$ super Yang-Mills theory coupled to an $\sprm(0)$ factor. As argued in Section \ref{sec:2M5s_RHS_trivial}, the infinite coupling Higgs branch has two phases, related by discrete gauging. More fundamentally, the infinite coupling Higgs moduli are obtained by an $\surm(3)$ hyper-K\"ahler quotient of the minimal nilpotent orbit closure of $E_8$ --- the result is a nilpotent orbit closure of $E_6$ (or a $\Z_2$ cover thereof). We confirm this conclusion with a number of different reasonings. The magnetic quiver derived is in fact a novel construction of this moduli space and, simultaneously, represents a physical construction thereof.
Once the $\sprm(0)$ factor is coupled to a larger gauge group, which admits charged hypermultiplets, the Higgs branch description becomes more intricate. Nonetheless, the magnetic quiver allows for an analysis.

Another class of intriguing  Higgs moduli spaces arises from coupling the ``clusters'' of $(-2)(-3)$ or $(-2)(-2)(-3)$ curves to an $\sprm(0)$ factor. The minimal configurations support an $\surm(2)\times G_2$ product gauge group. Again, the physical intuition indicates that the infinite coupling Higgs branches exhibit an $F_4$ global symmetry. But in fact more is true, $\Higgs_\infty$ of the $\surm(2)\times G_2\times \sprm(0)$ theory on $(-2)(-2)(-3)(-1)$ curves is the  20 dimensional nilpotent orbit closure of $F_4$. Based on a combination of techniques --- brane systems, 6d quivers, magnetic quivers, and quiver subtraction --- we derived the infinite coupling Higgs branch phase diagram and found agreement with the known $F_4(a_3)$ Hasse diagram from the mathematics literature.
The $\surm(2)\times G_2\times \sprm(0)$ theory on  $(-2)(-3)(-1)$ curves is related to the $F_4$ nilpotent orbit by an $S_3$ discrete symmetry.

Lastly, by picking non-trivial boundary conditions on both sides we find interesting Higgs branches of 6d theories. There are two scenarios to distinguish: the boundary conditions may overlap, but there is at least one $\sprm(0)$ factor remaining, or the boundary conditions overlap without an $\sprm(0)$ part. The first scenario has been exemplified in Section \ref{sec:non-overlapping_bc}, and the infinite coupling Higgs branches are still related to $\clorbit{E_8}^{\min}$ by gauging various subgroups of $E_8$.
For the minimal $\surm(3)\times \sprm(0)\times\surm(3)$ quiver theory, the infinite coupling Higgs branches are precisely given by the $\surm(3)\times\surm(3)$ hyper-K\"ahler quotient of $\clorbit{E_8}^{\min}$ (or a further $S_2$ or $S_3$ quotient thereof). However, for non-minimal gauge groups the moduli spaces are rather intricate such that one has to rely on the magnetic quivers for an explicit description.
The second scenario has been addressed in Section \ref{sec:overlapping_bc}. The collapse of all $(-1)$ curves leads to theories purely defined on $(-2)$ curves. Here, two interesting phenomena occur. Firstly, those 6d theories have an alternative unitary magnetic quiver construction, which allows to independently verify the predictions obtained from the orthosymplectic quivers. Secondly, some of these brane configurations are known to produce a mismatch in the anomaly polynomial, compared to the expectation for the effective 6d theory. In terms of the magnetic quivers, we observe that the Coulomb branches contain a free sector. Upon removing these moduli, the Hilbert series agrees with the results of the unitary magnetic quivers.

\paragraph{Outlook.}
As noted in \cite{Mekareeya:2016yal,Hassler:2019eso}, negative branes extend the brane system construction and give rise to new effective theories.
In this note, we have reiterated and expanded this reasoning by demonstrating that a host of physical properties can be extracted from such brane systems and reveal interesting Higgs branch geometries. It is however an open challenge to derive the identifications of Table \ref{tab:negative_branes} purely from the brane system.

On a different note, the magnetic quiver approach is currently limited to special partitions of $\sorm(2n)$, for the same reasons as in the 3d $\Ncal=4$ case. Even for special partitions, some $T_\rho[\sorm(2n)]$ tails contain ``bad'' $\sprm(k)$ nodes, which renders the models incomputable by means of the monopole formula. One might try to extend monopole formula techniques to accommodate for such cases, as proposed in \cite{Akhond:2022jts}.

It has been argued \cite{Distler:2022yse} that a given theory labelled by two very even partitions of $\sorm(4k)$ may in fact give rise to two distinct theories that differ in their Higgs branch spectrum. While these partitions are special, their $T_\rho[\sorm(4k)]$ tails suffer from bad nodes. Hence, magnetic quivers are not yet sensitive to this question, but it would be interesting to remedy this circumstance.

\paragraph{Acknowledgements.}
We are indebted to Santiago Cabrera for collaboration in an early stage. 
We would like to thank Antoine Bourget, Julius Grimminger, Daniel Juteau, Rudolph Kalveks, Noppadal Mekareeya, Paul-Konstantin Oehlmann, Tom Rudelius, Washington Taylor, and Zhenghao Zhong for discussion and correspondence.
We are also grateful to Rudolph Kalveks for invaluable help with \texttt{Mathematica}. M.S. thanks Jie Gu and Ryo Suzuki for use of their computing facilities.
We thank the Simons Center for Geometry and Physics, Stony Brook University for the hospitality and the partial support during the initial stage of this work at the Simons Summer workshop 2018.
A.H. thanks the Simons Center for Geometry and Physics, Stony Brook University for the hospitality and the partial support during the final stage of this work at the Simons Summer workshop 2022.
A.H. was supported by STFC grants ST/P000762/1 and ST/T000791/1.
M.S. was further supported by the  Austrian Science Fund (FWF) grant P28590, the National Natural Science Foundation of China (grant no.
11950410497), and the China Postdoctoral Science Foundation (grant no. 2019M650616).
M.S. is grateful for the warm hospitality of Fudan University, Department of Physics during various stages of this work.
\appendix
    \section{Background material}
\label{app:background}
\subsection{Brane creation and annihilation}
\label{app:orientifolds}
Following \cite{Hanany:1996ie}, in a system of D$p$-D$(p{+}2)$-\NS\ branes, 
D$p$ brane 
creation or annihilation happens whenever a \NS\ passes through an D$(p{+}2)$. 
In 
the presence of O$p$ planes, which carry non-trivial brane charge, a \NS\ brane 
can pass through an D$(p{+}2)$ with or without creation of an additional D$p$ 
brane. 
To begin with, recall \cite{Evans:1997hk,Hanany:1999sj,Hanany:2000fq}
\begin{compactitem}
 \item An O$p^\pm$ becomes an O$p^\mp$ when passing through a half \NS ; 
likewise, $\widetilde{\text{O}p}^\pm$ turns into $\widetilde{\text{O}p}^\mp$.
\item An O$p^\pm$ becomes an $\widetilde{\text{O}p}^\pm$ when passing through 
a half D$(p{+}2)$, and vice versa.
\end{compactitem}
According to \cite{Hanany:1999sj,Feng:2000eq}, the charges of the O$p$ planes 
(in unites of the physical D$p$ branes) are given by 
\begin{align}
 \text{charge}(\text{O}p^{\pm}) = \pm 2^{p-5} 
 \; , \quad
 \text{charge}(\widetilde{\text{O}p}^{-}) = \frac{1}{2}- 2^{p-5}
 \; , \quad
 \text{charge}(\widetilde{\text{O}p}^{+}) =  2^{p-5} \; .
 \label{eq:charges_orientifold}
\end{align}
Following the conventions of \cite{Gaiotto:2008ak}, the different 
orientifolds are denoted by:
\begin{alignat}{2}
\Osm \; \&\; 2n\cdot \frac{1}{2}\Ds &: \quad 
  \raisebox{-.5\height}{
  \includegraphics[page=1]{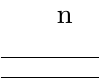}
	} 
\quad ,& \qquad \qquad 
	\Osmt\; \&\; 2n\cdot \frac{1}{2}\Ds &: \quad 
  \raisebox{-.5\height}{
  \includegraphics[page=2]{figures/figures_app_backgrd.pdf}
	} 
	\quad , \\
\Osp\; \&\; 2n\cdot \frac{1}{2}\Ds &: \quad 
  \raisebox{-.5\height}{
  \includegraphics[page=3]{figures/figures_app_backgrd.pdf}
	} 
\quad ,& \qquad \qquad 
	\Ospt\; \&\; 2n\cdot \frac{1}{2}\Ds &: \quad 
  \raisebox{-.5\height}{
  \includegraphics[page=4]{figures/figures_app_backgrd.pdf}
	} 
	\quad ,
\end{alignat}
i.e.\ 
O$6^-$ empty line, $\widetilde{\text{O}6}^-$ solid line, O$6^+$ dotted line, 
$\widetilde{\text{O}6}^+$ dashed line.

Next, there are four scenarios for brane creation and annihilation. These follow 
from preservation of the linking number before and after the transition.
The linking numbers $l_{\NS\ }$ for half \NS\ or $l_{\text{D}(p+2) }$ for half 
D$(p{+}2)$ are 
defined as \cite{Hanany:1996ie}
\begin{subequations}
\label{eq:linking_numbers}
\begin{align}
l_{\NS\ } &= \frac{1}{2} \left( R_{\text{D}(p{+}2)} - L_{\text{D}(p{+}2)} 
\right) + \left( L_{\text{D}p} - R_{\text{D}p} \right) \,,\\
l_{\text{D}(p+2) } &= \frac{1}{2} \left( R_{\NS\ } - L_{\NS\ } \right) + \left( 
L_{\text{D}p} - R_{\text{D}p} \right) \,,
\end{align}
\end{subequations}
where $L_X$, $R_X$ denote the total number of branes of type $X$ to the left 
or right, respectively.
Note that the O$p$ planes contribute to $L_{Dp}$ and $R_{Dp}$ according to 
\eqref{eq:charges_orientifold}; naturally, 
half \NS\ or half D$(p{+}2)$ branes contribute with charge 
$\frac{1}{2}$ to the numbers $L$ and $R$, respectively.
It then follows that
\begin{subequations}
\begin{align}
\raisebox{-.5\height}{
  \includegraphics[page=5]{figures/figures_app_backgrd.pdf}
	}
	\qquad
	&\leftrightarrow
	\qquad
	\raisebox{-.5\height}{
	  \includegraphics[page=6]{figures/figures_app_backgrd.pdf}
	} 
	\label{eq:brane_creation_1}
	\\
\raisebox{-.5\height}{
  \includegraphics[page=7]{figures/figures_app_backgrd.pdf}
	}
	\qquad
	&\leftrightarrow
	\qquad
	\raisebox{-.5\height}{
	  \includegraphics[page=8]{figures/figures_app_backgrd.pdf}
	} 
		\label{eq:brane_creation_2}
	\\
\raisebox{-.5\height}{
  \includegraphics[page=9]{figures/figures_app_backgrd.pdf}
	}
	\qquad
	&\leftrightarrow
	\qquad
	\raisebox{-.5\height}{
	  \includegraphics[page=10]{figures/figures_app_backgrd.pdf}
	} 
		\label{eq:brane_creation_3}
	\\
\raisebox{-.5\height}{
  \includegraphics[page=11]{figures/figures_app_backgrd.pdf}
	}
	\qquad
	&\leftrightarrow
	\qquad
	\raisebox{-.5\height}{
	  \includegraphics[page=12]{figures/figures_app_backgrd.pdf}
	} 
		\label{eq:brane_creation_4}
\end{align}
\label{eq:brane_creation_all}
\end{subequations}
by requiring that all linking numbers \eqref{eq:linking_numbers} remain 
constant.
%
%
\subsection{Global symmetry for orthosymplectic quiver}
\label{app:global_sym}
Following \cite{Gaiotto:2008ak}, there are conditions upon which 
orthogonal and symplectic gauge nodes in a $3$d $\Ncal=4$ gauge theory have \emph{positive balance}, \emph{zero balance}, or \emph{negative balance}. Gauge nodes with zero balance, also called \emph{balanced} gauge nodes, are expected to have monopole operators of spin 1 under 
the R-charge that lead to symmetry enhancement. 

An $\sorm(k)$ (or $\orm(k)$) gauge theory coupled to fundamental 
hypermultiplets with $\usprm(2n)$ flavour symmetry is called
 \begin{equation}
 \text{\emph{positively balanced} if } \quad   n > k-1\, , \qquad \text{and \emph{balanced} 
if}\quad   n =k-1 \,.
\label{eq:balanced_SO}
\end{equation}

Analogously, an $\usprm(2l)=\sprm(l)$ gauge theory coupled to fundamental 
hypermultiplets with $\orm(2n)$ flavour symmetry is called
\begin{equation}
 \text{\emph{positively balanced} if } \quad   n > 2l+1 \,, \qquad 
 \text{ and \emph{balanced} if } \quad   n =2l+1 \,.
\end{equation}
Considering an orthosymplectic quiver, i.e.\ a linear quiver with alternating 
orthogonal and symplectic gauge nodes, a chain of $p$ balanced nodes gives rise 
to the following enhanced Coulomb branch symmetry:
\begin{compactitem}
 \item An $\sorm(p+1)$ symmetry, if there are no $\sorm(2)$ (or $\orm(2)$) 
gauge nodes at the ends.
\item An $\sorm(p+2)$ symmetry, if there is an $\sorm(2)$ (or $\orm(2)$) 
gauge node at one of the two ends.
\item An $\sorm(p+3)$ symmetry, if there is an $\sorm(2)$ (or $\orm(2)$) 
gauge node at each end.
\end{compactitem}

\subsection{Notation}
The magnetic quivers in this work are composed of unitary gauge nodes $\urm(n)$, special orthogonal gauge nodes $\sorm(k)$, and symplectic gauge nodes $\sprm(l)$. The magnetic lattices and dressing factors have been detailed in \cite{Cremonesi:2013lqa}. For unframed orthosymplectic magnetic quivers with product gauge group $G=\prod_I \sorm(2n_i) \times \prod_J \sprm(k_j)$, there exists a trivially acting $\Z_2^{\mathrm{diag}}\subset G$. As discussed in \cite{Bourget:2020xdz}, this $\Z_2^{\mathrm{diag}}$ has to be removed from the gauge group. Thus, the magnetic lattice of $G\slash \Z_2^{\mathrm{diag}}$ is $\left(\bigoplus_{I} \Z^{n_I} \oplus
\bigoplus_{J} \Z^{k_J} \right) \cup 
\left(\bigoplus_{I} (\Z+\frac{1}{2})^{n_I} \oplus
\bigoplus_{J} (\Z+\frac{1}{2})^{k_J} \right)$.

    \section{Examples}
    \label{sec:examples}
    \subsection{\texorpdfstring{Magnetic quivers for $A$-type boundary conditions}{Magnetic quivers for A-type boundary conditions}}
\label{app:A-type_bc}
Recall a few preliminaries: Denote the A-type ADHM quiver for $n$ $\surm(k)$ instantons on 
$\C^2$ by 
 \begin{align}
  Y^A_{n,k} 
  =
  \raisebox{-.5\height}{
    \includegraphics[page=1]{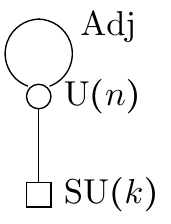}
	}  
\end{align}
 Here, the conventions of \cite{Cremonesi:2014uva} are used for the 3d $\Ncal=4$ $T_{\rho}^\sigma[\surm(k)]$ theories. 
\begin{myStat}[A-type]
\label{prop:A-type}
The magnetic quiver for the infinite gauge coupling phase of 
$T_{\surm(k)}^{n}(\rho_L,\rho_R)$ is given by 
\begin{subequations}
 \begin{align}
 \magQuiv_{\rho_L,\rho_R} &= \left( T_{\rho_L}[\surm(k)] \times Y^A_{n,k} 
\times 
T_{\rho_R}[\surm(k)] \right) \slash \slash \slash \surm(k) \\
&=	\raisebox{-.5\height}{
\includegraphics[page=2]{figures/figures_app_SU4.pdf}
}
\end{align}
and the integers $\{a_i\}_{i=1}{\ell}$, $\{b_j\}_{j=1}^{\ell^\prime}$ are determined by the partitions $\rho_L$, $\rho_R$, respectively. See e.g.\ \cite{Cremonesi:2014uva}. Then the equality of moduli spaces 
\begin{align}
 \Higgs^{6d}_{\infty}\left(T_{\surm(k)}^{n}(\rho_L,\rho_R)\right) = 
\Coulomb^{3d}\left(\magQuiv_{\rho_L,\rho_R}\right)
\end{align}
\end{subequations}
holds. The Higgs branch dimension at infinite coupling is
\begin{align}
 \dim_\HH\   \Higgs^{6d}_{\infty}\left(T_{\surm(k)}^{n}(\rho_L,\rho_R)\right) 
  = 
 \dim_\HH\ \Coulomb^{3d}\left(\magQuiv_{\rho_L,\rho_R}\right)
 &=
n+  \dim\ \surm(k)  - \dim_\HH\ \clorbit{\rho_L}- \dim_\HH\ \clorbit{\rho_R} \\
&= n + \rank\ \surm(k) + \dim_{\HH} \slice{\Ncal,\rho_L} 
+ \dim_{\HH} \slice{\Ncal,\rho_R} \,. \notag
\end{align}
\end{myStat}
Here, $\slice{\Ncal,\rho}$ denotes the intersection of the transverse slice to the orbit $\orbit{\rho}$ with the nilpotent cone $\Ncal$. Recalling the Coulomb branches $\Coulomb\left(T_{\rho}[\surm(k)] \right)=\slice{\Ncal,\rho}$, the dimension formula is straightforward. 
Statement \ref{prop:A-type} can immediately be generalised to describe all phases of the $T_{\surm(k)}^{n}(\rho_L,\rho_R)$ theories via using the discrete gauging proposal \cite{Hanany:2018vph} and its manifestation on the Coulomb branches of magnetic quivers \cite{Hanany:2018cgo,Hanany:2018dvd}. Appendix \ref{app:Examples_SU4} exemplifies the $\surm(4)$ case.

\subsection{Examples: \texorpdfstring{$\surm(4)$}{SU(4)}}
\label{app:Examples_SU4}
Starting from the M-theory setting $n$ \Mf\ branes on $\C^2\slash \Z_4$, the 
corresponding Type IIA description has the advantage that boundary conditions 
of \Ds\ on \De\ branes can be introduced additionally. Focusing on $\surm(4)$ 
examples, one considers the boundary conditions displayed in Table \ref{tab:SU4_examples}.

For the A-type case, the magnetic quiver is available for finite as well as 
infinite coupling. The transition between both phases is realised via 
\emph{discrete gauging} \cite{Hanany:2018vph}. The magnetic quivers are related 
via an operation on the bouquet of $n$ $\uo$-nodes, see 
\cite{Hanany:2018cgo,Hanany:2018dvd}. Hence, Table \ref{tab:SU4_examples} only details the infinite coupling magnetic quiver.

\begin{table}[t]
    \centering
    \scalebox{0.975}{
    \begin{tabular}{lrrr}
    \toprule 
    data & \multicolumn{1}{c}{branes system} & \multicolumn{1}{c}{6d electric theory} & \multicolumn{1}{c}{magnetic quiver} \\ \midrule
 $\substack{\rho_L =  (1^4)  \\ \rho_R = (1^4)}$      &
    \raisebox{-.5\height}{
    \includegraphics[page=3]{figures/figures_app_SU4.pdf}
	} &
	\raisebox{-.5\height}{
	\includegraphics[page=4]{figures/figures_app_SU4.pdf}
	}
	&
	\raisebox{-.5\height}{
	\includegraphics[page=5]{figures/figures_app_SU4.pdf}
	}
          \\
$\substack{ \rho_L =(2,1^2)\\ \rho_R = (1^4)}$ &
\raisebox{-.5\height}{
\includegraphics[page=6]{figures/figures_app_SU4.pdf}
	} 
	&
	\raisebox{-.5\height}{
	\includegraphics[page=7]{figures/figures_app_SU4.pdf}
	}
	&
	\raisebox{-.5\height}{
	\includegraphics[page=8]{figures/figures_app_SU4.pdf}
	} \\
 $\substack{\rho_L =(2^2) \\ \rho_R = (1^4)}$  & 
 \raisebox{-.5\height}{
 \includegraphics[page=9]{figures/figures_app_SU4.pdf}
	}
 & 
 \raisebox{-.5\height}{
 \includegraphics[page=10]{figures/figures_app_SU4.pdf}
	}
 & 
 \raisebox{-.5\height}{
 \includegraphics[page=11]{figures/figures_app_SU4.pdf}
	}\\
$\substack{\rho_L =(3,1) \\ \rho_R = (1^4)}$  &
\raisebox{-.5\height}{
\includegraphics[page=12]{figures/figures_app_SU4.pdf}
	}
& 
\raisebox{-.5\height}{
\includegraphics[page=13]{figures/figures_app_SU4.pdf}
	}
&
\raisebox{-.5\height}{
\includegraphics[page=14]{figures/figures_app_SU4.pdf}
	}
\\
$\substack{ \rho_L =(4)\\ \rho_R = (1^4)}$  &
\raisebox{-.5\height}{
\includegraphics[page=15]{figures/figures_app_SU4.pdf}
	}
&
\raisebox{-.5\height}{
\includegraphics[page=16]{figures/figures_app_SU4.pdf}
	}
&
\raisebox{-.5\height}{
\includegraphics[page=17]{figures/figures_app_SU4.pdf}
	} 
	\\
	\bottomrule
    \end{tabular}
}
    \caption{Magnetic quivers for infinite coupling Higgs branch for $n$ M5 branes on an $A_3$ singularity $\C^2 \slash \Z_4$ with boundary conditions $\rho_{L,R}$. In the magnetic quiver, the contribution from $\rho_L$ is coloured in black, while contributions from $\rho_R$ and $2k$ is coloured in \textcolor{blue}{blue}. As summarised in Appendix \ref{app:global_sym}, the global symmetry on the Coulomb branch can be deduced from the balanced nodes, which are indicated by a \textcolor{red!50}{red filling}.}
    \label{tab:SU4_examples}
\end{table}

    \subsection{\texorpdfstring{Magnetic quivers for $D$-type boundary conditions}{Magnetic quivers for D-type boundary conditions}}
\label{app:D-type_bc}
A few preliminaries are required: Partitions of classical Lie algebras other than $\surmL(n)$ require a 
more careful treatment. In particular, a map that takes a partition of $\gfrak$ 
to a partition of the GNO-dual $\GNOfrak$  is necessary, see for 
instance \cite[Sec.\ 6]{Cabrera:2017njm}. Generically, for classical $\gfrak$ 
the \emph{Barbasch-Vogan} map acts as 
\begin{align}
 d_{\mathrm{BV}} : \left\{ \substack{\text{partitions} \\ \text{of } \gfrak} 
\right\}
 \to  \left\{\substack{\text{special partitions} \\ \text{of } \GNOfrak} 
\right\}
\end{align}
Fortunately, for (GNO) self-dual algebras like $\surmL(n)$ and $\sormL(2n)$, 
the Barbasch-Vogan map reduces to the \emph{Lusztig-Spaltenstein} map defined 
via
\begin{align}
d_{\mathrm{LS}}(\rho)
= \begin{cases}
   \rho^T \, ,  & \gfrak = \surmL(n) \\
   (\rho^T)_D \,, & \gfrak = \sormL(2n)
\end{cases}
 \,,
\end{align}
where $(\cdot)^T$ denotes transposition and $(\cdot)_D$ D-collapse. Partitions 
are called \emph{special} if $d_{\mathrm{LS}}^2 (\rho) = \rho$. Hence, all 
A-type partitions are special, but there exist D-type partitions that are 
non-special.

The $D$-type ADHM quiver for $n$ $\sorm(2k)$ instantons on $\C^2$ is 
denoted by
\begin{align}
  Y^D_{n,2k} 
  =
  \raisebox{-.5\height}{
  \includegraphics[page=1]{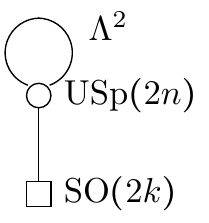}
	}  
\end{align}
Here, the conventions of \cite{Cremonesi:2014uva} are used for $3$d $\Ncal=4$
$T_\rho^\sigma[\sorm(2k)]$ theories, with $\rho, \sigma$ two \emph{special} 
D-type partitions 
of $2k$. It is enough to restrict to $T_\rho[\sorm(2k)] \equiv 
T_\rho^{(1^{2k})}[\sorm(2k)]$. 
The precise statement becomes
\begin{myStat}[D-type]
\label{prop:D-type} 
Let $\rho_{L,R}$ be two special D-type partitions of $2k$. 
The magnetic quiver for the infinite gauge coupling phase of
$T_{\sorm(2k)}^{n}(\rho_L,\rho_R)$ is 
\begin{subequations}
\begin{align}
 \magQuiv_{\rho_L,\rho_R} &= \left( T_{\rho_L}[\sorm(2k)] \times Y^D_{n,2k} 
\times 
T_{\rho_R}[\sorm(2k)] \right) \slash \slash \slash (\sorm(2k) \slash \Z_2) \\
&=\raisebox{-.5\height}{
\includegraphics[page=2]{figures/figures_app_SO12.pdf}
} 
\end{align}
and the integers $\{a_i\}_{i=1}{\ell}$, $\{b_j\}_{j=1}^{\ell^\prime}$ are determined by the partitions $\rho_L$, $\rho_R$, respectively. See e.g.\ \cite{Cremonesi:2014uva}. Then the equality of moduli spaces 
\begin{align}
 \Higgs^{6d}_{\infty}\left(T_{\sorm(2k)}^{n}(\rho_L,\rho_R) \right) = 
\Coulomb^{3d}(\magQuiv_{\rho_L,\rho_R})
\end{align} 
\end{subequations}
holds. The Higgs branch dimension at infinite coupling is
\begin{align}
 \dim_\HH\   \Higgs^{6d}_{\infty}\left(T_{\sorm(2k)}^{n}(\rho_L,\rho_R) \right) 
  = 
 \dim_\HH\ \Coulomb^{3d}(\magQuiv_{\rho_L,\rho_R})
 &=
n+  \dim\ \sorm(2k)  - \dim_\HH\ \clorbit{\rho_L}- \dim_\HH\ \clorbit{\rho_R}\\
&= n + \rank\ \sorm(2k) + \dim_{\HH} \slice{\Ncal,\rho_L} 
+ \dim_{\HH} \slice{\Ncal,\rho_R}  \notag
\,.
\end{align}
\end{myStat}
Again, $\slice{\Ncal,\rho}$ denotes the intersection of the transverse slice to the orbit $\orbit{\rho}$ with the nilpotent cone $\Ncal$. Recalling that $\Coulomb\left(T_{\rho}[\sorm(2k)] \right)=\slice{\Ncal,\rho}$, the dimension formula sums over the dimensions of the legs plus the rank of the central node.
The discrete quotient analysis of \cite{Hanany:2018cgo} allows to extend Statement \ref{prop:D-type} to further phases of $T_{\sorm(2k)}^{n}(\rho_L,\rho_R)$. In Appendix \ref{app:Examples_SO12}, examples for $\sorm(12)$ are discussed in detail.

\subsection{Examples: \texorpdfstring{$\sorm(12)$}{SO(12)}}
\label{app:Examples_SO12}
By symmetry of the brane configuration one may consider, say, only the 
left-hand-side and work out the effects of the boundary conditions 
corresponding 
to $\rho_L$ without worrying about the right-hand-side.
In order words, vary the left-hand-side partition $\rho\equiv \rho_L$ while keeping the 
right-hand-side partition trivial $\rho_R=(1^{12})$. In the following examples, the magnetic 
quiver at the infinite coupling point is displayed by using the results of \cite{Cabrera:2019dob}.
\paragraph{Non-special partitions:}
Following for instance \cite{Heckman:2016ssk}, one can consider all D-type 
partitions as legitimate boundary conditions. The electric quiver can be deduced 
by the results of \cite{Mekareeya:2016yal}. However, the magnetic quiver 
reproduces the correct Higgs branch only for \emph{special} partitions. 
Nevertheless, one can still work out the magnetic quiver to any non-special 
partition and emphasis the short-comings.
\begin{itemize}
 \item $\sorm(12)$ has the following non-special partitions: 
 \begin{align}
  (3,2^2,1^5)\,, \qquad 
  (3,2^4,1)\,, \qquad 
  (5,2^2,1^3)\,, \qquad 
  (7,2^2,1)
 \end{align}
 \item Since non-special means $d_{\mathrm{LS}}^2 (\rho)\neq \rho$, one 
computes the $d_{\mathrm{LS}}^2$ action to be
\begin{subequations}
\label{eq:LS_squared}
\begin{alignat}{2}
 d_{\mathrm{LS}}^2(3,2^2,1^5) &= (3^2,1^6) 
 \;,\quad &
 d_{\mathrm{LS}}^2(3,2^4,1) &= (3^2,2^2,1^2) 
 \;, \\
 d_{\mathrm{LS}}^2(5,2^2,1^3) &= (5,3,1^4) 
\;,\quad &
 d_{\mathrm{LS}}^2(7,2^2,1) &= (7,3,1^2) \,.
\end{alignat}
\end{subequations}
\item Similar to the remark around \cite[eq.\ 
(6.2)]{Cabrera:2017njm} for $3$d $\Ncal=4$ theories, it is known that the Type 
IIB \Dthree-\Dfive-\NS\ brane construction for non-special partitions does 
not yield the desired moduli spaces. 
In more detail, the brane construction for a non-special partitions yields the 
world-volume theory labelled by $d_{\mathrm{LS}}^2(\rho)$ instead. 
\end{itemize}
Therefore, the expectation for a magnetic quiver associated to boundary 
conditions of a non-special partition $\rho$ is that it is identical to 
$d_{\mathrm{LS}}^2(\rho)$, which is not the correct moduli space.

\paragraph{Notation.}
In the magnetic quiver, the contribution from $\rho_L$ is coloured in 
black, while contributions from $\rho_R$ and $2k$ is coloured in \textcolor{blue}{blue}. As summarised in Appendix \ref{app:global_sym}, the 
global symmetry on the Coulomb branch can be deduced from the balanced nodes, 
which are indicated by a \textcolor{red!50}{red filling}. Gauge nodes which are 
\emph{bad} are indicated by a \textcolor{gray!50}{grey filling}.
%
%
\subsubsection{partition \texorpdfstring{$(1^{12})$}{111111111111}}
This case is covered in \cite{Cabrera:2019dob}. The Brane configuration and 6d electric theory read
\begin{align}
 \raisebox{-.5\height}{
 \includegraphics[page=3]{figures/figures_app_SO12.pdf}
	}
	\qquad \longrightarrow \qquad 
\raisebox{-.5\height}{
\includegraphics[page=4]{figures/figures_app_SO12.pdf}
	} 
	+\substack{
	2n{-}4\text{ nodes} \\
	Sp(2)/SO(12) }
\end{align}
compute the Higgs branch dimension
\begin{align}
    \dim(\Higgs_{f}) =38 \;, \quad 
        \dim(\Higgs_\infty) 
    = n+66  
    = \dim(\Higgs_f) + 29+(n-1)\,. 
\end{align}
The finite coupling magnetic quiver is given by
\begin{align}
 &\raisebox{-.5\height}{
 \includegraphics[page=5]{figures/figures_app_SO12.pdf}
	} 
	\label{eq:magQuiv_1^12_finite} 
 \qquad 
 \begin{cases}
 G_J &= \sorm(12) \times \sorm(12)\\
     \dim\ \Coulomb &= 38 = \dim(\Higgs_f) \,.
 \end{cases}
\end{align}
\paragraph{Infinite coupling.}
Moving the $12$ \De\ passed the left most half \NS\ brane, accounting for brane 
creating as summarised in Appendix \ref{app:orientifolds}, one arrives at
\begin{align}
 \raisebox{-.5\height}{
 \includegraphics[page=6]{figures/figures_app_SO12.pdf}
	}
\end{align}
and the magnetic quiver at the CFT fixed point is 
\begin{align}
 &\raisebox{-.5\height}{
 \includegraphics[page=7]{figures/figures_app_SO12.pdf}
	} 
	\label{eq:magQuiv_1^12_infinite} 
\qquad 
\begin{cases}
G_J &= \sorm(12) \times \sorm(12) \\
 \dim\ \Coulomb &= n+66 = \dim\  \Higgs_\infty
\end{cases}
\end{align}
\paragraph{Transitions.}
The difference between finite coupling \eqref{eq:magQuiv_1^12_finite} and the CFT fixed point \eqref{eq:magQuiv_1^12_infinite} is one $E_8$ transition and $(n-1)$ $D_4$ transitions, see \cite{Cabrera:2019dob}.
%
%
\subsubsection{partition \texorpdfstring{$(2^2,1^{8})$}{2211111111}}
The brane configuration and 6d electric theory read 
\begin{align}
 \raisebox{-.5\height}{
  \includegraphics[page=8]{figures/figures_app_SO12.pdf}
	}
	\quad \longrightarrow \quad
 \raisebox{-.5\height}{
 \includegraphics[page=9]{figures/figures_app_SO12.pdf}
	} 
	+\substack{
	2n{-}4\text{ nodes} \\
	Sp(2)/SO(12) }
\end{align}
One computes the Higgs branch dimension to be
\begin{align}
    \dim(\Higgs_{f})  = 29 
    \;,\quad 
    \dim(\Higgs_\infty) 
    = n+57  =
    \dim(\Higgs_f) +29 +(n-1)\,.\
    \end{align}
The finite coupling magnetic quiver is given by
\begin{align}
 &\raisebox{-.5\height}{
 \includegraphics[page=10]{figures/figures_app_SO12.pdf}
	} 
	\label{eq:magQuiv_2^2-1^8_finite} 
	\qquad 
	\begin{cases}
	G_J &= \sorm(8) \times \sorm(12) \\
    \dim\ \Coulomb &= 29 = \dim(\Higgs_f)	    
	\end{cases}
\end{align}
\paragraph{Infinite coupling.}
Moving the $10$ \De\ passed the two left most half \NS\ branes, accounting for 
brane 
creating as summarised in Appendix \ref{app:orientifolds}, one arrives at
\begin{align}
 \raisebox{-.5\height}{
 \includegraphics[page=11]{figures/figures_app_SO12.pdf}
	}
\end{align}
and the magnetic quiver becomes
\begin{align}
 &\raisebox{-.5\height}{
  \includegraphics[page=12]{figures/figures_app_SO12.pdf}
	} 
	\label{eq:magQuiv_2^2-1^8_infinite} 
\qquad 
\begin{cases}
 G_J &= \sorm(8) \times \sorm(12) \\
 \dim\ \Coulomb &= n+57 = \dim\ \Higgs_\infty
\end{cases}
\end{align}
\paragraph{Transitions.}
The difference between \eqref{eq:magQuiv_2^2-1^8_infinite} and \eqref{eq:magQuiv_2^2-1^8_finite} are one $E_8$ transition and $n-1$ $D_4$ transitions. For instance, \eqref{eq:magQuiv_2^2-1^8_infinite} before the $n-1$ $D_4$ transitions (and subsequent discrete gauging) is a magnetic quiver of the form
\begin{align}
 \raisebox{-.5\height}{
   \includegraphics[page=13]{figures/figures_app_SO12.pdf}
	} 
	\label{eq:magQuiv_2^2-1^8_intermediate}
\end{align}
from which a quiver subtraction of the $E_8$ quiver leads to \eqref{eq:magQuiv_2^2-1^8_finite}.
%
%
\subsubsection{partition \texorpdfstring{$(3,1^{9})$}{3111111111}}
The brane configuration and the 6d electric theory read
\begin{align}
 \raisebox{-.5\height}{
    \includegraphics[page=14]{figures/figures_app_SO12.pdf}
	}
	\quad \longrightarrow \quad
 \raisebox{-.5\height}{
    \includegraphics[page=15]{figures/figures_app_SO12.pdf}
	} 
	+\substack{
	2n{-}4\text{ nodes} \\
	Sp(2)/SO(12) }
\end{align}
One computes the Higgs branch dimension
\begin{align}
    \dim(\Higgs_{f})  = 28 
    \;,\quad 
    \dim(\Higgs_\infty) 
    = n+56  
    =\dim(\Higgs_f) +29+(n-1)\,. 
\end{align}
The finite coupling magnetic quiver is given by
\begin{align}
 &\raisebox{-.5\height}{
     \includegraphics[page=16]{figures/figures_app_SO12.pdf}
	} 
	\label{eq:magQuiv_3-1^9_finite} 
	\qquad
	\begin{cases}
	G_J &= \sorm(9)\times \sorm(12) \\
    \dim\ \Coulomb &= 28 = \dim(\Higgs_f) 	    
	\end{cases}
\end{align}
\paragraph{Infinite coupling.}
Moving the $10$ \De\ passed the three left most half \NS\ branes, accounting 
for 
brane creating as summarised in Appendix \ref{app:orientifolds}, one arrives at
\begin{align}
 \raisebox{-.5\height}{
      \includegraphics[page=17]{figures/figures_app_SO12.pdf}
	}
\end{align}
and the magnetic quiver becomes
\begin{align}
 &\raisebox{-.5\height}{
       \includegraphics[page=18]{figures/figures_app_SO12.pdf}
	} 
	\label{eq:magQuiv_3-1^9_infinite} 
	\qquad
	\begin{cases}
G_J &= \sorm(9) \times \sorm(12) \\  
 \dim\ \Coulomb  &= n+56 = \dim (\Higgs_\infty)
	\end{cases}
\end{align}
\paragraph{Transitions.}
The difference between \eqref{eq:magQuiv_3-1^9_infinite} and \eqref{eq:magQuiv_3-1^9_finite} is given by one $E_8$ transition and $(n-1)$ $D_4$ transitions. This is straightforwardly verified by quiver subtraction, analogously to \eqref{eq:magQuiv_2^2-1^8_intermediate}.
%
%
\subsubsection{partition \texorpdfstring{$(2^4,1^{4})$}{22221111}}
The brane configuration and 6d electric theory read
\begin{align}
 \raisebox{-.5\height}{
        \includegraphics[page=19]{figures/figures_app_SO12.pdf}
	}
	\quad \longrightarrow \quad
 \raisebox{-.5\height}{
    \includegraphics[page=20]{figures/figures_app_SO12.pdf}
	} 
	+\substack{
	2n{-}4\text{ nodes} \\
	Sp(2)/SO(12) }
\end{align}
compute Higgs branch dimension
\begin{align}
    \dim(\Higgs_{f}) &= 24 
    \;,\quad 
    \dim(\Higgs_\infty) 
    = n+52 
    =\dim(\Higgs_f) +29 +(n-1)\,. 
\end{align}

The finite coupling magnetic quiver is given by
\begin{align}
 &\raisebox{-.5\height}{
     \includegraphics[page=21]{figures/figures_app_SO12.pdf}
	} 
	\label{eq:magQuiv_2^4-1^4_finite} 
	\qquad
	\begin{cases}
	G_J &= \sorm(12)\\
	\dim\ \Coulomb &=24=\dim(\Higgs_f)     
	\end{cases}
\end{align}
\paragraph{Infinite Coupling.}
Moving the $8$ \De\ passed the two left most half \NS\ branes, accounting for brane creating as summarised in Appendix \ref{app:orientifolds}, one arrives at
\begin{align}
 \raisebox{-.5\height}{
      \includegraphics[page=22]{figures/figures_app_SO12.pdf}
	}
\end{align}
and the magnetic quiver becomes
\begin{align}
 &\raisebox{-.5\height}{
       \includegraphics[page=23]{figures/figures_app_SO12.pdf}
	} 
	\label{eq:magQuiv_2^4-1^4_infinite} 
	\qquad
	\begin{cases}
G_J &= \sorm(4) \times \sorm(12)  \\  
 \dim\ \Coulomb  &= n+52  = \dim\Higgs_\infty
	\end{cases}
\end{align}
\paragraph{Transitions.}
The difference between \eqref{eq:magQuiv_2^4-1^4_infinite} and \eqref{eq:magQuiv_2^4-1^4_finite} is given by one $E_8$ transition and $(n-1)$ $D_4$ transitions. This is straightforwardly verified by quiver subtraction, analogously to \eqref{eq:magQuiv_2^2-1^8_intermediate}.
%
%
\subsubsection{partition \texorpdfstring{$(3,2^2,1^{5})$}{32211111}}
Brane configuration
\begin{align}
 \raisebox{-.5\height}{
        \includegraphics[page=24]{figures/figures_app_SO12.pdf}
	}
	\quad \longrightarrow \quad
  \raisebox{-.5\height}{
        \includegraphics[page=25]{figures/figures_app_SO12.pdf}
	} 
	+\substack{
	2n{-}4\text{ nodes} \\
	Sp(2)/SO(12) }
\end{align}
compute Higgs branch dimension
\begin{align}
    \dim(\Higgs_{f})  = 22 
    \;,\quad 
    \dim(\Higgs_\infty) 
    = n+50 
    =\dim(\Higgs_f) +29+(n-1)\,. 
\end{align}
The finite coupling magnetic quiver is given by
\begin{align}
 &\raisebox{-.5\height}{
    \includegraphics[page=26]{figures/figures_app_SO12.pdf}
	} 
	\label{eq:magQuiv_3-2^2-1^5_finite} 
	\qquad 
\begin{cases}
G_J &= \sorm(2) \times\sorm(12) \\
 \dim\ \Coulomb &= 21 = \dim(\Higgs_f) -1
\end{cases}
\end{align}
which is one less than the classical Higgs branch. Note that \eqref{eq:magQuiv_3-2^2-1^5_finite} is the same as the finite coupling magnetic quiver \eqref{eq:magQuiv_3^2-1^6_finite} for partition $(3^2,1^6)$, because $(3,2^2,1^5)$ is a non-special partition.
\paragraph{Infinite Coupling.}
Moving the $8$ \De\ passed the three left most half \NS\ branes, accounting for 
brane 
creating as summarised in Appendix \ref{app:orientifolds}, one arrives at
\begin{align}
 \raisebox{-.5\height}{
    \includegraphics[page=27]{figures/figures_app_SO12.pdf}
	}
\end{align}
and the magnetic quiver becomes
\begin{align}
 \raisebox{-.5\height}{
     \includegraphics[page=28]{figures/figures_app_SO12.pdf}
	} 
	\label{eq:magQuiv_3-2^2-1^5_infinite}
	\qquad 
	\begin{cases}
	 \dim\Coulomb = n+ 49 = \dim\Higgs_\infty -1
	\end{cases}
\end{align}
However, $(3,2^2,1^5)$ is a non-special partition and one can see that the 
resulting magnetic quiver is identical to the one of $(3^2,1^6)$, because 
of \eqref{eq:LS_squared}. Therefore, the magnetic quiver will not capture the 
Higgs branch correctly, as for instance seen by inspecting the dimension and 
global symmetry.
%
%
\subsubsection{partition \texorpdfstring{$(2^{6})$}{222222}}
The brane configuration and 6d theory are 
\begin{align}
 \raisebox{-.5\height}{
      \includegraphics[page=29]{figures/figures_app_SO12.pdf}
	}
	\quad \longrightarrow \quad
 \raisebox{-.5\height}{
 \includegraphics[page=30]{figures/figures_app_SO12.pdf}
	} 
	+\substack{
	2n{-}4\text{ nodes} \\
	Sp(2)/SO(12) }
\end{align}
compute Higgs branch dimension
\begin{align} 
    \dim(\Higgs_{f})  =32
    \;,\quad 
    \dim(\Higgs_\infty)
    = n+51 
    = \dim(\Higgs_f) +(n-2)+21\,.
\end{align}
using \cite[eq.\ (9)]{Danielsson:1997kt} and $\dim(\mathrm{spin}_{\sorm(12)})=32$. The difference in dimension is consistent with the fact that there are $n-2$ curves of self-intersection $-4$, giving rise to 1-dimensional $D_4$ transitions. Also, there is one $-3$ curves, which contributes $21$ new hypermultiplets after collapse.

\paragraph{Infinite coupling.}
Moving the $6$ \De\ passed the two left most half \NS\ branes, accounting for 
brane 
creating as summarised in Appendix \ref{app:orientifolds}, one arrives at
\begin{align}
 \raisebox{-.5\height}{
  \includegraphics[page=31]{figures/figures_app_SO12.pdf}
	}
\end{align}
and the magnetic quiver becomes
\begin{align}
 &\raisebox{-.5\height}{
   \includegraphics[page=32]{figures/figures_app_SO12.pdf}
	} 
	\label{eq:magQuiv_2^6_infinite} 
	\qquad 
\begin{cases}
 G_J &=\sorm(12)   \\
 \dim\Coulomb &= n+51 = \dim\Higgs_\infty 
\end{cases}
\end{align}
%
%
\subsubsection{partition \texorpdfstring{$(3,2^4,1)$}{322221}}
The brane configuration and the 6d theory are
\begin{align}
 \raisebox{-.5\height}{
    \includegraphics[page=33]{figures/figures_app_SO12.pdf}
	}
	\quad \longrightarrow \quad
 \raisebox{-.5\height}{
     \includegraphics[page=34]{figures/figures_app_SO12.pdf}
	} 
	+\substack{
	2n{-}4\text{ nodes} \\
	Sp(2)/SO(12) }
\end{align}
compute Higgs branch dimension
\begin{align}
    \dim(\Higgs_{f})  =29
    \;,\quad 
    \dim(\Higgs_\infty) 
    = n+48 
    =\dim(\Higgs_f) +(n-2)+21\,,
\end{align}
using \cite[eq.\ (9)]{Danielsson:1997kt} and $\dim(\mathrm{spin}_{\sorm(11)})=32$. Again, the difference in dimension stems from $(n-2)$ $-4$ curves, each with a $D_4$ transition, and one $-3$ curves, with $21$ new hypermultiplets.
\paragraph{Infinite coupling.}
Moving the $6$ \De\ passed the three left most half \NS\ branes, accounting for brane creating as summarised in Appendix \ref{app:orientifolds}, one arrives at
\begin{align}
 \raisebox{-.5\height}{
      \includegraphics[page=35]{figures/figures_app_SO12.pdf}
	}
\end{align}
and the magnetic quiver becomes
\begin{align}
 \raisebox{-.5\height}{
       \includegraphics[page=36]{figures/figures_app_SO12.pdf}
	} 
	\label{eq:magQuiv_3-2^4-1_infinite}
	\qquad 
	\begin{cases}
	 \dim\Coulomb= n+46 = \dim\Higgs_\infty -2
	\end{cases}
\end{align}
However, $(3,2^4,1)$ is a non-special partition and one can see that the 
resulting magnetic quiver is identical to the one of $(3^2,2^2,1^2)$, because 
of \eqref{eq:LS_squared}. Therefore, the magnetic quiver will not capture the 
Higgs branch correctly, as for instance seen by inspecting the dimension and 
global symmetry.
%
%
\subsubsection{partition \texorpdfstring{$(3^2,1^6)$}{33111111}}
The brane configuration and the 6d theory read
\begin{align}
 \raisebox{-.5\height}{
        \includegraphics[page=37]{figures/figures_app_SO12.pdf}
	}
	\quad \longrightarrow \quad
 \raisebox{-.5\height}{
         \includegraphics[page=38]{figures/figures_app_SO12.pdf}
	} 
	+\substack{
	2n{-}4\text{ nodes} \\
	Sp(2)/SO(12) }
\end{align}
compute Higgs branch dimension
\begin{align}
    \dim(\Higgs_{f}) =21
    \;,\quad 
    \dim(\Higgs_\infty) 
    = n+49 
    =\dim \Higgs_f +29+(n+1)\,. 
\end{align}
The finite coupling magnetic quiver is given by
\begin{align}
 &\raisebox{-.5\height}{
          \includegraphics[page=39]{figures/figures_app_SO12.pdf}
	} 
	\label{eq:magQuiv_3^2-1^6_finite} 
	\qquad
\begin{cases}
G_J &= \sorm(2) \times \sorm(12) \\
     \dim\ \Coulomb &= 21 = \dim(\Higgs_f)  \,.
\end{cases}
\end{align}
\paragraph{Infinite coupling.}
Moving the $8$ \De\ passed the three left most half \NS\ branes, accounting for brane creating as summarised in Appendix \ref{app:orientifolds}, one arrives at
\begin{align}
 \raisebox{-.5\height}{
    \includegraphics[page=40]{figures/figures_app_SO12.pdf}
	}
\end{align}
and the magnetic quiver becomes
\begin{align}
 &\raisebox{-.5\height}{
     \includegraphics[page=41]{figures/figures_app_SO12.pdf}
	} 
	\label{eq:magQuiv_3^2-1^6_infinite} 
	\qquad
\begin{cases}
 G_J &= \sorm(6) \times \sorm(2) \times \sorm(12)  \\ 
 \dim\ \Coulomb &=  n+49 = \dim\ \Higgs_\infty 
\end{cases}
\end{align}
\paragraph{Transitions.}
The difference between \eqref{eq:magQuiv_3^2-1^6_infinite} and \eqref{eq:magQuiv_3^2-1^6_finite} is given by one $E_8$ transition and $(n-1)$ $D_4$ transitions. This is straightforwardly verified by quiver subtraction, analogously to \eqref{eq:magQuiv_2^2-1^8_intermediate}.
%
%
\subsubsection{partition \texorpdfstring{$(3^2,2^2,1^2)$}{332211}}
The brane configuration and the 6d electric theory read
\begin{align}
 \raisebox{-.5\height}{
 \includegraphics[page=42]{figures/figures_app_SO12.pdf}
	}
	\quad \longrightarrow \quad
 \raisebox{-.5\height}{
 \includegraphics[page=43]{figures/figures_app_SO12.pdf}
	} 
	+\substack{
	2n{-}4\text{ nodes} \\
	Sp(2)/SO(12) }
\end{align}
compute Higgs branch dimension
\begin{align}
    \dim(\Higgs_{f})  =27
    \;,\quad 
    \dim(\Higgs_\infty) 
    = n+46 
    = \dim(\Higgs_f) +(n-2) +21\,, 
\end{align}
using \cite[eq.\ (9)]{Danielsson:1997kt} and $\dim(\mathrm{Spin}_{\sorm(10)})=16$. The difference in dimension can be traced back to $(n-2)$ $-4$ curves, each with a $D_4$ transition, and one $-3$ curves, with $21$ new hypermultiplets.
\paragraph{Infinite coupling.}
Moving the $6$ \De\ passed the three left most half \NS\ branes, accounting for brane creating as summarised in Appendix \ref{app:orientifolds}, one arrives at
\begin{align}
 \raisebox{-.5\height}{
 \includegraphics[page=44]{figures/figures_app_SO12.pdf}
	}
\end{align}
and the magnetic quiver becomes
\begin{align}
 &\raisebox{-.5\height}{
 \includegraphics[page=45]{figures/figures_app_SO12.pdf}
	} 
	\label{eq:magQuiv_3^2-2^2-1^2_infinite} 
	\qquad 
\begin{cases}
 G_J &= \sorm(2) \times \sorm(12) \\
 \dim\ \Coulomb &= n+46 = \dim\ \Higgs_\infty
\end{cases}
\end{align}
%
%
\subsubsection{partition \texorpdfstring{$(5,1^7)$}{5,1,1,1,1,1,1,1}}
The brane configuration and the 6d electric theory read
\begin{align}
 &\raisebox{-.5\height}{
 \includegraphics[page=46]{figures/figures_app_SO12.pdf}
	} \\
	&\quad \longrightarrow \quad
 \raisebox{-.5\height}{
 \includegraphics[page=47]{figures/figures_app_SO12.pdf}
	} 
	+\substack{
	2n{-}6\text{ nodes} \\
	Sp(2)/SO(12) } \notag
\end{align}
compute Higgs branch dimension
\begin{align}
    \dim(\Higgs_{f})  =20
    \;,\quad 
    \dim(\Higgs_\infty) 
    = n+48 
    = \dim(\Higgs_f) +29 +(n-1)\,.
\end{align}
The finite coupling magnetic quiver is given by
\begin{align}
 &\raisebox{-.5\height}{
 \includegraphics[page=48]{figures/figures_app_SO12.pdf}
	} 
	\label{eq:magQuiv_5-1^7_finite} 
	\qquad 
\begin{cases}
G_J &= \sorm(12) \\
    \dim\ \Coulomb =& 20 = \dim(\Higgs_f)  
\end{cases}
\end{align}
\paragraph{Infinite coupling.}
Moving the $8$ \De\ passed the three left most half \NS\ branes, accounting for brane creating as summarised in Appendix \ref{app:orientifolds}, one arrives at
\begin{align}
 \raisebox{-.5\height}{
 \includegraphics[page=49]{figures/figures_app_SO12.pdf}
	}
\end{align}
and the magnetic quiver becomes
\begin{align}
 &\raisebox{-.5\height}{
  \includegraphics[page=50]{figures/figures_app_SO12.pdf}
	} 
	\label{eq:magQuiv_5-1^7_infinite} 
\qquad 
\begin{cases}
 G_J &= \sorm(7) \times \sorm(12) \\ 
 \dim\ \Coulomb &=  n+48 = \dim\ \Higgs_\infty 
\end{cases}
\end{align}
\paragraph{Transitions.}
The difference between \eqref{eq:magQuiv_5-1^7_infinite} and \eqref{eq:magQuiv_5-1^7_finite} is given by one $E_8$ transition and $(n-1)$ $D_4$ transitions. This is straightforwardly verified by quiver subtraction, analogously to \eqref{eq:magQuiv_2^2-1^8_intermediate}.
%
%
\subsubsection{partition \texorpdfstring{$(3^3,1^3)$}{3,3,3,1,1,1}}
The brane configuration and the 6d electric theory read
\begin{align}
 \raisebox{-.5\height}{
   \includegraphics[page=51]{figures/figures_app_SO12.pdf}
	}
	\qquad \longrightarrow \qquad
 \raisebox{-.5\height}{
    \includegraphics[page=52]{figures/figures_app_SO12.pdf}
	} 
	+\substack{
	2n{-}4\text{ nodes} \\
	Sp(2)/SO(12) }
\end{align}
compute Higgs branch dimension
\begin{align}
    \dim(\Higgs_{f})  =26
    \;,\quad 
    \dim(\Higgs_\infty) 
    = n+45 
    =\dim(\Higgs_f) +(n-2)+21\,,
\end{align}
using \cite[eq.\ (9)]{Danielsson:1997kt} and $\dim(\mathrm{Spin}_{\sorm(9)})=16$. Again, the increase in Higgs branch dimension is due to $n-2$ $D_4$ transitions plus one collapse of a $-3$ curve.
\paragraph{Infinite coupling.}
Moving the $6$ \De\ passed the three left most half \NS\ branes, accounting for 
brane 
creating as summarised in Appendix \ref{app:orientifolds}, one arrives at
\begin{align}
 \raisebox{-.5\height}{
     \includegraphics[page=53]{figures/figures_app_SO12.pdf}
	}
\end{align}
and the magnetic quiver becomes
\begin{align}
 &\raisebox{-.5\height}{
      \includegraphics[page=54]{figures/figures_app_SO12.pdf}
	} 
	\label{eq:magQuiv_3^3-1^3_infinite} 
	\qquad 
\begin{cases}
 G_J &= \sorm(3) \times \sorm(3) \times \sorm(12) \\ 
 \dim\ \Coulomb &=  n+45 = \dim\ \Higgs_\infty
\end{cases}
\end{align}
%
%
\subsubsection{partition \texorpdfstring{$(3^4)$}{3,3,3,3}}
The brane configuration and the 6d electric theory read
\begin{align}
 \raisebox{-.5\height}{
 \includegraphics[page=55]{figures/figures_app_SO12.pdf}
	}
	\quad \longrightarrow \quad
 \raisebox{-.5\height}{
  \includegraphics[page=56]{figures/figures_app_SO12.pdf}
	} 
	+\substack{
	2n{-}4\text{ nodes} \\
	Sp(2)/SO(12) }
\end{align}
compute Higgs branch dimension
\begin{align}
    \dim(\Higgs_{f}) =25
    \;,\quad 
    \dim(\Higgs_\infty) 
    = n+44  
    = \dim(\Higgs_f) +(n-2)+21\,,
\end{align}
using that $\sorm(7)$ \emph{can only be Higgsed} to $\surm(3)$. Also \cite[above eq.\ (9)]{Danielsson:1997kt} and $\dim(\mathrm{Spin}_{\sorm(7)})=8$ has been used. The by now familiar difference of $(n-2)+21$ is due to $(n-2)$ $D_4$ transitions and $21$ new hypermultiplets from a single $-3$ curve.
\paragraph{Infinite coupling.}
Moving the $4$ \De\ passed the three left most half \NS\ branes, accounting for 
brane 
creating as summarised in Appendix \ref{app:orientifolds}, one arrives at
\begin{align}
 \raisebox{-.5\height}{
   \includegraphics[page=57]{figures/figures_app_SO12.pdf}
	}
\end{align}
and the magnetic quiver becomes
\begin{align}
 &\raisebox{-.5\height}{
    \includegraphics[page=58]{figures/figures_app_SO12.pdf}
	} 
	\label{eq:magQuiv_3^4_infinite} 
	\qquad 
	\begin{cases}
G_J &= \sorm(4) \times \sorm(12) \\ 
 \dim\ \Coulomb &= n+44 = \dim\ \Higgs_\infty \,.	
	\end{cases}
\end{align}
%
%
\subsubsection{partition \texorpdfstring{$(5,2^2,1^3)$}{522111}}
The brane configuration and 6d electric theory read 
\begin{align}
 &\raisebox{-.5\height}{
     \includegraphics[page=59]{figures/figures_app_SO12.pdf}
	} \\
	&\quad \longrightarrow \quad
  \raisebox{-.5\height}{
  \includegraphics[page=60]{figures/figures_app_SO12.pdf}
	} 
	+\substack{
	2n{-}6\text{ nodes} \\
	Sp(2)/SO(12) } \notag
\end{align}
compute Higgs branch dimension
\begin{align}
    \dim(\Higgs_{f}) =25
    \;,\quad 
    \dim(\Higgs_\infty) 
    = n+44 
    = \dim(\Higgs_f) + (n+19)\,,
\end{align}
using \cite[eq.\ (9)]{Danielsson:1997kt} and $\dim(\mathrm{Spin}_{\sorm(9)})=16$. The jump in Higgs branch dimensions follows from the $(n-2)$ $-4$ curves and a single $-3$ curve. 
\paragraph{Infinite coupling.}
Moving the $6$ \De\ passed the six left most half \NS\ branes, accounting for 
brane 
creating as summarised in Appendix \ref{app:orientifolds}, one arrives at
\begin{align}
 \raisebox{-.5\height}{
   \includegraphics[page=61]{figures/figures_app_SO12.pdf}
	}
\end{align}
and the magnetic quiver becomes
\begin{align}
 \raisebox{-.5\height}{
 \includegraphics[page=62]{figures/figures_app_SO12.pdf}
	} 
	\label{eq:magQuiv_5-2^2-1^3_infinite}
	\qquad 
	\begin{cases}
	     \dim\ \Coulomb =n+43 = \dim\ \Higgs_\infty -1
	\end{cases}
\end{align}
However, $(5,2^2,1^3)$ is a non-special partition and one can see that the 
resulting magnetic quiver is identical to the one of $(5,3,1^4)$, because 
of \eqref{eq:LS_squared}. Therefore, the magnetic quiver will not capture the 
Higgs branch correctly, as for instance seen by inspecting the dimension and 
global symmetry.
%
%
\subsubsection{partition \texorpdfstring{$(4^2,1^4)$}{441111}}
The brane configuration and the 6d electric quiver read
\begin{align}
 &\raisebox{-.5\height}{
 \includegraphics[page=63]{figures/figures_app_SO12.pdf}
	} \\
	&\quad \longrightarrow \quad
  \raisebox{-.5\height}{
  \includegraphics[page=64]{figures/figures_app_SO12.pdf}
	} 
	+\substack{
	2n{-}6\text{ nodes} \\
	Sp(2)/SO(12) } \notag
\end{align}
compute Higgs branch dimension
\begin{align}
    \dim(\Higgs_{f})  =25
    \;,\quad 
    \dim(\Higgs_\infty) 
    = n+44 
    = \dim(\Higgs_f) +(n-2)+21 \,,
\end{align}
using \cite[eq.\ (9)]{Danielsson:1997kt} and $\dim(\mathrm{Spin}_{\sorm(8)})=8$. As above, the $(n-2)+21$ new dimensions are accounted for by the $(n-2)$ $D_4$ transitions and $21$ new hypermultiplets from the single $-3$ curve.
\paragraph{Infinite coupling.}
Moving the $6$ \De\ passed the four left most half \NS\ branes, accounting for 
brane 
creating as summarised in Appendix \ref{app:orientifolds}, one arrives at
\begin{align}
 \raisebox{-.5\height}{
   \includegraphics[page=65]{figures/figures_app_SO12.pdf}
	}
\end{align}
and the magnetic quiver becomes
\begin{align}
 &\raisebox{-.5\height}{
    \includegraphics[page=66]{figures/figures_app_SO12.pdf}
	} 
	\label{eq:magQuiv_4^2-2^4_infinite} 
	\qquad 
\begin{cases}
 G_J &= \sorm(4) \times \sorm(12) \\
 \dim\ \Coulomb&= n+44 = \dim\ \Higgs_\infty \;.
\end{cases}
\end{align}
%
%
\subsubsection{partition \texorpdfstring{$(4^2,2^2)$}{4422}}
The brane configuration and the 6d electric theory read
\begin{align}
 &\raisebox{-.5\height}{
     \includegraphics[page=67]{figures/figures_app_SO12.pdf}
	} \\
	&\quad \longrightarrow \quad
  \raisebox{-.5\height}{
       \includegraphics[page=68]{figures/figures_app_SO12.pdf}
	} 
	+\substack{
	2n{-}6\text{ nodes} \\
	Sp(2)/SO(12) } \notag
\end{align}
compute Higgs branch dimension
\begin{align}
    \dim(\Higgs_{f})  =24
    \;,\quad 
    \dim(\Higgs_\infty) 
    = n+43 
    = \dim(\Higgs_f) + (n-2)+21\,, 
\end{align}
assuming that $\sorm(7)$ \emph{can only be Higgsed} to $\surm(3)$. Also, \cite[above eq.\ (9)]{Danielsson:1997kt} and $\dim(\mathrm{Spin}_{\sorm(7)})=8$ has been used. The increase in dimension  follows from the same logic as above.
\paragraph{Infinite coupling.}
Moving the $4$ \De\ passed the four left most half \NS\ branes, accounting for 
brane 
creating as summarised in Appendix \ref{app:orientifolds}, one arrives at
\begin{align}
 \raisebox{-.5\height}{
        \includegraphics[page=69]{figures/figures_app_SO12.pdf}
	}
\end{align}
and the magnetic quiver becomes
\begin{align}
 &\raisebox{-.5\height}{
\includegraphics[page=70]{figures/figures_app_SO12.pdf}
	}  
	\qquad 
	\begin{cases}
G_J &=  \sorm(12) \\ 
 \dim\ \Coulomb &= n+43 = \dim\ \Higgs_\infty \,.	 
	\end{cases}
\end{align}
%
%
\subsubsection{partition \texorpdfstring{$(4^2,3,1)$}{4431}}
The brane configuration and the 6d electric theory read
\begin{align}
 &\raisebox{-.5\height}{
 \includegraphics[page=71]{figures/figures_app_SO12.pdf}
	}\\
	&\quad \longrightarrow \quad
  \raisebox{-.5\height}{
  \includegraphics[page=72]{figures/figures_app_SO12.pdf}
	} 
	+\substack{
	2n{-}6\text{ nodes} \\
	Sp(2)/SO(12) } \notag
\end{align}
compute Higgs branch dimension
\begin{align}
    \dim(\Higgs_{f})  =23
    \;,\quad 
    \dim(\Higgs_\infty) 
    = n+42  
    =\dim(\Higgs_f) +(n-2)+21 \,,
\end{align}
using that $G_2$ \emph{can only be Higgsed} to $\surm(3)$. The jump in Higgs branch dimension is clear.
\paragraph{Infinite coupling.}
Moving the $4$ \De\ passed the four left most half \NS\ branes, accounting for 
brane 
creating as summarised in Appendix \ref{app:orientifolds}, one arrives at
\begin{align}
 \raisebox{-.5\height}{
   \includegraphics[page=73]{figures/figures_app_SO12.pdf}
	}
\end{align}
and the magnetic quiver becomes
\begin{align}
 &\raisebox{-.5\height}{
   \includegraphics[page=74]{figures/figures_app_SO12.pdf}
	} 
	\label{eq:magQuiv_4^2-3-1_infinite} 
\qquad 
\begin{cases}
 G_J &=  \sorm(12)  \\
 \dim\ \Coulomb &= n+42 = \dim\ \Higgs_\infty \;.
\end{cases}
\end{align}
%
%
\subsubsection{partition \texorpdfstring{$(5,3,1^4)$}{531111}}
The brane configuration and the 6d electric theory read
\begin{align}
 &\raisebox{-.5\height}{
    \includegraphics[page=75]{figures/figures_app_SO12.pdf}
	} \\
	&\quad \rightarrow \quad
\raisebox{-.5\height}{
    \includegraphics[page=76]{figures/figures_app_SO12.pdf}
	} 
	+\substack{
	2n{-}6\text{ nodes} \\
	Sp(2)/SO(12) } \notag
\end{align}
compute Higgs branch dimension
\begin{align}
    \dim(\Higgs_{f}) =24
    \;,\quad 
    \dim(\Higgs_\infty) 
    = n+43 
    = \dim(\Higgs_f) +(n-2)+21\,, 
\end{align}
using \cite[eq.\ (9)]{Danielsson:1997kt} and $\dim(\mathrm{Spin}_{\sorm(8)})=8$. Again, the jump in dimensions is easily accounted for.
\paragraph{Infinite coupling.}
Moving the $6$ \De\ passed the five left most half \NS\ branes, accounting for 
brane 
creating as summarised in Appendix \ref{app:orientifolds}, one arrives at
\begin{align}
 \raisebox{-.5\height}{
     \includegraphics[page=77]{figures/figures_app_SO12.pdf}
	}
\end{align}
and the magnetic quiver becomes
\begin{align}
 &\raisebox{-.5\height}{
      \includegraphics[page=78]{figures/figures_app_SO12.pdf}
	} 
	\label{eq:magQuiv_5-3-1^4_infinite} 
	\qquad 
\begin{cases}
 G_J &= \sorm(4) \times \sorm(12) \\
 \dim\ \Coulomb &= n+43 =\dim\ \Higgs_\infty \;.
\end{cases}
\end{align}
%
%
\subsubsection{partition \texorpdfstring{$(5,3,2^2)$}{5322}}
The brane configuration and the 6d electric theory read
\begin{align}
 &\raisebox{-.5\height}{
    \includegraphics[page=79]{figures/figures_app_SO12.pdf}
	} \\
	&\quad \longrightarrow \quad
 \raisebox{-.5\height}{
     \includegraphics[page=80]{figures/figures_app_SO12.pdf}
	} 
	+\substack{
	2n{-}6\text{ nodes} \\
	Sp(2)/SO(12) } \qquad
\end{align}
compute Higgs branch dimension
\begin{align}
    \dim(\Higgs_{f})  =23
    \;,\quad 
    \dim(\Higgs_\infty) 
    = n+42 
    = \dim(\Higgs_f) + (n-2)+21\,,
\end{align}
using \cite[above eq.\ (9)]{Danielsson:1997kt} as well as $\dim(\mathrm{Spin}_{\sorm(7)})=8$, and recalling that $\sorm(7)$ \emph{can only be Higgsed} to $\surm(3)$.
\paragraph{Infinite coupling.}
Moving the $4$ \De\ passed the five left most half \NS\ branes, accounting for 
brane 
creating as summarised in Appendix \ref{app:orientifolds}, one arrives at
\begin{align}
 \raisebox{-.5\height}{
      \includegraphics[page=81]{figures/figures_app_SO12.pdf}
	}
\end{align}
and the magnetic quiver becomes
\begin{align}
 &\raisebox{-.5\height}{
       \includegraphics[page=82]{figures/figures_app_SO12.pdf}
	} 
	\label{eq:magQuiv_5-3-2^2_infinite} 
	\qquad 
	\begin{cases}
G_J &= \sorm(12)  \\
 \dim\ \Coulomb &= n+42 = \dim\ \Higgs_\infty \;.
	\end{cases}
\end{align}
%
%
\subsubsection{partition \texorpdfstring{$(5,3^2,1)$}{5,3,3,1}}
The brane configuration and the 6d electric theory read
\begin{align}
 &\raisebox{-.5\height}{
        \includegraphics[page=83]{figures/figures_app_SO12.pdf}
	}\\
	&\quad \longrightarrow \quad
  \raisebox{-.5\height}{
          \includegraphics[page=84]{figures/figures_app_SO12.pdf}
	} 
	+\substack{
	2n{-}6\text{ nodes} \\
	Sp(2)/SO(12) } \notag
\end{align}
compute Higgs branch dimension
\begin{align}
    \dim(\Higgs_{f}) =22
    \;,\quad 
    \dim(\Higgs_\infty) 
    = n+41 
    = \dim(\Higgs_f) +(n-2)+21\,. \notag
\end{align}
assuming that $G_2$ \emph{can only be Higgsed} to $\surm(3)$.
\paragraph{Infinite coupling.}
Moving the $4$ \De\ passed the five left most half \NS\ branes, accounting for 
brane creating as summarised in Appendix \ref{app:orientifolds}, one arrives at
\begin{align}
 \raisebox{-.5\height}{
\includegraphics[page=85]{figures/figures_app_SO12.pdf}
	}
\end{align}
and the magnetic quiver becomes
\begin{align}
 &\raisebox{-.5\height}{
 \includegraphics[page=86]{figures/figures_app_SO12.pdf}
	} 
	\label{eq:magQuiv_5-3^2-1_infinite} 
	\qquad 
	\begin{cases}
G_J &= \sorm(2) \times \sorm(12) \\
 \dim\ \Coulomb  &= n+41 =\dim\ \Higgs_\infty \;.	
	\end{cases}
\end{align}
%
%
\subsubsection{partition \texorpdfstring{$(5^2,1^2)$}{5,5,1,1}}
The brane configuration and the electric theory read
\begin{align}
 &\raisebox{-.5\height}{
  \includegraphics[page=87]{figures/figures_app_SO12.pdf}
	} \\
	&\quad \longrightarrow \quad
 \raisebox{-.5\height}{
   \includegraphics[page=88]{figures/figures_app_SO12.pdf}
	} 
	+\substack{
	2n{-}6\text{ nodes} \\
	Sp(2)/SO(12) } \notag
\end{align}
compute Higgs branch dimension
\begin{align}
    \dim(\Higgs_{f})  =21
    \;,\quad 
    \dim(\Higgs_\infty) 
    = n+40  
    = \dim(\Higgs_f) +(n-2)+21\,,
\end{align}
assuming that $\surm(3)$ \emph{cannot be Higgsed} any further.
\paragraph{Infinite coupling.}
Moving the $4$ \De\ passed the five left most half \NS\ branes, accounting for 
brane 
creating as summarised in Appendix \ref{app:orientifolds}, one arrives at
\begin{align}
 \raisebox{-.5\height}{
    \includegraphics[page=89]{figures/figures_app_SO12.pdf}
	}
\end{align}
and the magnetic quiver becomes
\begin{align}
 &\raisebox{-.5\height}{
 \includegraphics[page=90]{figures/figures_app_SO12.pdf}
	} 
	\label{eq:magQuiv_5^2-1^2_infinite} 
	\qquad 
\begin{cases}
G_J &= \sorm(2) \times \sorm(12) \\
 \dim\ \Coulomb  &= n+40 =\dim\ \Higgs_\infty \;.
\end{cases}
\end{align}
%
%
\subsubsection{partition \texorpdfstring{$(6^2)$}{66}}
The brane configuration and the 6d electric theory read
\begin{align}
 &\raisebox{-.5\height}{
  \includegraphics[page=91]{figures/figures_app_SO12.pdf}
	} \\
	&\quad \longrightarrow \quad
 \raisebox{-.5\height}{
   \includegraphics[page=92]{figures/figures_app_SO12.pdf}
	} 
	+\substack{
	2n{-}8\text{ nodes} \\
	Sp(2)/SO(12) } \notag
\end{align}
compute Higgs branch dimension
\begin{align}
    \dim(\Higgs_{f})  =21
    \;,\quad 
    \dim(\Higgs_\infty) 
    = n+39 
    = \dim(\Higgs_f) +(n-3)+21\,,
\end{align}
using that $\sorm(7)$ \emph{can only be Higgsed} to $\surm(3)$. The increase in Higgs branch dimension can be accounted for as follows: there are $n-3$ $-4$ curves, each with a 1-dimensional $D_4$ transition, plus a single $-3$ curves, which yields $21$ additional degrees of freedom.
\paragraph{Infinite coupling.}
Moving the $2$ \De\ passed the six left most half \NS\ branes, accounting for 
brane 
creating as summarised in Appendix \ref{app:orientifolds}, one arrives at
\begin{align}
 \raisebox{-.5\height}{
    \includegraphics[page=93]{figures/figures_app_SO12.pdf}
	}
\end{align}
and the magnetic quiver becomes
\begin{align}
 &\raisebox{-.5\height}{
     \includegraphics[page=94]{figures/figures_app_SO12.pdf}
	} 
	\label{eq:magQuiv_6^2_infinite} 
	\qquad 
\begin{cases}
 G_J &=  \sorm(12) \\
 \dim\ \Coulomb &= n+39 = \dim\ \Higgs_\infty \;.
\end{cases}
\end{align}
%
%
\subsubsection{partition \texorpdfstring{$(7,1^5)$}{7,1,1,1,1,1}}
The brane configuration and 6d theory are
\begin{align}
 &\raisebox{-.5\height}{
      \includegraphics[page=95]{figures/figures_app_SO12.pdf}
	} \\
	&\quad \longrightarrow \quad
 \raisebox{-.5\height}{
       \includegraphics[page=96]{figures/figures_app_SO12.pdf}
	} 
	+\substack{
	2n{-}8\text{ nodes} \\
	Sp(2)/SO(12) } 
	\notag
\end{align}
compute Higgs branch dimension
\begin{align}
    \dim(\Higgs_{f})  =23
    \;,\quad 
    \dim(\Higgs_\infty) 
    = n+42 
    = \dim(\Higgs_f) +n+19\,,
\end{align}
using that $\sorm(7)$ can only be Higgsed to $\surm(3)$ and using \cite[above eq.\ (9)]{Danielsson:1997kt}.
\paragraph{Infinite coupling.}
Moving the $6$ \De\ passed the seven left most half \NS\ branes, accounting for 
brane creating as summarised in Appendix \ref{app:orientifolds}, one arrives at
\begin{align}
 \raisebox{-.5\height}{
 \includegraphics[page=97]{figures/figures_app_SO12.pdf}
	}
\end{align}
and the magnetic quiver becomes
\begin{align}
 &\raisebox{-.5\height}{
  \includegraphics[page=98]{figures/figures_app_SO12.pdf}
	} 
	\label{eq:magQuiv_7-1^5_infinite} 
	\qquad 
\begin{cases}
 G_J &= \sorm(5) \times \sorm(12) \\
 \dim\ \Coulomb &= n+42 = \dim\ \Higgs_\infty \;.
\end{cases}
 \end{align}
%
%
\subsubsection{partition \texorpdfstring{$(7,2^2,1)$}{7221}}
The brane configuration and the 6d theory are
\begin{align}
 &\raisebox{-.5\height}{
   \includegraphics[page=99]{figures/figures_app_SO12.pdf}
	} \\
	&\quad \longrightarrow \quad
 \raisebox{-.5\height}{
 \includegraphics[page=100]{figures/figures_app_SO12.pdf}
	} 
	+\substack{
	2n{-}8\text{ nodes} \\
	Sp(2)/SO(12) } \notag
\end{align}
compute Higgs branch dimension
\begin{align}
    \dim(\Higgs_{f})  =21
    \;,\quad 
    \dim(\Higgs_\infty) 
    = n+40 
    =\dim(\Higgs_f) + n+19\,,
\end{align}
using that $G_2$ \emph{can only be Higgsed} to $\surm(3)$.
\paragraph{Infinite coupling.}
Moving the $4$ \De\ passed the seven left most half \NS\ branes, accounting for 
brane 
creating as summarised in Appendix \ref{app:orientifolds}, one arrives at
\begin{align}
 \raisebox{-.5\height}{
 \includegraphics[page=101]{figures/figures_app_SO12.pdf}
	}
\end{align}
and the magnetic quiver becomes
\begin{align}
 \raisebox{-.5\height}{
  \includegraphics[page=102]{figures/figures_app_SO12.pdf}
	} 
	\label{eq:magQuiv_7-2^2-1_infinite}
	\qquad 
	\begin{cases}
    \dim\Coulomb= n+39 = \dim\Higgs_\infty -1	
	\end{cases}
\end{align}
However, $(7,2^2,1)$ is a non-special partition and one can see that the 
resulting magnetic quiver is identical to the one of $(7,3,1^2)$, because 
of \eqref{eq:LS_squared}. Therefore, the magnetic quiver will not capture the 
Higgs branch correctly, as for instance seen by inspecting the dimension and 
global symmetry.
%
%
\subsubsection{partition \texorpdfstring{$(7,3,1^2)$}{7,3,1,1}}
The brane configuration and the electric theory are
\begin{align}
 &\raisebox{-.5\height}{
   \includegraphics[page=103]{figures/figures_app_SO12.pdf}
	} \\
	&\quad \longrightarrow \quad
 \raisebox{-.5\height}{
 \includegraphics[page=104]{figures/figures_app_SO12.pdf}
	} 
	+\substack{
	2n{-}8\text{ nodes} \\
	Sp(2)/SO(12) } \notag
\end{align}
compute Higgs branch dimension
\begin{align}
    \dim(\Higgs_{f})  =20
    \;,\quad 
    \dim(\Higgs_\infty) 
    = n+39 
    = \dim(\Higgs_f) +n+19\,,
\end{align}
using that $\surm(3)$ \emph{cannot be Higgsed} further.
\paragraph{Infinite coupling.}
Moving the $4$ \De\ passed the seven left most half \NS\ branes, accounting for 
brane 
creating as summarised in Appendix \ref{app:orientifolds}, one arrives at
\begin{align}
 \raisebox{-.5\height}{
  \includegraphics[page=105]{figures/figures_app_SO12.pdf}
	}
\end{align}
and the magnetic quiver becomes
\begin{align}
 &\raisebox{-.5\height}{
   \includegraphics[page=106]{figures/figures_app_SO12.pdf}
	} 
	\label{eq:magQuiv_7-3-1^2_infinite}
	\qquad 
	\begin{cases}
G_J &=  \sorm(12) \\
 \dim\ \Coulomb &= n+39 = \dim\ \Higgs_\infty \;.	
	\end{cases}
\end{align}
%
%
\subsubsection{partition \texorpdfstring{$(7,5)$}{7,5}}
The brane configuration and the 6d theory are given by
\begin{align}
 &\raisebox{-.5\height}{
    \includegraphics[page=107]{figures/figures_app_SO12.pdf}
	} \\
 &\quad \longrightarrow \quad
  \raisebox{-.5\height}{
      \includegraphics[page=108]{figures/figures_app_SO12.pdf}
	} 
	+\substack{
	2n{-}8\text{ nodes} \\
	Sp(2)/SO(12) } \notag
\end{align}
compute Higgs branch dimension
\begin{align}
    \dim(\Higgs_{f}) =20
    \;,\quad 
    \dim(\Higgs_\infty) 
    = n+38 
    = \dim(\Higgs_f) +n +18\,,
\end{align}
using that $\sorm(7)$ cannot be Higgsed further and using \cite[above eq.\ (9)]{Danielsson:1997kt}.
\paragraph{Infinite coupling.}
Moving the $2$ \De\ passed the seven left most half \NS\ branes, accounting for 
brane 
creating as summarised in Appendix \ref{app:orientifolds}, one arrives at
\begin{align}
 \raisebox{-.5\height}{
       \includegraphics[page=109]{figures/figures_app_SO12.pdf}
	}
\end{align}
and the magnetic quiver becomes
\begin{align}
 &\raisebox{-.5\height}{
        \includegraphics[page=110]{figures/figures_app_SO12.pdf}
	} 
	\label{eq:magQuiv_7-5_infinite}
	\qquad 
	\begin{cases}
G_J &=  \sorm(12) \\
 \dim\Coulomb  &= n+38 = \dim\Higgs_\infty	
	\end{cases}
\end{align}
%
%
\subsubsection{partition \texorpdfstring{$(9,1^3)$}{9,1,1,1}}
The brane configuration and the electric theory are given by
\begin{align}
 &\raisebox{-.5\height}{
         \includegraphics[page=111]{figures/figures_app_SO12.pdf}
	} \\
	&\quad \longrightarrow \quad
 \raisebox{-.5\height}{
\includegraphics[page=112]{figures/figures_app_SO12.pdf}
	} 
	+\substack{
	2n{-}10\text{ nodes} \\
	Sp(2)/SO(12) } \notag
\end{align}
compute Higgs branch dimension
\begin{align}
    \dim(\Higgs_{f}) =20
    \;,\quad 
    \dim(\Higgs_\infty) 
    = n+38 
    = \dim(\Higgs_f) +n+18\,,
\end{align}
using that $\sorm(7)$ can only be Higgsed to $\surm(3)$ and \cite[above eq.\ (9)]{Danielsson:1997kt}.
\paragraph{Infinite coupling.}
Moving the $4$ \De\ passed the nine left most half \NS\ branes, accounting for 
brane 
creating as summarised in Appendix \ref{app:orientifolds}, one arrives at
\begin{align}
 \raisebox{-.5\height}{
 \includegraphics[page=113]{figures/figures_app_SO12.pdf}
	}
\end{align}
and the magnetic quiver becomes
\begin{align}
 &\raisebox{-.5\height}{
  \includegraphics[page=114]{figures/figures_app_SO12.pdf}
	} 
	\label{eq:magQuiv_9-1^3_infinite} 
	\qquad 
	\begin{cases}
G_J &= \sorm(3)\times \sorm(12) \\
 \dim\Coulomb &= n+38 = \dim\Higgs_\infty \;.	
	\end{cases}
\end{align}
%
%
\subsubsection{partition \texorpdfstring{$(9,3)$}{9,3}}
The brane configuration and the 6d theory are given by
\begin{align}
 &\raisebox{-.5\height}{
   \includegraphics[page=115]{figures/figures_app_SO12.pdf}
	} \\
	&\quad \longrightarrow \quad
 \raisebox{-.5\height}{
    \includegraphics[page=116]{figures/figures_app_SO12.pdf}
	} 
	+\substack{
	2n{-}10\text{ nodes} \\
	Sp(2)/SO(12) } \notag
\end{align}
compute Higgs branch dimension
\begin{align}
    \dim(\Higgs_{f})  =19
    \;,\quad 
    \dim(\Higgs_\infty) 
    = n+37 
    =\dim(\Higgs_f) +n+18\,,
\end{align}
using that $G_2$ can only be Higgsed to $\surm(3)$.
\paragraph{Infinite coupling.}
Moving the $2$ \De\ passed the left most half \NS\ branes, accounting for 
brane 
creating as summarised in Appendix \ref{app:orientifolds}, one arrives at
\begin{align}
 \raisebox{-.5\height}{
     \includegraphics[page=117]{figures/figures_app_SO12.pdf}
	}
\end{align}
and the magnetic quiver becomes
\begin{align}
 &\raisebox{-.5\height}{
      \includegraphics[page=118]{figures/figures_app_SO12.pdf}
	} 
	\label{eq:magQuiv_9-3_infinite} 
	\qquad 
\begin{cases}
 G_J &=  \sorm(12) \\
 \dim\ \Coulomb &= n+37 =\dim\ \Higgs_\infty \,.
\end{cases}
\end{align}
%
%
\subsubsection{partition \texorpdfstring{$(11,1)$}{11,1}}
The brane configuration and the electric theory read
\begin{align}
 &\raisebox{-.5\height}{
       \includegraphics[page=119]{figures/figures_app_SO12.pdf}
	} \\
	&\quad \longrightarrow \quad
  \raisebox{-.5\height}{
         \includegraphics[page=120]{figures/figures_app_SO12.pdf}
	} 
	+\substack{
	2n{-}12\text{ nodes} \\
	Sp(2)/SO(12) } \notag
\end{align}
compute Higgs branch dimension
\begin{align}
    \dim(\Higgs_{f})  =19
    \;,\quad 
    \dim(\Higgs_\infty) 
    = n+36  
    =\dim(\Higgs_f) +n+17\,,
\end{align}
using that $G_2$ can only be Higgsed to $\surm(3)$.
\paragraph{Infinite coupling.}
Moving the $2$ \De\ passed the left most half \NS\ branes, accounting for 
brane 
creating as summarised in Appendix \ref{app:orientifolds}, one arrives at
\begin{align}
 \raisebox{-.5\height}{
    \includegraphics[page=121]{figures/figures_app_SO12.pdf}
	}
\end{align}
and the magnetic quiver becomes
\begin{align}
 &\raisebox{-.5\height}{
     \includegraphics[page=122]{figures/figures_app_SO12.pdf}
	} 
	\label{eq:magQuiv_11-1_infinite} 
	\qquad 
\begin{cases}
 G_J &= \sorm(12) \\
 \dim\Coulomb &= n+36 = \dim\Higgs_\infty \;.
\end{cases}
\end{align}
%
%
 \bibliographystyle{JHEP}     
 {\footnotesize{\bibliography{references}}}

\end{document}